\documentclass[a4paper,UKenglish]{lipics-v2016}

\usepackage[links]{latex/agda}
 
\usepackage{microtype}

\title{Modelling Bitcoin in Agda\footnote{Work supported by 
EPSRC grant EP/G033374/1
\emph{Theory and Applications of Induction Recursion}; CORCON FP7 Marie Curie International Research Project, PIRSES-GA-2013-612638;
COMPUTAL FP7 Marie Curie International Research Project, 
PIRSES-GA-2011-294962; 
CA COST Action CA15123 (EUTYPES)}}
\titlerunning{Modelling Bitcoin in Agda} 

\author[1]{Anton Setzer}
\affil[1]{Dept. of Computer Science, Swansea University, Singleton Park, Swansea SA2 8PP, UK\\
  \texttt{a.g.setzer@swansea.ac.uk}}
\authorrunning{A. Setzer} 

\Copyright{Anton Setzer}


\EventEditors{}
\EventLongTitle{}
\EventShortTitle{}
\EventAcronym{}
\EventYear{}
\EventDate{}
\EventLocation{}
\EventLogo{}
\SeriesVolume{}
\ArticleNo{}

\usepackage{tabularx}
\usepackage{float}
\usepackage{textgreek}
\usepackage{amssymb}

\newcommand{\eg}{e.g.}
\newcommand{\myparagraph}[1]{\smallskip\par\noindent{\bf #1}}
\newcommand{\Nbb}{\mathbb{N}}

\newcommand{\face}[6]
{\frame%
{\frametitle{#1}
\begin{center}
\begin{tabular}{@{}ll}
\begin{tabular}{@{}l}
\begin{tabular}{@{}l}\emph{#1} (#2)\end{tabular}
\\
\\
{\small\begin{tabular}{@{}l}#3\end{tabular}}
\\ 
\\
{\small\begin{tabular}{@{}l}#4\end{tabular}}
\end{tabular}
&
\begin{tabular}{@{}l}
{\includegraphics[height=#6]{#5}}
\end{tabular}
\end{tabular}
\end{center}
}}

\newcommand{\facetwopic}[8] 
{\frame%
{\frametitle{#1}
\begin{center}
\begin{tabular}{@{}ll}
\begin{tabular}{@{}l}
\begin{tabular}{@{}l}\emph{#1} (#2)\end{tabular}
\\
\\
{\small\begin{tabular}{@{}l}#3\end{tabular}}
\\ 
\\
{\small\begin{tabular}{@{}l}#4\end{tabular}}
\end{tabular}
&
\begin{tabular}{@{}l}
\only<1| handout:0>{\includegraphics[height=#6]{#5}}
\only<2->{\includegraphics[height=#8]{#7}}
\end{tabular}
\end{tabular}
\end{center}
}}

\newcommand{\facel}[6] 
{\frame%
{\frametitle{#1}
\begin{center}
\begin{tabular}{@{}ll}
\begin{tabular}{@{}l}
{\includegraphics[height=#6]{#5}}
\end{tabular}
&
\begin{tabular}{@{}l}
\begin{tabular}{@{}l}\emph{#1} (#2)\end{tabular}
\\
\\
{\small\begin{tabular}{@{}l}#3\end{tabular}}
\\ 
\\
{\small\begin{tabular}{@{}l}#4\end{tabular}}
\end{tabular}
\end{tabular}
\end{center}
}}

\newcommand{\facenopic}[4] 
{\frame%
{\frametitle{#1}
\begin{center}
\begin{tabular}{l}
\begin{tabular}{l}\emph{#1} (#2)\end{tabular}
\\
\\
{\small\begin{tabular}{l}#3\end{tabular}}
\\ 
\\
{\small\begin{tabular}{l}#4\end{tabular}}
\end{tabular}
\end{center}
}}

\newlength{\picheight}
\newcommand{\fourfaces}[8]
{%
\noindent%
\begin{tabular}[b]{@{}l@{}}
#1\\
{\includegraphics[height=\picheight]{#2}}
\end{tabular}
\hfill
\begin{tabular}[b]{@{}r@{}}
#3\\
{\includegraphics[height=\picheight]{#4}}
\end{tabular}\\[1ex]
\begin{tabular}[b]{@{}l@{}}
{\includegraphics[height=\picheight]{#6}}\\
#5
\end{tabular}
\hfill
\begin{tabular}[b]{@{}r@{}}
{\includegraphics[height=\picheight]{#8}}\\
#7
\end{tabular}
}

\newcommand{\fifthface}[2]
{%
\hspace*{\fill}\makebox[0ex]{%
\raisebox{50mm}[0pt][0pt]{%
\begin{tabular}[t]{@{}c@{}}
#1\\
{\includegraphics[height=\picheight]{#2}}
\end{tabular}%
}}\hspace*{\fill}%
}

\newcommand{\pictxttop}[3]
{\begin{tabular}[b]{@{}c@{}}
#1\\
{\includegraphics[height=#3]{#2}}
\end{tabular}}

\newcommand{\pictxttopl}[3]
{\begin{tabular}[b]{@{}l@{}}
#1\\
{\includegraphics[height=#3]{#2}}
\end{tabular}}

\newcommand{\pictxttopr}[3]
{\begin{tabular}[b]{@{}r@{}}
#1\\
{\includegraphics[height=#3]{#2}}
\end{tabular}}

\newcommand{\pictxtbot}[3]
{\begin{tabular}[t]{@{}c@{}}
{\includegraphics[height=#3]{#2}}\\
#1
\end{tabular}}

\newcommand{\pictxtbotl}[3]
{\begin{tabular}[t]{@{}l@{}}
{\includegraphics[height=#3]{#2}}\\
#1
\end{tabular}}

\DeclareUnicodeCharacter{"1D9C}{\ensuremath{{}^{\mathsf{c}}}}
\DeclareUnicodeCharacter{"1D62}{\ensuremath{{}_i}}
\DeclareUnicodeCharacter{"2092}{\ensuremath{{}_o}}
\DeclareUnicodeCharacter{"2099}{\ensuremath{{}_{n}}} 
\DeclareUnicodeCharacter{"2237}{\ensuremath{::}}
\DeclareUnicodeCharacter{"2238}{\ensuremath{\dotminus}}
\DeclareUnicodeCharacter{"2115}{\ensuremath{\Nbb}}

\newcommand{\agdaField}{} 
\newcommand{\dotminus}{{- \kern - 4.75pt {}^\cdot \kern 3pt}}

\begin{document}

\maketitle
\begin{abstract}
We present two models of the block chain of Bitcoin
in the interactive theorem prover
Agda. 
The first one is based on a simple model of
bank accounts, while having transactions with multiple inputs and outputs. 
The second model models transactions,
which refer directly to unspent transaction outputs, rather than
user accounts. The resulting blockchain
gives rise to a transaction tree.
That model is formalised using an extended form of induction-recursion,
one of the unique features of Agda.
The set of transaction trees and transactions
is defined inductively, while simultaneously recursively defining the list of 
unspent transaction  outputs.
Both structures model standard transactions, coinbase transactions, 
transaction fees, 
the exact message to be signed by those spending money
in a transaction,  block rewards,  
blocks, and the blockchain,
and the second structure models as well
maturation time for coinbase transactions and Merkle trees. 
Hashing and cryptographic operations  and their correctness are dealt
with abstractly by postulating corresponding operations. 
An indication is given how the correctness of this model could be specified
and proven in Agda.

\end{abstract}

\nocite{setzer:types2017budapest:abstract}


\AgdaHide{
\begin{code}%
\>[0]\AgdaKeyword{module}\AgdaSpace{}%
\AgdaModule{libraries.listLib}\AgdaSpace{}%
\AgdaKeyword{where}\<%
\\
\\
\>[0]\AgdaKeyword{open}\AgdaSpace{}%
\AgdaKeyword{import}\AgdaSpace{}%
\AgdaModule{Data.List}\<%
\\
\>[0]\AgdaKeyword{open}\AgdaSpace{}%
\AgdaKeyword{import}\AgdaSpace{}%
\AgdaModule{Data.Fin}\AgdaSpace{}%
\AgdaKeyword{hiding}\AgdaSpace{}%
\AgdaSymbol{(}\AgdaFunction{\_+\_}\AgdaSymbol{)}\<%
\\
\>[0]\AgdaKeyword{open}\AgdaSpace{}%
\AgdaKeyword{import}\AgdaSpace{}%
\AgdaModule{Data.Nat}\<%
\\
\>[0]\AgdaKeyword{open}\AgdaSpace{}%
\AgdaKeyword{import}\AgdaSpace{}%
\AgdaModule{Data.Bool}\<%
\\
\>[0]\AgdaKeyword{open}\AgdaSpace{}%
\AgdaKeyword{import}\AgdaSpace{}%
\AgdaModule{Data.Product}\<%
\\
\>[0]\AgdaKeyword{open}\AgdaSpace{}%
\AgdaKeyword{import}\AgdaSpace{}%
\AgdaModule{Data.Unit.Base}\<%
\\
\>[0]\AgdaKeyword{open}\AgdaSpace{}%
\AgdaKeyword{import}\AgdaSpace{}%
\AgdaModule{Function}\<%
\\
\>[0]\AgdaKeyword{open}\AgdaSpace{}%
\AgdaKeyword{import}\AgdaSpace{}%
\AgdaModule{Relation.Binary.PropositionalEquality}\<%
\\
\\
\\
\>[0]\AgdaFunction{mapL}\AgdaSpace{}%
\AgdaSymbol{:}\AgdaSpace{}%
\AgdaSymbol{\{}\AgdaBound{X}\AgdaSpace{}%
\AgdaBound{Y}\AgdaSpace{}%
\AgdaSymbol{:}\AgdaSpace{}%
\AgdaPrimitiveType{Set}\AgdaSymbol{\}(}\AgdaBound{\,f\,}\AgdaSpace{}%
\AgdaSymbol{:}\AgdaSpace{}%
\AgdaBound{X}\AgdaSpace{}%
\AgdaSymbol{→}\AgdaSpace{}%
\AgdaBound{Y}\AgdaSymbol{)(}\AgdaBound{l}\AgdaSpace{}%
\AgdaSymbol{:}\AgdaSpace{}%
\AgdaDatatype{List}\AgdaSpace{}%
\AgdaBound{X}\AgdaSymbol{)}\AgdaSpace{}%
\AgdaSymbol{→}\AgdaSpace{}%
\AgdaDatatype{List}\AgdaSpace{}%
\AgdaBound{Y}\<%
\\
\>[0]\AgdaFunction{mapL}\AgdaSpace{}%
\AgdaBound{\,f\,}\AgdaSpace{}%
\AgdaInductiveConstructor{[]}%
\>[16]\AgdaSymbol{=}\AgdaSpace{}%
\AgdaInductiveConstructor{[]}\<%
\\
\>[0]\AgdaFunction{mapL}\AgdaSpace{}%
\AgdaBound{\,f\,}\AgdaSpace{}%
\AgdaSymbol{(}\AgdaBound{x}\AgdaSpace{}%
\AgdaInductiveConstructor{∷}\AgdaSpace{}%
\AgdaBound{l}\AgdaSymbol{)}%
\>[16]\AgdaSymbol{=}\AgdaSpace{}%
\AgdaBound{\,f\,}\AgdaSpace{}%
\AgdaBound{x}\AgdaSpace{}%
\AgdaInductiveConstructor{∷}\AgdaSpace{}%
\AgdaFunction{mapL}\AgdaSpace{}%
\AgdaBound{\,f\,}\AgdaSpace{}%
\AgdaBound{l}\<%
\\
\\
\>[0]\AgdaFunction{corLengthMapL}\AgdaSpace{}%
\AgdaSymbol{:}\AgdaSpace{}%
\AgdaSymbol{\{}\AgdaBound{X}\AgdaSpace{}%
\AgdaBound{Y}\AgdaSpace{}%
\AgdaSymbol{:}\AgdaSpace{}%
\AgdaPrimitiveType{Set}\AgdaSymbol{\}(}\AgdaBound{\,f\,}\AgdaSpace{}%
\AgdaSymbol{:}\AgdaSpace{}%
\AgdaBound{X}\AgdaSpace{}%
\AgdaSymbol{→}\AgdaSpace{}%
\AgdaBound{Y}\AgdaSymbol{)(}\AgdaBound{l}\AgdaSpace{}%
\AgdaSymbol{:}\AgdaSpace{}%
\AgdaDatatype{List}\AgdaSpace{}%
\AgdaBound{X}\AgdaSymbol{)}\AgdaSpace{}%
\AgdaSymbol{→}\AgdaSpace{}%
\AgdaFunction{length}\AgdaSpace{}%
\AgdaSymbol{(}\AgdaFunction{mapL}\AgdaSpace{}%
\AgdaBound{\,f\,}\AgdaSpace{}%
\AgdaBound{l}\AgdaSymbol{)}\AgdaSpace{}%
\AgdaDatatype{≡}\AgdaSpace{}%
\AgdaFunction{length}\AgdaSpace{}%
\AgdaBound{l}\<%
\\
\>[0]\AgdaFunction{corLengthMapL}\AgdaSpace{}%
\AgdaBound{\,f\,}\AgdaSpace{}%
\AgdaInductiveConstructor{[]}\AgdaSpace{}%
\AgdaSymbol{=}\AgdaSpace{}%
\AgdaInductiveConstructor{refl}\<%
\\
\>[0]\AgdaFunction{corLengthMapL}\AgdaSpace{}%
\AgdaBound{\,f\,}\AgdaSpace{}%
\AgdaSymbol{(}\AgdaBound{x}\AgdaSpace{}%
\AgdaInductiveConstructor{∷}\AgdaSpace{}%
\AgdaBound{l}\AgdaSymbol{)}\AgdaSpace{}%
\AgdaSymbol{=}\AgdaSpace{}%
\AgdaFunction{cong}\AgdaSpace{}%
\AgdaInductiveConstructor{suc}\AgdaSpace{}%
\AgdaSymbol{(}\AgdaFunction{corLengthMapL}\AgdaSpace{}%
\AgdaBound{\,f\,}%
\>[53]\AgdaBound{l}\AgdaSymbol{)}\<%
\\
\\
\\
\\
\>[0]\AgdaFunction{nth}\AgdaSpace{}%
\AgdaSymbol{:}\AgdaSpace{}%
\AgdaSymbol{\{}\AgdaBound{X}\AgdaSpace{}%
\AgdaSymbol{:}\AgdaSpace{}%
\AgdaPrimitiveType{Set}\AgdaSymbol{\}(}\AgdaBound{l}\AgdaSpace{}%
\AgdaSymbol{:}\AgdaSpace{}%
\AgdaDatatype{List}\AgdaSpace{}%
\AgdaBound{X}\AgdaSymbol{)}\AgdaSpace{}%
\AgdaSymbol{(}\AgdaBound{i}\AgdaSpace{}%
\AgdaSymbol{:}\AgdaSpace{}%
\AgdaDatatype{Fin}\AgdaSpace{}%
\AgdaSymbol{(}\AgdaFunction{length}\AgdaSpace{}%
\AgdaBound{l}\AgdaSymbol{))}\AgdaSpace{}%
\AgdaSymbol{→}\AgdaSpace{}%
\AgdaBound{X}\<%
\\
\>[0]\AgdaFunction{nth}\AgdaSpace{}%
\AgdaInductiveConstructor{[]}\AgdaSpace{}%
\AgdaSymbol{()}\<%
\\
\>[0]\AgdaFunction{nth}\AgdaSpace{}%
\AgdaSymbol{(}\AgdaBound{x}\AgdaSpace{}%
\AgdaInductiveConstructor{∷}\AgdaSpace{}%
\AgdaBound{l}\AgdaSymbol{)}\AgdaSpace{}%
\AgdaInductiveConstructor{zero}\AgdaSpace{}%
\AgdaSymbol{=}\AgdaSpace{}%
\AgdaBound{x}\<%
\\
\>[0]\AgdaFunction{nth}\AgdaSpace{}%
\AgdaSymbol{(}\AgdaBound{x}\AgdaSpace{}%
\AgdaInductiveConstructor{∷}\AgdaSpace{}%
\AgdaBound{l}\AgdaSymbol{)}\AgdaSpace{}%
\AgdaSymbol{(}\AgdaInductiveConstructor{suc}\AgdaSpace{}%
\AgdaBound{i}\AgdaSymbol{)}\AgdaSpace{}%
\AgdaSymbol{=}\AgdaSpace{}%
\AgdaFunction{nth}\AgdaSpace{}%
\AgdaBound{l}\AgdaSpace{}%
\AgdaBound{i}\<%
\\
\\
\>[0]\AgdaFunction{delFromList}\AgdaSpace{}%
\AgdaSymbol{:}\AgdaSpace{}%
\AgdaSymbol{\{}\AgdaBound{X}\AgdaSpace{}%
\AgdaSymbol{:}\AgdaSpace{}%
\AgdaPrimitiveType{Set}\AgdaSymbol{\}(}\AgdaBound{l}\AgdaSpace{}%
\AgdaSymbol{:}\AgdaSpace{}%
\AgdaDatatype{List}\AgdaSpace{}%
\AgdaBound{X}\AgdaSymbol{)(}\AgdaBound{i}\AgdaSpace{}%
\AgdaSymbol{:}\AgdaSpace{}%
\AgdaDatatype{Fin}\AgdaSpace{}%
\AgdaSymbol{(}\AgdaFunction{length}\AgdaSpace{}%
\AgdaBound{l}\AgdaSymbol{))}\AgdaSpace{}%
\AgdaSymbol{→}\AgdaSpace{}%
\AgdaDatatype{List}\AgdaSpace{}%
\AgdaBound{X}\<%
\\
\>[0]\AgdaFunction{delFromList}\AgdaSpace{}%
\AgdaInductiveConstructor{[]}\AgdaSpace{}%
\AgdaSymbol{()}\<%
\\
\>[0]\AgdaFunction{delFromList}\AgdaSpace{}%
\AgdaSymbol{(}\AgdaBound{x}\AgdaSpace{}%
\AgdaInductiveConstructor{∷}\AgdaSpace{}%
\AgdaBound{l}\AgdaSymbol{)}\AgdaSpace{}%
\AgdaInductiveConstructor{zero}\AgdaSpace{}%
\AgdaSymbol{=}\AgdaSpace{}%
\AgdaBound{l}\<%
\\
\>[0]\AgdaFunction{delFromList}\AgdaSpace{}%
\AgdaSymbol{(}\AgdaBound{x}\AgdaSpace{}%
\AgdaInductiveConstructor{∷}\AgdaSpace{}%
\AgdaBound{l}\AgdaSymbol{)}\AgdaSpace{}%
\AgdaSymbol{(}\AgdaInductiveConstructor{suc}\AgdaSpace{}%
\AgdaBound{i}\AgdaSymbol{)}\AgdaSpace{}%
\AgdaSymbol{=}\AgdaSpace{}%
\AgdaBound{x}\AgdaSpace{}%
\AgdaInductiveConstructor{∷}\AgdaSpace{}%
\AgdaFunction{delFromList}\AgdaSpace{}%
\AgdaBound{l}\AgdaSpace{}%
\AgdaBound{i}\<%
\\
\\
\>[0]\AgdaComment{{-}{-} an index of (delFromList l i) is mapped to an index of l}\<%
\\
\>[0]\AgdaFunction{delFromListIndexToOrigIndex}\AgdaSpace{}%
\AgdaSymbol{:}\AgdaSpace{}%
\AgdaSymbol{\{}\AgdaBound{X}%
\>[147I]\AgdaSymbol{:}%
\>[148I]\AgdaPrimitiveType{Set}\AgdaSymbol{\}(}\AgdaBound{l}\AgdaSpace{}%
\AgdaSymbol{:}\AgdaSpace{}%
\AgdaDatatype{List}\AgdaSpace{}%
\AgdaBound{X}\AgdaSymbol{)(}\AgdaBound{i}\AgdaSpace{}%
\AgdaSymbol{:}\AgdaSpace{}%
\AgdaDatatype{Fin}\AgdaSpace{}%
\AgdaSymbol{(}\AgdaFunction{length}\AgdaSpace{}%
\AgdaBound{l}\AgdaSymbol{))}\<%
\\
\>[147I][@{}l@{\AgdaIndent{0}}]\<[148I]%
\>[35]\AgdaSymbol{(}\AgdaBound{j}\AgdaSpace{}%
\AgdaSymbol{:}\AgdaSpace{}%
\AgdaDatatype{Fin}\AgdaSpace{}%
\AgdaSymbol{(}\AgdaFunction{length}\AgdaSpace{}%
\AgdaSymbol{(}\AgdaFunction{delFromList}\AgdaSpace{}%
\AgdaBound{l}\AgdaSpace{}%
\AgdaBound{i}\AgdaSymbol{)))}\<%
\\
\>[147I][@{}l@{\AgdaIndent{0}}]%
\>[35]\AgdaSymbol{→}\AgdaSpace{}%
\AgdaDatatype{Fin}\AgdaSpace{}%
\AgdaSymbol{(}\AgdaFunction{length}\AgdaSpace{}%
\AgdaBound{l}\AgdaSymbol{)}\<%
\\
\>[0]\AgdaFunction{delFromListIndexToOrigIndex}\AgdaSpace{}%
\AgdaInductiveConstructor{[]}\AgdaSpace{}%
\AgdaSymbol{()}\AgdaSpace{}%
\AgdaBound{j}\<%
\\
\>[0]\AgdaFunction{delFromListIndexToOrigIndex}\AgdaSpace{}%
\AgdaSymbol{(}\AgdaBound{x}\AgdaSpace{}%
\AgdaInductiveConstructor{∷}\AgdaSpace{}%
\AgdaBound{l}\AgdaSymbol{)}\AgdaSpace{}%
\AgdaInductiveConstructor{zero}\AgdaSpace{}%
\AgdaBound{j}\AgdaSpace{}%
\AgdaSymbol{=}\AgdaSpace{}%
\AgdaInductiveConstructor{suc}\AgdaSpace{}%
\AgdaBound{j}\<%
\\
\>[0]\AgdaFunction{delFromListIndexToOrigIndex}\AgdaSpace{}%
\AgdaSymbol{(}\AgdaBound{x}\AgdaSpace{}%
\AgdaInductiveConstructor{∷}\AgdaSpace{}%
\AgdaBound{l}\AgdaSymbol{)}\AgdaSpace{}%
\AgdaSymbol{(}\AgdaInductiveConstructor{suc}\AgdaSpace{}%
\AgdaBound{i}\AgdaSymbol{)}\AgdaSpace{}%
\AgdaInductiveConstructor{zero}\AgdaSpace{}%
\AgdaSymbol{=}\AgdaSpace{}%
\AgdaInductiveConstructor{zero}\<%
\\
\>[0]\AgdaFunction{delFromListIndexToOrigIndex}\AgdaSpace{}%
\AgdaSymbol{(}\AgdaBound{x}\AgdaSpace{}%
\AgdaInductiveConstructor{∷}\AgdaSpace{}%
\AgdaBound{l}\AgdaSymbol{)}\AgdaSpace{}%
\AgdaSymbol{(}\AgdaInductiveConstructor{suc}\AgdaSpace{}%
\AgdaBound{i}\AgdaSymbol{)}\AgdaSpace{}%
\AgdaSymbol{(}\AgdaInductiveConstructor{suc}\AgdaSpace{}%
\AgdaBound{j}\AgdaSymbol{)}\AgdaSpace{}%
\AgdaSymbol{=}\AgdaSpace{}%
\AgdaInductiveConstructor{suc}\AgdaSpace{}%
\AgdaSymbol{(}\AgdaFunction{delFromListIndexToOrigIndex}\AgdaSpace{}%
\AgdaBound{l}\AgdaSpace{}%
\AgdaBound{i}\AgdaSpace{}%
\AgdaBound{j}\AgdaSymbol{)}\<%
\\
\\
\>[0]\AgdaFunction{correctNthDelFromList}%
\>[197I]\AgdaSymbol{:}%
\>[198I]\AgdaSymbol{\{}\AgdaBound{X}\AgdaSpace{}%
\AgdaSymbol{:}\AgdaSpace{}%
\AgdaPrimitiveType{Set}\AgdaSymbol{\}(}\AgdaBound{l}\AgdaSpace{}%
\AgdaSymbol{:}\AgdaSpace{}%
\AgdaDatatype{List}\AgdaSpace{}%
\AgdaBound{X}\AgdaSymbol{)(}\AgdaBound{i}\AgdaSpace{}%
\AgdaSymbol{:}\AgdaSpace{}%
\AgdaDatatype{Fin}\AgdaSpace{}%
\AgdaSymbol{(}\AgdaFunction{length}\AgdaSpace{}%
\AgdaBound{l}\AgdaSymbol{))}\<%
\\
\>[197I][@{}l@{\AgdaIndent{0}}]\<[198I]%
\>[24]\AgdaSymbol{(}\AgdaBound{j}\AgdaSpace{}%
\AgdaSymbol{:}\AgdaSpace{}%
\AgdaDatatype{Fin}\AgdaSpace{}%
\AgdaSymbol{(}\AgdaFunction{length}\AgdaSpace{}%
\AgdaSymbol{(}\AgdaFunction{delFromList}\AgdaSpace{}%
\AgdaBound{l}\AgdaSpace{}%
\AgdaBound{i}\AgdaSymbol{)))}\<%
\\
\>[197I][@{}l@{\AgdaIndent{0}}]%
\>[24]\AgdaSymbol{→}\AgdaSpace{}%
\AgdaFunction{nth}\AgdaSpace{}%
\AgdaSymbol{(}\AgdaFunction{delFromList}\AgdaSpace{}%
\AgdaBound{l}\AgdaSpace{}%
\AgdaBound{i}\AgdaSymbol{)}\AgdaSpace{}%
\AgdaBound{j}\AgdaSpace{}%
\AgdaDatatype{≡}\AgdaSpace{}%
\AgdaFunction{nth}\AgdaSpace{}%
\AgdaBound{l}\AgdaSpace{}%
\AgdaSymbol{(}\AgdaFunction{delFromListIndexToOrigIndex}\AgdaSpace{}%
\AgdaBound{l}\AgdaSpace{}%
\AgdaBound{i}\AgdaSpace{}%
\AgdaBound{j}\AgdaSymbol{)}\<%
\\
\>[0]\AgdaFunction{correctNthDelFromList}\AgdaSpace{}%
\AgdaInductiveConstructor{[]}\AgdaSpace{}%
\AgdaSymbol{()}\AgdaSpace{}%
\AgdaBound{j}\<%
\\
\>[0]\AgdaFunction{correctNthDelFromList}\AgdaSpace{}%
\AgdaSymbol{(}\AgdaBound{x}\AgdaSpace{}%
\AgdaInductiveConstructor{∷}\AgdaSpace{}%
\AgdaBound{l}\AgdaSymbol{)}\AgdaSpace{}%
\AgdaInductiveConstructor{zero}\AgdaSpace{}%
\AgdaBound{j}\AgdaSpace{}%
\AgdaSymbol{=}\AgdaSpace{}%
\AgdaInductiveConstructor{refl}\<%
\\
\>[0]\AgdaFunction{correctNthDelFromList}\AgdaSpace{}%
\AgdaSymbol{(}\AgdaBound{x}\AgdaSpace{}%
\AgdaInductiveConstructor{∷}\AgdaSpace{}%
\AgdaBound{l}\AgdaSymbol{)}\AgdaSpace{}%
\AgdaSymbol{(}\AgdaInductiveConstructor{suc}\AgdaSpace{}%
\AgdaBound{i}\AgdaSymbol{)}\AgdaSpace{}%
\AgdaInductiveConstructor{zero}\AgdaSpace{}%
\AgdaSymbol{=}\AgdaSpace{}%
\AgdaInductiveConstructor{refl}\<%
\\
\>[0]\AgdaFunction{correctNthDelFromList}\AgdaSpace{}%
\AgdaSymbol{(}\AgdaBound{x}\AgdaSpace{}%
\AgdaInductiveConstructor{∷}\AgdaSpace{}%
\AgdaBound{l}\AgdaSymbol{)}\AgdaSpace{}%
\AgdaSymbol{(}\AgdaInductiveConstructor{suc}\AgdaSpace{}%
\AgdaBound{i}\AgdaSymbol{)}\AgdaSpace{}%
\AgdaSymbol{(}\AgdaInductiveConstructor{suc}\AgdaSpace{}%
\AgdaBound{j}\AgdaSymbol{)}\AgdaSpace{}%
\AgdaSymbol{=}\AgdaSpace{}%
\AgdaFunction{correctNthDelFromList}\AgdaSpace{}%
\AgdaBound{l}\AgdaSpace{}%
\AgdaBound{i}\AgdaSpace{}%
\AgdaBound{j}\<%
\\
\\
\>[0]\AgdaComment{{-}{-}\textbackslash{}listLib}\<%
\end{code}
} 

\newcommand{\listLibconcatListIndextoOriginIndices}{
\begin{code}%
\>[0]\AgdaFunction{concatListIndex2OriginIndices}%
\>[256I]\AgdaSymbol{:}%
\>[257I]\AgdaSymbol{\{}\AgdaBound{X}\AgdaSpace{}%
\AgdaBound{Y}\AgdaSpace{}%
\AgdaSymbol{:}\AgdaSpace{}%
\AgdaPrimitiveType{Set}\AgdaSymbol{\}(}\AgdaBound{l}\AgdaSpace{}%
\AgdaBound{l'}\AgdaSpace{}%
\AgdaSymbol{:}\AgdaSpace{}%
\AgdaDatatype{List}\AgdaSpace{}%
\AgdaBound{X}\AgdaSymbol{)}\<%
\\
\>[256I][@{}l@{\AgdaIndent{0}}]\<[257I]%
\>[32]\AgdaSymbol{(}\AgdaBound{\,f\,}%
\>[37]\AgdaSymbol{:}%
\>[40]\AgdaDatatype{Fin}\AgdaSpace{}%
\AgdaSymbol{(}\AgdaFunction{length}\AgdaSpace{}%
\AgdaBound{l}\AgdaSymbol{)}\AgdaSpace{}%
\AgdaSymbol{→}\AgdaSpace{}%
\AgdaBound{Y}\AgdaSymbol{)(}\AgdaBound{f'}%
\>[64]\AgdaSymbol{:}%
\>[67]\AgdaDatatype{Fin}\AgdaSpace{}%
\AgdaSymbol{(}\AgdaFunction{length}\AgdaSpace{}%
\AgdaBound{l'}\AgdaSymbol{)}\AgdaSpace{}%
\AgdaSymbol{→}\AgdaSpace{}%
\AgdaBound{Y}\AgdaSymbol{)}\<%
\\
\>[256I][@{}l@{\AgdaIndent{0}}]%
\>[32]\AgdaSymbol{(}\AgdaBound{i}%
\>[37]\AgdaSymbol{:}\AgdaSpace{}%
\AgdaDatatype{Fin}\AgdaSpace{}%
\AgdaSymbol{(}\AgdaFunction{length}\AgdaSpace{}%
\AgdaSymbol{(}\AgdaBound{l}\AgdaSpace{}%
\AgdaFunction{\ensuremath{+\!\!+}}\AgdaSpace{}%
\AgdaBound{l'}\AgdaSymbol{)))}%
\>[64]\AgdaSymbol{→}\AgdaSpace{}%
\AgdaBound{Y}\<%
\end{code}
} 

\AgdaHide{
\begin{code}%
\>[0]\AgdaFunction{concatListIndex2OriginIndices}\AgdaSpace{}%
\AgdaInductiveConstructor{[]}\AgdaSpace{}%
\AgdaBound{l'}\AgdaSpace{}%
\AgdaBound{\,f\,}\AgdaSpace{}%
\AgdaBound{f'}\AgdaSpace{}%
\AgdaBound{i}\AgdaSpace{}%
\AgdaSymbol{=}\AgdaSpace{}%
\AgdaBound{f'}\AgdaSpace{}%
\AgdaBound{i}\<%
\\
\>[0]\AgdaFunction{concatListIndex2OriginIndices}\AgdaSpace{}%
\AgdaSymbol{(}\AgdaBound{x}\AgdaSpace{}%
\AgdaInductiveConstructor{∷}\AgdaSpace{}%
\AgdaBound{l}\AgdaSymbol{)}\AgdaSpace{}%
\AgdaBound{l'}\AgdaSpace{}%
\AgdaBound{\,f\,}\AgdaSpace{}%
\AgdaBound{f'}\AgdaSpace{}%
\AgdaInductiveConstructor{zero}\AgdaSpace{}%
\AgdaSymbol{=}\AgdaSpace{}%
\AgdaBound{\,f\,}\AgdaSpace{}%
\AgdaInductiveConstructor{zero}\<%
\\
\>[0]\AgdaFunction{concatListIndex2OriginIndices}\AgdaSpace{}%
\AgdaSymbol{(}\AgdaBound{x}\AgdaSpace{}%
\AgdaInductiveConstructor{∷}\AgdaSpace{}%
\AgdaBound{l}\AgdaSymbol{)}\AgdaSpace{}%
\AgdaBound{l'}\AgdaSpace{}%
\AgdaBound{\,f\,}\AgdaSpace{}%
\AgdaBound{f'}\AgdaSpace{}%
\AgdaSymbol{(}\AgdaInductiveConstructor{suc}\AgdaSpace{}%
\AgdaBound{i}\AgdaSymbol{)}\AgdaSpace{}%
\AgdaSymbol{=}\<%
\\
\>[0][@{}l@{\AgdaIndent{0}}]%
\>[9]\AgdaFunction{concatListIndex2OriginIndices}\AgdaSpace{}%
\AgdaBound{l}\AgdaSpace{}%
\AgdaBound{l'}\AgdaSpace{}%
\AgdaSymbol{(}\AgdaBound{\,f\,}\AgdaSpace{}%
\AgdaFunction{∘}\AgdaSpace{}%
\AgdaInductiveConstructor{suc}\AgdaSymbol{)}\AgdaSpace{}%
\AgdaBound{f'}\AgdaSpace{}%
\AgdaBound{i}\<%
\\
\\
\\
\>[0]\AgdaFunction{corCconcatListIndex2OriginIndices}%
\>[313I]\AgdaSymbol{:}%
\>[314I]\AgdaSymbol{\{}\AgdaBound{X}\AgdaSpace{}%
\AgdaBound{Y}\AgdaSpace{}%
\AgdaSymbol{:}\AgdaSpace{}%
\AgdaPrimitiveType{Set}\AgdaSymbol{\}(}\AgdaBound{l}\AgdaSpace{}%
\AgdaBound{l'}\AgdaSpace{}%
\AgdaSymbol{:}\AgdaSpace{}%
\AgdaDatatype{List}\AgdaSpace{}%
\AgdaBound{X}\AgdaSymbol{)}\<%
\\
\>[313I][@{}l@{\AgdaIndent{0}}]\<[314I]%
\>[36]\AgdaSymbol{(}\AgdaBound{\,f\,}\AgdaSpace{}%
\AgdaSymbol{:}\AgdaSpace{}%
\AgdaBound{X}\AgdaSpace{}%
\AgdaSymbol{→}\AgdaSpace{}%
\AgdaBound{Y}\AgdaSymbol{)}\<%
\\
\>[313I][@{}l@{\AgdaIndent{0}}]%
\>[36]\AgdaSymbol{(}\AgdaBound{g}\AgdaSpace{}%
\AgdaSymbol{:}\AgdaSpace{}%
\AgdaDatatype{Fin}\AgdaSpace{}%
\AgdaSymbol{(}\AgdaFunction{length}\AgdaSpace{}%
\AgdaBound{l}\AgdaSymbol{)}\AgdaSpace{}%
\AgdaSymbol{→}\AgdaSpace{}%
\AgdaBound{Y}\AgdaSymbol{)}\<%
\\
\>[313I][@{}l@{\AgdaIndent{0}}]%
\>[36]\AgdaSymbol{(}\AgdaBound{g'}\AgdaSpace{}%
\AgdaSymbol{:}\AgdaSpace{}%
\AgdaDatatype{Fin}\AgdaSpace{}%
\AgdaSymbol{(}\AgdaFunction{length}\AgdaSpace{}%
\AgdaBound{l'}\AgdaSymbol{)}\AgdaSpace{}%
\AgdaSymbol{→}\AgdaSpace{}%
\AgdaBound{Y}\AgdaSymbol{)}\<%
\\
\>[313I][@{}l@{\AgdaIndent{0}}]%
\>[36]\AgdaSymbol{(}\AgdaBound{cor1}\AgdaSpace{}%
\AgdaSymbol{:}\AgdaSpace{}%
\AgdaSymbol{(}\AgdaBound{i}\AgdaSpace{}%
\AgdaSymbol{:}\AgdaSpace{}%
\AgdaDatatype{Fin}\AgdaSpace{}%
\AgdaSymbol{(}\AgdaFunction{length}\AgdaSpace{}%
\AgdaBound{l}\AgdaSymbol{))}\AgdaSpace{}%
\AgdaSymbol{→}\AgdaSpace{}%
\AgdaBound{\,f\,}\AgdaSpace{}%
\AgdaSymbol{(}\AgdaFunction{nth}\AgdaSpace{}%
\AgdaBound{l}\AgdaSpace{}%
\AgdaBound{i}\AgdaSymbol{)}\AgdaSpace{}%
\AgdaDatatype{≡}\AgdaSpace{}%
\AgdaBound{g}\AgdaSpace{}%
\AgdaBound{i}\AgdaSymbol{)}\<%
\\
\>[313I][@{}l@{\AgdaIndent{0}}]%
\>[36]\AgdaSymbol{(}\AgdaBound{cor2}\AgdaSpace{}%
\AgdaSymbol{:}\AgdaSpace{}%
\AgdaSymbol{(}\AgdaBound{i'}\AgdaSpace{}%
\AgdaSymbol{:}\AgdaSpace{}%
\AgdaDatatype{Fin}\AgdaSpace{}%
\AgdaSymbol{(}\AgdaFunction{length}\AgdaSpace{}%
\AgdaBound{l'}\AgdaSymbol{))}\AgdaSpace{}%
\AgdaSymbol{→}\AgdaSpace{}%
\AgdaBound{\,f\,}\AgdaSpace{}%
\AgdaSymbol{(}\AgdaFunction{nth}\AgdaSpace{}%
\AgdaBound{l'}\AgdaSpace{}%
\AgdaBound{i'}\AgdaSymbol{)}\AgdaSpace{}%
\AgdaDatatype{≡}\AgdaSpace{}%
\AgdaBound{g'}\AgdaSpace{}%
\AgdaBound{i'}\AgdaSymbol{)}\<%
\\
\>[313I][@{}l@{\AgdaIndent{0}}]%
\>[36]\AgdaSymbol{(}\AgdaBound{i}\AgdaSpace{}%
\AgdaSymbol{:}\AgdaSpace{}%
\AgdaDatatype{Fin}\AgdaSpace{}%
\AgdaSymbol{(}\AgdaFunction{length}\AgdaSpace{}%
\AgdaSymbol{(}\AgdaBound{l}\AgdaSpace{}%
\AgdaFunction{\ensuremath{+\!\!+}}\AgdaSpace{}%
\AgdaBound{l'}\AgdaSymbol{)))}\<%
\\
\>[313I][@{}l@{\AgdaIndent{0}}]%
\>[36]\AgdaSymbol{→}%
\>[372I]\AgdaBound{\,f\,}\AgdaSpace{}%
\AgdaSymbol{(}\AgdaFunction{nth}\AgdaSpace{}%
\AgdaSymbol{(}\AgdaBound{l}\AgdaSpace{}%
\AgdaFunction{\ensuremath{+\!\!+}}\AgdaSpace{}%
\AgdaBound{l'}\AgdaSymbol{)}\AgdaSpace{}%
\AgdaBound{i}\AgdaSymbol{)}\<%
\\
\>[372I][@{}l@{\AgdaIndent{0}}]%
\>[39]\AgdaDatatype{≡}\AgdaSpace{}%
\AgdaFunction{concatListIndex2OriginIndices}\AgdaSpace{}%
\AgdaBound{l}\AgdaSpace{}%
\AgdaBound{l'}\AgdaSpace{}%
\AgdaBound{g}\AgdaSpace{}%
\AgdaBound{g'}\AgdaSpace{}%
\AgdaBound{i}\<%
\\
\>[0]\AgdaFunction{corCconcatListIndex2OriginIndices}\AgdaSpace{}%
\AgdaInductiveConstructor{[]}\AgdaSpace{}%
\AgdaBound{l'}\AgdaSpace{}%
\AgdaBound{\,f\,}\AgdaSpace{}%
\AgdaBound{g}\AgdaSpace{}%
\AgdaBound{g'}\AgdaSpace{}%
\AgdaBound{cor1}\AgdaSpace{}%
\AgdaBound{cor2}\AgdaSpace{}%
\AgdaBound{i}\AgdaSpace{}%
\AgdaSymbol{=}\AgdaSpace{}%
\AgdaBound{cor2}\AgdaSpace{}%
\AgdaBound{i}\<%
\\
\>[0]\AgdaFunction{corCconcatListIndex2OriginIndices}\AgdaSpace{}%
\AgdaSymbol{(}\AgdaBound{x}\AgdaSpace{}%
\AgdaInductiveConstructor{∷}\AgdaSpace{}%
\AgdaBound{l}\AgdaSymbol{)}\AgdaSpace{}%
\AgdaBound{l'}\AgdaSpace{}%
\AgdaBound{\,f\,}\AgdaSpace{}%
\AgdaBound{g}\AgdaSpace{}%
\AgdaBound{g'}\AgdaSpace{}%
\AgdaBound{cor1}\AgdaSpace{}%
\AgdaBound{cor2}\AgdaSpace{}%
\AgdaInductiveConstructor{zero}\AgdaSpace{}%
\AgdaSymbol{=}\AgdaSpace{}%
\AgdaBound{cor1}\AgdaSpace{}%
\AgdaInductiveConstructor{zero}\<%
\\
\>[0]\AgdaFunction{corCconcatListIndex2OriginIndices}\AgdaSpace{}%
\AgdaSymbol{(}\AgdaBound{x}\AgdaSpace{}%
\AgdaInductiveConstructor{∷}\AgdaSpace{}%
\AgdaBound{l}\AgdaSymbol{)}\AgdaSpace{}%
\AgdaBound{l'}\AgdaSpace{}%
\AgdaBound{\,f\,}\AgdaSpace{}%
\AgdaBound{g}\AgdaSpace{}%
\AgdaBound{g'}\AgdaSpace{}%
\AgdaBound{cor1}\AgdaSpace{}%
\AgdaBound{cor2}\AgdaSpace{}%
\AgdaSymbol{(}\AgdaInductiveConstructor{suc}\AgdaSpace{}%
\AgdaBound{i}\AgdaSymbol{)}\AgdaSpace{}%
\AgdaSymbol{=}\<%
\\
\>[0][@{}l@{\AgdaIndent{0}}]%
\>[4]\AgdaFunction{corCconcatListIndex2OriginIndices}\AgdaSpace{}%
\AgdaBound{l}\AgdaSpace{}%
\AgdaBound{l'}\AgdaSpace{}%
\AgdaBound{\,f\,}\AgdaSpace{}%
\AgdaSymbol{(}\AgdaBound{g}\AgdaSpace{}%
\AgdaFunction{∘}\AgdaSpace{}%
\AgdaInductiveConstructor{suc}\AgdaSymbol{)}\AgdaSpace{}%
\AgdaBound{g'}\AgdaSpace{}%
\AgdaSymbol{(}\AgdaBound{cor1}\AgdaSpace{}%
\AgdaFunction{∘}\AgdaSpace{}%
\AgdaInductiveConstructor{suc}\AgdaSymbol{)}\AgdaSpace{}%
\AgdaBound{cor2}\AgdaSpace{}%
\AgdaBound{i}\<%
\\
\\
\\
\>[0]\AgdaComment{{-}{-}\textbackslash{}listLib}\<%
\end{code}
} 

\newcommand{\listLiblistOfElementsOfFin}{
\begin{code}%
\>[0]\AgdaFunction{listOfElementsOfFin}\AgdaSpace{}%
\AgdaSymbol{:}\AgdaSpace{}%
\AgdaSymbol{(}\AgdaBound{n}\AgdaSpace{}%
\AgdaSymbol{:}\AgdaSpace{}%
\AgdaDatatype{ℕ}\AgdaSymbol{)}\AgdaSpace{}%
\AgdaSymbol{→}\AgdaSpace{}%
\AgdaDatatype{List}\AgdaSpace{}%
\AgdaSymbol{(}\AgdaDatatype{Fin}\AgdaSpace{}%
\AgdaBound{n}\AgdaSymbol{)}\<%
\\
\>[0]\AgdaFunction{listOfElementsOfFin}\AgdaSpace{}%
\AgdaInductiveConstructor{zero}\AgdaSpace{}%
\AgdaSymbol{=}\AgdaSpace{}%
\AgdaInductiveConstructor{[]}\<%
\\
\>[0]\AgdaFunction{listOfElementsOfFin}\AgdaSpace{}%
\AgdaSymbol{(}\AgdaInductiveConstructor{suc}\AgdaSpace{}%
\AgdaBound{n}\AgdaSymbol{)}\AgdaSpace{}%
\AgdaSymbol{=}\AgdaSpace{}%
\AgdaInductiveConstructor{zero}\AgdaSpace{}%
\AgdaInductiveConstructor{∷}\AgdaSpace{}%
\AgdaSymbol{(}\AgdaFunction{mapL}\AgdaSpace{}%
\AgdaInductiveConstructor{suc}\AgdaSpace{}%
\AgdaSymbol{(}\AgdaFunction{listOfElementsOfFin}\AgdaSpace{}%
\AgdaBound{n}\AgdaSymbol{))}\<%
\end{code}
} 

\AgdaHide{
\begin{code}%
\>[0]\<%
\\
\>[0]\AgdaFunction{corListOfElementsOfFinLength}\AgdaSpace{}%
\AgdaSymbol{:}\AgdaSpace{}%
\AgdaSymbol{(}\AgdaBound{n}\AgdaSpace{}%
\AgdaSymbol{:}\AgdaSpace{}%
\AgdaDatatype{ℕ}\AgdaSymbol{)}\AgdaSpace{}%
\AgdaSymbol{→}\AgdaSpace{}%
\AgdaFunction{length}\AgdaSpace{}%
\AgdaSymbol{(}\AgdaFunction{listOfElementsOfFin}\AgdaSpace{}%
\AgdaBound{n}\AgdaSymbol{)}\AgdaSpace{}%
\AgdaDatatype{≡}\AgdaSpace{}%
\AgdaBound{n}\<%
\\
\>[0]\AgdaFunction{corListOfElementsOfFinLength}\AgdaSpace{}%
\AgdaInductiveConstructor{zero}\AgdaSpace{}%
\AgdaSymbol{=}\AgdaSpace{}%
\AgdaInductiveConstructor{refl}\<%
\\
\>[0]\AgdaFunction{corListOfElementsOfFinLength}\AgdaSpace{}%
\AgdaSymbol{(}\AgdaInductiveConstructor{suc}\AgdaSpace{}%
\AgdaBound{n}\AgdaSymbol{)}\AgdaSpace{}%
\AgdaSymbol{=}\AgdaSpace{}%
\AgdaFunction{cong}\AgdaSpace{}%
\AgdaInductiveConstructor{suc}\AgdaSpace{}%
\AgdaFunction{cor3}\<%
\\
\>[0][@{}l@{\AgdaIndent{0}}]%
\>[4]\AgdaKeyword{where}\<%
\\
\>[4][@{}l@{\AgdaIndent{0}}]%
\>[9]\AgdaFunction{cor1}\AgdaSpace{}%
\AgdaSymbol{:}\AgdaSpace{}%
\AgdaFunction{length}\AgdaSpace{}%
\AgdaSymbol{(}\AgdaFunction{mapL}\AgdaSpace{}%
\AgdaSymbol{\{}\AgdaArgument{Y}\AgdaSpace{}%
\AgdaSymbol{=}\AgdaSpace{}%
\AgdaDatatype{Fin}\AgdaSpace{}%
\AgdaSymbol{(}\AgdaInductiveConstructor{suc}\AgdaSpace{}%
\AgdaBound{n}\AgdaSymbol{)\}}\AgdaSpace{}%
\AgdaSymbol{(λ}\AgdaSpace{}%
\AgdaBound{i}\AgdaSpace{}%
\AgdaSymbol{→}\AgdaSpace{}%
\AgdaInductiveConstructor{suc}\AgdaSpace{}%
\AgdaBound{i}\AgdaSymbol{)}\AgdaSpace{}%
\AgdaSymbol{(}\AgdaFunction{listOfElementsOfFin}\AgdaSpace{}%
\AgdaBound{n}\AgdaSymbol{))}\AgdaSpace{}%
\AgdaDatatype{≡}\AgdaSpace{}%
\AgdaFunction{length}\AgdaSpace{}%
\AgdaSymbol{(}\AgdaFunction{listOfElementsOfFin}\AgdaSpace{}%
\AgdaBound{n}\AgdaSymbol{)}\<%
\\
\>[4][@{}l@{\AgdaIndent{0}}]%
\>[9]\AgdaFunction{cor1}\AgdaSpace{}%
\AgdaSymbol{=}\AgdaSpace{}%
\AgdaFunction{corLengthMapL}\AgdaSpace{}%
\AgdaInductiveConstructor{suc}\AgdaSpace{}%
\AgdaSymbol{(}\AgdaFunction{listOfElementsOfFin}\AgdaSpace{}%
\AgdaBound{n}\AgdaSymbol{)}\<%
\\
\\
\>[4][@{}l@{\AgdaIndent{0}}]%
\>[9]\AgdaFunction{cor2}\AgdaSpace{}%
\AgdaSymbol{:}\AgdaSpace{}%
\AgdaFunction{length}\AgdaSpace{}%
\AgdaSymbol{(}\AgdaFunction{listOfElementsOfFin}\AgdaSpace{}%
\AgdaBound{n}\AgdaSymbol{)}\AgdaSpace{}%
\AgdaDatatype{≡}\AgdaSpace{}%
\AgdaBound{n}\<%
\\
\>[4][@{}l@{\AgdaIndent{0}}]%
\>[9]\AgdaFunction{cor2}\AgdaSpace{}%
\AgdaSymbol{=}\AgdaSpace{}%
\AgdaFunction{corListOfElementsOfFinLength}\AgdaSpace{}%
\AgdaBound{n}\<%
\\
\\
\>[4][@{}l@{\AgdaIndent{0}}]%
\>[9]\AgdaFunction{cor3}\AgdaSpace{}%
\AgdaSymbol{:}\AgdaSpace{}%
\AgdaFunction{length}\AgdaSpace{}%
\AgdaSymbol{(}\AgdaFunction{mapL}\AgdaSpace{}%
\AgdaSymbol{\{}\AgdaArgument{Y}\AgdaSpace{}%
\AgdaSymbol{=}\AgdaSpace{}%
\AgdaDatatype{Fin}\AgdaSpace{}%
\AgdaSymbol{(}\AgdaInductiveConstructor{suc}\AgdaSpace{}%
\AgdaBound{n}\AgdaSymbol{)\}}\AgdaSpace{}%
\AgdaSymbol{(λ}\AgdaSpace{}%
\AgdaBound{i}\AgdaSpace{}%
\AgdaSymbol{→}\AgdaSpace{}%
\AgdaInductiveConstructor{suc}\AgdaSpace{}%
\AgdaBound{i}\AgdaSymbol{)}\AgdaSpace{}%
\AgdaSymbol{(}\AgdaFunction{listOfElementsOfFin}\AgdaSpace{}%
\AgdaBound{n}\AgdaSymbol{))}\AgdaSpace{}%
\AgdaDatatype{≡}\AgdaSpace{}%
\AgdaBound{n}\<%
\\
\>[4][@{}l@{\AgdaIndent{0}}]%
\>[9]\AgdaFunction{cor3}\AgdaSpace{}%
\AgdaSymbol{=}\AgdaSpace{}%
\AgdaFunction{trans}\AgdaSpace{}%
\AgdaFunction{cor1}\AgdaSpace{}%
\AgdaFunction{cor2}\<%
\\
\\
\>[0]\AgdaComment{{-}{-} subtract list consists of elements from the list which are about to}\<%
\\
\>[0]\AgdaComment{{-}{-} be subtracted from it.}\<%
\\
\>[0]\AgdaComment{{-}{-} every element of the list can be subtracted only once}\<%
\\
\>[0]\AgdaComment{{-}{-} however since elements can occur multiple times they can still occur}\<%
\\
\>[0]\AgdaComment{{-}{-} multiple times (as many times as they occur in the list) from the list}\<%
\\
\\
\>[0]\AgdaComment{{-}{-}\textbackslash{}listLib}\<%
\end{code}
} 

\newcommand{\listLibSubList}{
\begin{code}%
\>[0]\AgdaKeyword{data}\AgdaSpace{}%
\AgdaDatatype{SubList}\AgdaSpace{}%
\AgdaSymbol{\{}\AgdaBound{X}\AgdaSpace{}%
\AgdaSymbol{:}\AgdaSpace{}%
\AgdaPrimitiveType{Set}\AgdaSymbol{\}}\AgdaSpace{}%
\AgdaSymbol{:}\AgdaSpace{}%
\AgdaSymbol{(}\AgdaBound{l}\AgdaSpace{}%
\AgdaSymbol{:}\AgdaSpace{}%
\AgdaDatatype{List}\AgdaSpace{}%
\AgdaBound{X}\AgdaSymbol{)}\AgdaSpace{}%
\AgdaSymbol{→}\AgdaSpace{}%
\AgdaPrimitiveType{Set}\AgdaSpace{}%
\AgdaKeyword{where}\<%
\\
\>[0][@{}l@{\AgdaIndent{0}}]%
\>[2]\AgdaInductiveConstructor{[]}%
\>[8]\AgdaSymbol{:}%
\>[11]\AgdaSymbol{\{}\AgdaBound{l}\AgdaSpace{}%
\AgdaSymbol{:}\AgdaSpace{}%
\AgdaDatatype{List}\AgdaSpace{}%
\AgdaBound{X}\AgdaSymbol{\}}\AgdaSpace{}%
\AgdaSymbol{→}\AgdaSpace{}%
\AgdaDatatype{SubList}\AgdaSpace{}%
\AgdaBound{l}\<%
\\
\>[0][@{}l@{\AgdaIndent{0}}]%
\>[2]\AgdaInductiveConstructor{cons}%
\>[8]\AgdaSymbol{:}%
\>[11]\AgdaSymbol{\{}\AgdaBound{l}\AgdaSpace{}%
\AgdaSymbol{:}\AgdaSpace{}%
\AgdaDatatype{List}\AgdaSpace{}%
\AgdaBound{X}\AgdaSymbol{\}(}\AgdaBound{i}\AgdaSpace{}%
\AgdaSymbol{:}\AgdaSpace{}%
\AgdaDatatype{Fin}\AgdaSpace{}%
\AgdaSymbol{(}\AgdaFunction{length}\AgdaSpace{}%
\AgdaBound{l}\AgdaSymbol{))(}\AgdaBound{o}\AgdaSpace{}%
\AgdaSymbol{:}\AgdaSpace{}%
\AgdaDatatype{SubList}\AgdaSpace{}%
\AgdaSymbol{(}\AgdaFunction{delFromList}\AgdaSpace{}%
\AgdaBound{l}\AgdaSpace{}%
\AgdaBound{i}\AgdaSymbol{))}\AgdaSpace{}%
\AgdaSymbol{→}\AgdaSpace{}%
\AgdaDatatype{SubList}\AgdaSpace{}%
\AgdaBound{l}\<%
\end{code}
} 

\AgdaHide{
\begin{code}%
\>[0]\<%
\\
\>[0]\AgdaComment{{-}{-}\textbackslash{}listLib}\<%
\end{code}
} 

\newcommand{\listLiblistMinusSubList}{
\begin{code}%
\>[0]\AgdaFunction{listMinusSubList}\AgdaSpace{}%
\AgdaSymbol{:}\AgdaSpace{}%
\AgdaSymbol{\{}\AgdaBound{X}\AgdaSpace{}%
\AgdaSymbol{:}\AgdaSpace{}%
\AgdaPrimitiveType{Set}\AgdaSymbol{\}(}\AgdaBound{l}\AgdaSpace{}%
\AgdaSymbol{:}\AgdaSpace{}%
\AgdaDatatype{List}\AgdaSpace{}%
\AgdaBound{X}\AgdaSymbol{)(}\AgdaBound{o}\AgdaSpace{}%
\AgdaSymbol{:}\AgdaSpace{}%
\AgdaDatatype{SubList}\AgdaSpace{}%
\AgdaBound{l}\AgdaSymbol{)}\AgdaSpace{}%
\AgdaSymbol{→}\AgdaSpace{}%
\AgdaDatatype{List}\AgdaSpace{}%
\AgdaBound{X}\<%
\\
\>[0]\AgdaFunction{listMinusSubList}\AgdaSpace{}%
\AgdaBound{l}\AgdaSpace{}%
\AgdaInductiveConstructor{[]}%
\>[31]\AgdaSymbol{=}\AgdaSpace{}%
\AgdaBound{l}\<%
\\
\>[0]\AgdaFunction{listMinusSubList}\AgdaSpace{}%
\AgdaBound{l}\AgdaSpace{}%
\AgdaSymbol{(}\AgdaInductiveConstructor{cons}\AgdaSpace{}%
\AgdaBound{i}\AgdaSpace{}%
\AgdaBound{o}\AgdaSymbol{)}%
\>[31]\AgdaSymbol{=}\AgdaSpace{}%
\AgdaFunction{listMinusSubList}\AgdaSpace{}%
\AgdaSymbol{(}\AgdaFunction{delFromList}\AgdaSpace{}%
\AgdaBound{l}\AgdaSpace{}%
\AgdaBound{i}\AgdaSymbol{)}\AgdaSpace{}%
\AgdaBound{o}\<%
\\
\\
\>[0]\AgdaFunction{subList2List}\AgdaSpace{}%
\AgdaSymbol{:}\AgdaSpace{}%
\AgdaSymbol{\{}\AgdaBound{X}\AgdaSpace{}%
\AgdaSymbol{:}\AgdaSpace{}%
\AgdaPrimitiveType{Set}\AgdaSymbol{\}\{}\AgdaBound{l}\AgdaSpace{}%
\AgdaSymbol{:}\AgdaSpace{}%
\AgdaDatatype{List}\AgdaSpace{}%
\AgdaBound{X}\AgdaSymbol{\}(}\AgdaBound{sl}\AgdaSpace{}%
\AgdaSymbol{:}\AgdaSpace{}%
\AgdaDatatype{SubList}\AgdaSpace{}%
\AgdaBound{l}\AgdaSymbol{)}\AgdaSpace{}%
\AgdaSymbol{→}\AgdaSpace{}%
\AgdaDatatype{List}\AgdaSpace{}%
\AgdaBound{X}\<%
\\
\>[0]\AgdaFunction{subList2List}\AgdaSpace{}%
\AgdaInductiveConstructor{[]}%
\>[34]\AgdaSymbol{=}\AgdaSpace{}%
\AgdaInductiveConstructor{[]}\<%
\\
\>[0]\AgdaFunction{subList2List}\AgdaSpace{}%
\AgdaSymbol{\{}\AgdaArgument{l}\AgdaSpace{}%
\AgdaSymbol{=}\AgdaSpace{}%
\AgdaBound{l}\AgdaSymbol{\}}\AgdaSpace{}%
\AgdaSymbol{(}\AgdaInductiveConstructor{cons}\AgdaSpace{}%
\AgdaBound{i}\AgdaSpace{}%
\AgdaBound{sl}\AgdaSymbol{)}%
\>[34]\AgdaSymbol{=}\AgdaSpace{}%
\AgdaFunction{nth}\AgdaSpace{}%
\AgdaBound{l}\AgdaSpace{}%
\AgdaBound{i}\AgdaSpace{}%
\AgdaInductiveConstructor{∷}\AgdaSpace{}%
\AgdaFunction{subList2List}\AgdaSpace{}%
\AgdaBound{sl}\<%
\end{code}
} 

\AgdaHide{
\begin{code}%
\>[0]\<%
\\
\>[0]\AgdaComment{{-}{-}\textbackslash{}listLib}\<%
\end{code}
} 

\newcommand{\listLibSubListplus}{
\begin{code}%
\>[0]\AgdaKeyword{data}\AgdaSpace{}%
\AgdaDatatype{SubList+}\AgdaSpace{}%
\AgdaSymbol{\{}\AgdaBound{X}\AgdaSpace{}%
\AgdaSymbol{:}\AgdaSpace{}%
\AgdaPrimitiveType{Set}\AgdaSymbol{\}}\AgdaSpace{}%
\AgdaSymbol{(}\AgdaBound{Y}\AgdaSpace{}%
\AgdaSymbol{:}\AgdaSpace{}%
\AgdaPrimitiveType{Set}\AgdaSymbol{)}\AgdaSpace{}%
\AgdaSymbol{:}\AgdaSpace{}%
\AgdaSymbol{(}\AgdaBound{l}\AgdaSpace{}%
\AgdaSymbol{:}\AgdaSpace{}%
\AgdaDatatype{List}\AgdaSpace{}%
\AgdaBound{X}\AgdaSymbol{)}\AgdaSpace{}%
\AgdaSymbol{→}\AgdaSpace{}%
\AgdaPrimitiveType{Set}\AgdaSpace{}%
\AgdaKeyword{where}\<%
\\
\>[0][@{}l@{\AgdaIndent{0}}]%
\>[2]\AgdaInductiveConstructor{[]}%
\>[7]\AgdaSymbol{:}%
\>[10]\AgdaSymbol{\{}\AgdaBound{l}\AgdaSpace{}%
\AgdaSymbol{:}\AgdaSpace{}%
\AgdaDatatype{List}\AgdaSpace{}%
\AgdaBound{X}\AgdaSymbol{\}}\AgdaSpace{}%
\AgdaSymbol{→}\AgdaSpace{}%
\AgdaDatatype{SubList+}\AgdaSpace{}%
\AgdaBound{Y}\AgdaSpace{}%
\AgdaBound{l}\<%
\\
\>[0][@{}l@{\AgdaIndent{0}}]%
\>[2]\AgdaInductiveConstructor{cons}%
\>[632I]\AgdaSymbol{:}%
\>[10]\AgdaSymbol{\{}\AgdaBound{l}\AgdaSpace{}%
\AgdaSymbol{:}\AgdaSpace{}%
\AgdaDatatype{List}\AgdaSpace{}%
\AgdaBound{X}\AgdaSymbol{\}(}\AgdaBound{i}\AgdaSpace{}%
\AgdaSymbol{:}\AgdaSpace{}%
\AgdaDatatype{Fin}\AgdaSpace{}%
\AgdaSymbol{(}\AgdaFunction{length}\AgdaSpace{}%
\AgdaBound{l}\AgdaSymbol{))(}\AgdaBound{y}\AgdaSpace{}%
\AgdaSymbol{:}\AgdaSpace{}%
\AgdaBound{Y}\AgdaSymbol{)(}\AgdaBound{o}\AgdaSpace{}%
\AgdaSymbol{:}\AgdaSpace{}%
\AgdaDatatype{SubList+}\AgdaSpace{}%
\AgdaBound{Y}\AgdaSpace{}%
\AgdaSymbol{(}\AgdaFunction{delFromList}\AgdaSpace{}%
\AgdaBound{l}\AgdaSpace{}%
\AgdaBound{i}\AgdaSymbol{))}\<%
\\
\>[632I][@{}l@{\AgdaIndent{0}}]%
\>[10]\AgdaSymbol{→}\AgdaSpace{}%
\AgdaDatatype{SubList+}\AgdaSpace{}%
\AgdaBound{Y}\AgdaSpace{}%
\AgdaBound{l}\<%
\end{code}
} 

\AgdaHide{
\begin{code}%
\>[0]\<%
\\
\>[0]\AgdaComment{{-}{-}\textbackslash{}listLib}\<%
\end{code}
} 

\newcommand{\listLiblistMinusSubListplus}{
\begin{code}%
\>[0]\AgdaFunction{listMinusSubList+}\AgdaSpace{}%
\AgdaSymbol{:}\AgdaSpace{}%
\AgdaSymbol{\{}\AgdaBound{X}\AgdaSpace{}%
\AgdaBound{Y}\AgdaSpace{}%
\AgdaSymbol{:}\AgdaSpace{}%
\AgdaPrimitiveType{Set}\AgdaSymbol{\}(}\AgdaBound{l}\AgdaSpace{}%
\AgdaSymbol{:}\AgdaSpace{}%
\AgdaDatatype{List}\AgdaSpace{}%
\AgdaBound{X}\AgdaSymbol{)(}\AgdaBound{o}\AgdaSpace{}%
\AgdaSymbol{:}\AgdaSpace{}%
\AgdaDatatype{SubList+}\AgdaSpace{}%
\AgdaBound{Y}\AgdaSpace{}%
\AgdaBound{l}\AgdaSymbol{)}\AgdaSpace{}%
\AgdaSymbol{→}\AgdaSpace{}%
\AgdaDatatype{List}\AgdaSpace{}%
\AgdaBound{X}\<%
\end{code}
} 

\AgdaHide{
\begin{code}%
\>[0]\<%
\\
\>[0]\AgdaFunction{listMinusSubList+}\AgdaSpace{}%
\AgdaBound{l}\AgdaSpace{}%
\AgdaInductiveConstructor{[]}\AgdaSpace{}%
\AgdaSymbol{=}\AgdaSpace{}%
\AgdaBound{l}\<%
\\
\>[0]\AgdaFunction{listMinusSubList+}\AgdaSpace{}%
\AgdaBound{l}\AgdaSpace{}%
\AgdaSymbol{(}\AgdaInductiveConstructor{cons}\AgdaSpace{}%
\AgdaBound{i}\AgdaSpace{}%
\AgdaBound{y}\AgdaSpace{}%
\AgdaBound{o}\AgdaSymbol{)}\AgdaSpace{}%
\AgdaSymbol{=}\AgdaSpace{}%
\AgdaFunction{listMinusSubList+}\AgdaSpace{}%
\AgdaSymbol{(}\AgdaFunction{delFromList}\AgdaSpace{}%
\AgdaBound{l}\AgdaSpace{}%
\AgdaBound{i}\AgdaSymbol{)}\AgdaSpace{}%
\AgdaBound{o}\<%
\\
\\
\\
\>[0]\AgdaComment{{-}{-} needed fromutxo2Msg}\<%
\\
\>[0]\AgdaComment{{-}{-}\textbackslash{}listLib}\<%
\end{code}
} 

\newcommand{\listLibsubListplustoList}{
\begin{code}%
\>[0]\AgdaFunction{subList+2List}\AgdaSpace{}%
\AgdaSymbol{:}\AgdaSpace{}%
\AgdaSymbol{\{}\AgdaBound{X}\AgdaSpace{}%
\AgdaBound{Y}\AgdaSpace{}%
\AgdaSymbol{:}\AgdaSpace{}%
\AgdaPrimitiveType{Set}\AgdaSymbol{\}\{}\AgdaBound{l}\AgdaSpace{}%
\AgdaSymbol{:}\AgdaSpace{}%
\AgdaDatatype{List}\AgdaSpace{}%
\AgdaBound{X}\AgdaSymbol{\}(}\AgdaBound{sl}\AgdaSpace{}%
\AgdaSymbol{:}\AgdaSpace{}%
\AgdaDatatype{SubList+}\AgdaSpace{}%
\AgdaBound{Y}\AgdaSpace{}%
\AgdaBound{l}\AgdaSymbol{)}\AgdaSpace{}%
\AgdaSymbol{→}\AgdaSpace{}%
\AgdaDatatype{List}\AgdaSpace{}%
\AgdaSymbol{(}\AgdaBound{X}\AgdaSpace{}%
\AgdaFunction{×}\AgdaSpace{}%
\AgdaBound{Y}\AgdaSymbol{)}\<%
\end{code}
} 

\AgdaHide{
\begin{code}%
\>[0]\<%
\\
\>[0]\AgdaFunction{subList+2List}\AgdaSpace{}%
\AgdaInductiveConstructor{[]}\AgdaSpace{}%
\AgdaSymbol{=}\AgdaSpace{}%
\AgdaInductiveConstructor{[]}\<%
\\
\>[0]\AgdaFunction{subList+2List}\AgdaSpace{}%
\AgdaSymbol{\{}\AgdaBound{X}\AgdaSymbol{\}}\AgdaSpace{}%
\AgdaSymbol{\{}\AgdaBound{Y}\AgdaSymbol{\}}\AgdaSpace{}%
\AgdaSymbol{\{}\AgdaBound{l}\AgdaSymbol{\}}\AgdaSpace{}%
\AgdaSymbol{(}\AgdaInductiveConstructor{cons}\AgdaSpace{}%
\AgdaBound{i}\AgdaSpace{}%
\AgdaBound{y}\AgdaSpace{}%
\AgdaBound{sl}\AgdaSymbol{)}\AgdaSpace{}%
\AgdaSymbol{=}\AgdaSpace{}%
\AgdaSymbol{(}\AgdaFunction{nth}\AgdaSpace{}%
\AgdaBound{l}\AgdaSpace{}%
\AgdaBound{i}\AgdaSpace{}%
\AgdaInductiveConstructor{,}\AgdaSpace{}%
\AgdaBound{y}\AgdaSymbol{)}\AgdaSpace{}%
\AgdaInductiveConstructor{∷}\AgdaSpace{}%
\AgdaFunction{subList+2List}\AgdaSpace{}%
\AgdaBound{sl}\<%
\\
\\
\>[0]\AgdaComment{{-}{-}\textbackslash{}listLib}\<%
\end{code}
} 

\newcommand{\listLiblistMinusSubListplustoOrgIndex}{
\begin{code}%
\>[0]\AgdaFunction{listMinusSubList+Index2OrgIndex}%
\>[717I]\AgdaSymbol{:}%
\>[718I]\AgdaSymbol{\{}\AgdaBound{X}\AgdaSpace{}%
\AgdaBound{Y}\AgdaSpace{}%
\AgdaSymbol{:}\AgdaSpace{}%
\AgdaPrimitiveType{Set}\AgdaSymbol{\}(}\AgdaBound{l}\AgdaSpace{}%
\AgdaSymbol{:}\AgdaSpace{}%
\AgdaDatatype{List}\AgdaSpace{}%
\AgdaBound{X}\AgdaSymbol{)(}\AgdaBound{o}\AgdaSpace{}%
\AgdaSymbol{:}\AgdaSpace{}%
\AgdaDatatype{SubList+}\AgdaSpace{}%
\AgdaBound{Y}\AgdaSpace{}%
\AgdaBound{l}\AgdaSymbol{)}\<%
\\
\>[717I][@{}l@{\AgdaIndent{0}}]\<[718I]%
\>[34]\AgdaSymbol{(}\AgdaBound{i}\AgdaSpace{}%
\AgdaSymbol{:}\AgdaSpace{}%
\AgdaDatatype{Fin}\AgdaSpace{}%
\AgdaSymbol{(}\AgdaFunction{length}\AgdaSpace{}%
\AgdaSymbol{(}\AgdaFunction{listMinusSubList+}\AgdaSpace{}%
\AgdaBound{l}\AgdaSpace{}%
\AgdaBound{o}\AgdaSymbol{)))}\AgdaSpace{}%
\AgdaSymbol{→}\AgdaSpace{}%
\AgdaDatatype{Fin}\AgdaSpace{}%
\AgdaSymbol{(}\AgdaFunction{length}\AgdaSpace{}%
\AgdaBound{l}\AgdaSymbol{)}\<%
\end{code}
} 

\AgdaHide{
\begin{code}%
\>[0]\<%
\\
\>[0]\AgdaFunction{listMinusSubList+Index2OrgIndex}\AgdaSpace{}%
\AgdaBound{l}\AgdaSpace{}%
\AgdaInductiveConstructor{[]}\AgdaSpace{}%
\AgdaBound{i}%
\>[52]\AgdaSymbol{=}\AgdaSpace{}%
\AgdaBound{i}\<%
\\
\>[0]\AgdaFunction{listMinusSubList+Index2OrgIndex}\AgdaSpace{}%
\AgdaBound{l}\AgdaSpace{}%
\AgdaSymbol{(}\AgdaInductiveConstructor{cons}\AgdaSpace{}%
\AgdaBound{i₁}\AgdaSpace{}%
\AgdaBound{y}\AgdaSpace{}%
\AgdaBound{o}\AgdaSymbol{)}\AgdaSpace{}%
\AgdaBound{i}%
\>[52]\AgdaSymbol{=}\<%
\\
\>[0][@{}l@{\AgdaIndent{0}}]%
\>[9]\AgdaFunction{delFromListIndexToOrigIndex}\AgdaSpace{}%
\AgdaBound{l}\AgdaSpace{}%
\AgdaBound{i₁}\AgdaSpace{}%
\AgdaSymbol{(}\AgdaFunction{listMinusSubList+Index2OrgIndex}\AgdaSpace{}%
\AgdaSymbol{(}\AgdaFunction{delFromList}\AgdaSpace{}%
\AgdaBound{l}\AgdaSpace{}%
\AgdaBound{i₁}\AgdaSymbol{)}\AgdaSpace{}%
\AgdaBound{o}\AgdaSpace{}%
\AgdaBound{i}\AgdaSymbol{)}\<%
\\
\\
\>[0]\AgdaFunction{corListMinusSubList+Index2OrgIndex}\AgdaSpace{}%
\AgdaSymbol{:}\AgdaSpace{}%
\AgdaSymbol{\{}\AgdaBound{X}\AgdaSpace{}%
\AgdaBound{Y}\AgdaSpace{}%
\AgdaSymbol{:}\AgdaSpace{}%
\AgdaPrimitiveType{Set}\AgdaSymbol{\}(}\AgdaBound{l}\AgdaSpace{}%
\AgdaSymbol{:}\AgdaSpace{}%
\AgdaDatatype{List}\AgdaSpace{}%
\AgdaBound{X}\AgdaSymbol{)(}\AgdaBound{o}\AgdaSpace{}%
\AgdaSymbol{:}\AgdaSpace{}%
\AgdaDatatype{SubList+}\AgdaSpace{}%
\AgdaBound{Y}\AgdaSpace{}%
\AgdaBound{l}\AgdaSymbol{)}\<%
\\
\>[0][@{}l@{\AgdaIndent{0}}]%
\>[34]\AgdaSymbol{(}\AgdaBound{i}\AgdaSpace{}%
\AgdaSymbol{:}\AgdaSpace{}%
\AgdaDatatype{Fin}\AgdaSpace{}%
\AgdaSymbol{(}\AgdaFunction{length}\AgdaSpace{}%
\AgdaSymbol{(}\AgdaFunction{listMinusSubList+}\AgdaSpace{}%
\AgdaBound{l}\AgdaSpace{}%
\AgdaBound{o}\AgdaSymbol{)))}\<%
\\
\>[0][@{}l@{\AgdaIndent{0}}]%
\>[34]\AgdaSymbol{→}%
\>[775I]\AgdaFunction{nth}\AgdaSpace{}%
\AgdaSymbol{(}\AgdaFunction{listMinusSubList+}\AgdaSpace{}%
\AgdaBound{l}\AgdaSpace{}%
\AgdaBound{o}\AgdaSymbol{)}\AgdaSpace{}%
\AgdaBound{i}\<%
\\
\>[775I][@{}l@{\AgdaIndent{0}}]%
\>[38]\AgdaDatatype{≡}\AgdaSpace{}%
\AgdaFunction{nth}\AgdaSpace{}%
\AgdaBound{l}\AgdaSpace{}%
\AgdaSymbol{(}\AgdaFunction{listMinusSubList+Index2OrgIndex}\AgdaSpace{}%
\AgdaBound{l}\AgdaSpace{}%
\AgdaBound{o}\AgdaSpace{}%
\AgdaBound{i}\AgdaSymbol{)}\<%
\\
\>[0]\AgdaFunction{corListMinusSubList+Index2OrgIndex}\AgdaSpace{}%
\AgdaBound{l}\AgdaSpace{}%
\AgdaInductiveConstructor{[]}\AgdaSpace{}%
\AgdaBound{i}\AgdaSpace{}%
\AgdaSymbol{=}\AgdaSpace{}%
\AgdaInductiveConstructor{refl}\<%
\\
\>[0]\AgdaFunction{corListMinusSubList+Index2OrgIndex}\AgdaSpace{}%
\AgdaInductiveConstructor{[]}\AgdaSpace{}%
\AgdaSymbol{(}\AgdaInductiveConstructor{cons}\AgdaSpace{}%
\AgdaSymbol{()}\AgdaSpace{}%
\AgdaBound{y}\AgdaSpace{}%
\AgdaBound{o}\AgdaSymbol{)}\AgdaSpace{}%
\AgdaBound{i}\<%
\\
\>[0]\AgdaFunction{corListMinusSubList+Index2OrgIndex}\AgdaSpace{}%
\AgdaSymbol{(}\AgdaBound{x}\AgdaSpace{}%
\AgdaInductiveConstructor{∷}\AgdaSpace{}%
\AgdaBound{l}\AgdaSymbol{)}\AgdaSpace{}%
\AgdaSymbol{(}\AgdaInductiveConstructor{cons}\AgdaSpace{}%
\AgdaInductiveConstructor{zero}\AgdaSpace{}%
\AgdaBound{y}\AgdaSpace{}%
\AgdaBound{o}\AgdaSymbol{)}\AgdaSpace{}%
\AgdaBound{i}\AgdaSpace{}%
\AgdaSymbol{=}\AgdaSpace{}%
\AgdaFunction{corListMinusSubList+Index2OrgIndex}\AgdaSpace{}%
\AgdaBound{l}\AgdaSpace{}%
\AgdaBound{o}\AgdaSpace{}%
\AgdaBound{i}\<%
\\
\>[0]\AgdaFunction{corListMinusSubList+Index2OrgIndex}\AgdaSpace{}%
\AgdaSymbol{(}\AgdaBound{x}\AgdaSpace{}%
\AgdaInductiveConstructor{∷}\AgdaSpace{}%
\AgdaBound{l}\AgdaSymbol{)}\AgdaSpace{}%
\AgdaSymbol{(}\AgdaInductiveConstructor{cons}\AgdaSpace{}%
\AgdaSymbol{(}\AgdaInductiveConstructor{suc}\AgdaSpace{}%
\AgdaBound{i₁}\AgdaSymbol{)}\AgdaSpace{}%
\AgdaBound{y}\AgdaSpace{}%
\AgdaBound{o}\AgdaSymbol{)}\AgdaSpace{}%
\AgdaBound{i}\AgdaSpace{}%
\AgdaSymbol{=}\AgdaSpace{}%
\AgdaFunction{trans}\AgdaSpace{}%
\AgdaFunction{eq1}\AgdaSpace{}%
\AgdaFunction{eq2}\<%
\\
\>[0][@{}l@{\AgdaIndent{0}}]%
\>[6]\AgdaKeyword{where}\<%
\\
\>[6][@{}l@{\AgdaIndent{0}}]%
\>[9]\AgdaFunction{eq1}%
\>[823I]\AgdaSymbol{:}%
\>[824I]\AgdaFunction{nth}\AgdaSpace{}%
\AgdaSymbol{(}\AgdaFunction{listMinusSubList+}\AgdaSpace{}%
\AgdaSymbol{(}\AgdaBound{x}\AgdaSpace{}%
\AgdaInductiveConstructor{∷}\AgdaSpace{}%
\AgdaFunction{delFromList}\AgdaSpace{}%
\AgdaBound{l}\AgdaSpace{}%
\AgdaBound{i₁}\AgdaSymbol{)}\AgdaSpace{}%
\AgdaBound{o}\AgdaSymbol{)}\AgdaSpace{}%
\AgdaBound{i}\AgdaSpace{}%
\AgdaDatatype{≡}\<%
\\
\>[823I][@{}l@{\AgdaIndent{0}}]\<[824I]%
\>[15]\AgdaFunction{nth}\AgdaSpace{}%
\AgdaSymbol{(}\AgdaBound{x}\AgdaSpace{}%
\AgdaInductiveConstructor{∷}\AgdaSpace{}%
\AgdaFunction{delFromList}\AgdaSpace{}%
\AgdaBound{l}\AgdaSpace{}%
\AgdaBound{i₁}\AgdaSymbol{)}\<%
\\
\>[15][@{}l@{\AgdaIndent{0}}]%
\>[18]\AgdaSymbol{(}\AgdaFunction{listMinusSubList+Index2OrgIndex}\AgdaSpace{}%
\AgdaSymbol{(}\AgdaBound{x}\AgdaSpace{}%
\AgdaInductiveConstructor{∷}\AgdaSpace{}%
\AgdaFunction{delFromList}\AgdaSpace{}%
\AgdaBound{l}\AgdaSpace{}%
\AgdaBound{i₁}\AgdaSymbol{)}\AgdaSpace{}%
\AgdaBound{o}\AgdaSpace{}%
\AgdaBound{i}\AgdaSymbol{)}\<%
\\
\>[6][@{}l@{\AgdaIndent{0}}]%
\>[9]\AgdaFunction{eq1}\AgdaSpace{}%
\AgdaSymbol{=}\AgdaSpace{}%
\AgdaFunction{corListMinusSubList+Index2OrgIndex}\AgdaSpace{}%
\AgdaSymbol{(}\AgdaBound{x}\AgdaSpace{}%
\AgdaInductiveConstructor{∷}\AgdaSpace{}%
\AgdaFunction{delFromList}\AgdaSpace{}%
\AgdaBound{l}\AgdaSpace{}%
\AgdaBound{i₁}\AgdaSymbol{)}\AgdaSpace{}%
\AgdaBound{o}\AgdaSpace{}%
\AgdaBound{i}\<%
\\
\\
\>[6][@{}l@{\AgdaIndent{0}}]%
\>[9]\AgdaFunction{eq2}%
\>[855I]\AgdaSymbol{:}%
\>[856I]\AgdaFunction{nth}\AgdaSpace{}%
\AgdaSymbol{(}\AgdaBound{x}\AgdaSpace{}%
\AgdaInductiveConstructor{∷}\AgdaSpace{}%
\AgdaFunction{delFromList}\AgdaSpace{}%
\AgdaBound{l}\AgdaSpace{}%
\AgdaBound{i₁}\AgdaSymbol{)}\<%
\\
\>[856I][@{}l@{\AgdaIndent{0}}]%
\>[18]\AgdaSymbol{(}\AgdaFunction{listMinusSubList+Index2OrgIndex}\AgdaSpace{}%
\AgdaSymbol{(}\AgdaBound{x}\AgdaSpace{}%
\AgdaInductiveConstructor{∷}\AgdaSpace{}%
\AgdaFunction{delFromList}\AgdaSpace{}%
\AgdaBound{l}\AgdaSpace{}%
\AgdaBound{i₁}\AgdaSymbol{)}\AgdaSpace{}%
\AgdaBound{o}\AgdaSpace{}%
\AgdaBound{i}\AgdaSymbol{)}\<%
\\
\>[855I][@{}l@{\AgdaIndent{0}}]\<[856I]%
\>[15]\AgdaDatatype{≡}%
\>[869I]\AgdaFunction{nth}\AgdaSpace{}%
\AgdaSymbol{(}\AgdaBound{x}\AgdaSpace{}%
\AgdaInductiveConstructor{∷}\AgdaSpace{}%
\AgdaBound{l}\AgdaSymbol{)}\<%
\\
\>[869I][@{}l@{\AgdaIndent{0}}]%
\>[18]\AgdaSymbol{(}\AgdaFunction{delFromListIndexToOrigIndex}\AgdaSpace{}%
\AgdaSymbol{(}\AgdaBound{x}\AgdaSpace{}%
\AgdaInductiveConstructor{∷}\AgdaSpace{}%
\AgdaBound{l}\AgdaSymbol{)}\AgdaSpace{}%
\AgdaSymbol{(}\AgdaInductiveConstructor{suc}\AgdaSpace{}%
\AgdaBound{i₁}\AgdaSymbol{)}\<%
\\
\>[869I][@{}l@{\AgdaIndent{0}}]%
\>[18]\AgdaSymbol{(}\AgdaFunction{listMinusSubList+Index2OrgIndex}\AgdaSpace{}%
\AgdaSymbol{(}\AgdaBound{x}\AgdaSpace{}%
\AgdaInductiveConstructor{∷}\AgdaSpace{}%
\AgdaFunction{delFromList}\AgdaSpace{}%
\AgdaBound{l}\AgdaSpace{}%
\AgdaBound{i₁}\AgdaSymbol{)}\AgdaSpace{}%
\AgdaBound{o}\AgdaSpace{}%
\AgdaBound{i}\AgdaSymbol{))}\<%
\\
\>[6][@{}l@{\AgdaIndent{0}}]%
\>[9]\AgdaFunction{eq2}%
\>[14]\AgdaSymbol{=}%
\>[885I]\AgdaFunction{correctNthDelFromList}\AgdaSpace{}%
\AgdaSymbol{(}\AgdaBound{x}\AgdaSpace{}%
\AgdaInductiveConstructor{∷}\AgdaSpace{}%
\AgdaBound{l}\AgdaSymbol{)}\AgdaSpace{}%
\AgdaSymbol{(}\AgdaInductiveConstructor{suc}\AgdaSpace{}%
\AgdaBound{i₁}\AgdaSymbol{)}\<%
\\
\>[14][@{}l@{\AgdaIndent{0}}]\<[885I]%
\>[16]\AgdaSymbol{((}\AgdaFunction{listMinusSubList+Index2OrgIndex}\AgdaSpace{}%
\AgdaSymbol{(}\AgdaBound{x}\AgdaSpace{}%
\AgdaInductiveConstructor{∷}\AgdaSpace{}%
\AgdaFunction{delFromList}\AgdaSpace{}%
\AgdaBound{l}\AgdaSpace{}%
\AgdaBound{i₁}\AgdaSymbol{)}\AgdaSpace{}%
\AgdaBound{o}\AgdaSpace{}%
\AgdaBound{i}\AgdaSymbol{))}\<%
\\
\\
\\
\>[0]\AgdaComment{{-}{-}\textbackslash{}listLib}\<%
\end{code}
} 

\newcommand{\listLibsublistplustoIndicesOriginalList}{
\begin{code}%
\>[0]\AgdaFunction{subList+2IndicesOriginalList}\AgdaSpace{}%
\AgdaSymbol{:}\AgdaSpace{}%
\AgdaSymbol{\{}\AgdaBound{X}\AgdaSpace{}%
\AgdaBound{Y}\AgdaSpace{}%
\AgdaSymbol{:}\AgdaSpace{}%
\AgdaPrimitiveType{Set}\AgdaSymbol{\}(}\AgdaBound{l}\AgdaSpace{}%
\AgdaSymbol{:}\AgdaSpace{}%
\AgdaDatatype{List}\AgdaSpace{}%
\AgdaBound{X}\AgdaSymbol{)(}\AgdaBound{sl}\AgdaSpace{}%
\AgdaSymbol{:}\AgdaSpace{}%
\AgdaDatatype{SubList+}\AgdaSpace{}%
\AgdaBound{Y}\AgdaSpace{}%
\AgdaBound{l}\AgdaSymbol{)}\AgdaSpace{}%
\AgdaSymbol{→}\AgdaSpace{}%
\AgdaDatatype{List}\AgdaSpace{}%
\AgdaSymbol{(}\AgdaDatatype{Fin}\AgdaSpace{}%
\AgdaSymbol{(}\AgdaFunction{length}\AgdaSpace{}%
\AgdaBound{l}\AgdaSymbol{)}\AgdaSpace{}%
\AgdaFunction{×}\AgdaSpace{}%
\AgdaBound{Y}\AgdaSymbol{)}\<%
\end{code}
} 

\AgdaHide{
\begin{code}%
\>[0]\AgdaFunction{subList+2IndicesOriginalList}\AgdaSpace{}%
\AgdaBound{l}\AgdaSpace{}%
\AgdaInductiveConstructor{[]}\AgdaSpace{}%
\AgdaSymbol{=}\AgdaSpace{}%
\AgdaInductiveConstructor{[]}\<%
\\
\>[0]\AgdaFunction{subList+2IndicesOriginalList}\AgdaSpace{}%
\AgdaSymbol{\{}\AgdaBound{X}\AgdaSymbol{\}}\AgdaSpace{}%
\AgdaSymbol{\{}\AgdaBound{Y}\AgdaSymbol{\}}\AgdaSpace{}%
\AgdaBound{l}\AgdaSpace{}%
\AgdaSymbol{(}\AgdaInductiveConstructor{cons}\AgdaSpace{}%
\AgdaBound{i}\AgdaSpace{}%
\AgdaBound{y}\AgdaSpace{}%
\AgdaBound{sl}\AgdaSymbol{)}\AgdaSpace{}%
\AgdaSymbol{=}\<%
\\
\>[0][@{}l@{\AgdaIndent{0}}]%
\>[6]\AgdaSymbol{(}\AgdaBound{i}\AgdaSpace{}%
\AgdaInductiveConstructor{,}\AgdaSpace{}%
\AgdaBound{y}\AgdaSymbol{)}\AgdaSpace{}%
\AgdaInductiveConstructor{∷}\AgdaSpace{}%
\AgdaFunction{mapL}\AgdaSpace{}%
\AgdaSymbol{(λ\{(}\AgdaBound{j}\AgdaSpace{}%
\AgdaInductiveConstructor{,}\AgdaSpace{}%
\AgdaBound{y}\AgdaSymbol{)}\AgdaSpace{}%
\AgdaSymbol{→}\AgdaSpace{}%
\AgdaSymbol{(}\AgdaFunction{delFromListIndexToOrigIndex}\AgdaSpace{}%
\AgdaBound{l}\AgdaSpace{}%
\AgdaBound{i}\AgdaSpace{}%
\AgdaBound{j}\AgdaSpace{}%
\AgdaInductiveConstructor{,}\AgdaSpace{}%
\AgdaBound{y}\AgdaSymbol{)\})}\AgdaSpace{}%
\AgdaFunction{res1}\<%
\\
\>[0][@{}l@{\AgdaIndent{0}}]%
\>[4]\AgdaKeyword{where}\<%
\\
\>[4][@{}l@{\AgdaIndent{0}}]%
\>[8]\AgdaFunction{res1}\AgdaSpace{}%
\AgdaSymbol{:}\AgdaSpace{}%
\AgdaDatatype{List}\AgdaSpace{}%
\AgdaSymbol{(}\AgdaDatatype{Fin}\AgdaSpace{}%
\AgdaSymbol{(}\AgdaFunction{length}\AgdaSpace{}%
\AgdaSymbol{(}\AgdaFunction{delFromList}\AgdaSpace{}%
\AgdaBound{l}\AgdaSpace{}%
\AgdaBound{i}\AgdaSymbol{))}\AgdaSpace{}%
\AgdaFunction{×}\AgdaSpace{}%
\AgdaBound{Y}\AgdaSymbol{)}\<%
\\
\>[4][@{}l@{\AgdaIndent{0}}]%
\>[8]\AgdaFunction{res1}\AgdaSpace{}%
\AgdaSymbol{=}\AgdaSpace{}%
\AgdaFunction{subList+2IndicesOriginalList}\AgdaSpace{}%
\AgdaSymbol{(}\AgdaFunction{delFromList}\AgdaSpace{}%
\AgdaBound{l}\AgdaSpace{}%
\AgdaBound{i}\AgdaSymbol{)}\AgdaSpace{}%
\AgdaBound{sl}\<%
\\
\\
\\
\>[0]\AgdaComment{{-}{-} \textbackslash{}listLib}\<%
\end{code}
} 

\newcommand{\listLibsumListViaf}{
\begin{code}%
\>[0]\AgdaFunction{sumListViaf}\AgdaSpace{}%
\AgdaSymbol{:}\AgdaSpace{}%
\AgdaSymbol{\{}\AgdaBound{X}\AgdaSpace{}%
\AgdaSymbol{:}\AgdaSpace{}%
\AgdaPrimitiveType{Set}\AgdaSymbol{\}}\AgdaSpace{}%
\AgdaSymbol{(}\AgdaBound{\,f\,}\AgdaSpace{}%
\AgdaSymbol{:}\AgdaSpace{}%
\AgdaBound{X}\AgdaSpace{}%
\AgdaSymbol{→}\AgdaSpace{}%
\AgdaDatatype{ℕ}\AgdaSymbol{)(}\AgdaBound{l}\AgdaSpace{}%
\AgdaSymbol{:}\AgdaSpace{}%
\AgdaDatatype{List}\AgdaSpace{}%
\AgdaBound{X}\AgdaSymbol{)}\AgdaSpace{}%
\AgdaSymbol{→}\AgdaSpace{}%
\AgdaDatatype{ℕ}\<%
\\
\>[0]\AgdaFunction{sumListViaf}\AgdaSpace{}%
\AgdaBound{\,f\,}\AgdaSpace{}%
\AgdaInductiveConstructor{[]}\AgdaSpace{}%
\AgdaSymbol{=}\AgdaSpace{}%
\AgdaNumber{0}\<%
\\
\>[0]\AgdaFunction{sumListViaf}\AgdaSpace{}%
\AgdaBound{\,f\,}\AgdaSpace{}%
\AgdaSymbol{(}\AgdaBound{x}\AgdaSpace{}%
\AgdaInductiveConstructor{∷}\AgdaSpace{}%
\AgdaBound{l}\AgdaSymbol{)}\AgdaSpace{}%
\AgdaSymbol{=}\AgdaSpace{}%
\AgdaBound{\,f\,}\AgdaSpace{}%
\AgdaBound{x}\AgdaSpace{}%
\AgdaPrimitive{+}\AgdaSpace{}%
\AgdaFunction{sumListViaf}\AgdaSpace{}%
\AgdaBound{\,f\,}\AgdaSpace{}%
\AgdaBound{l}\<%
\end{code}
} 

\AgdaHide{
\begin{code}%
\>[0]\<%
\\
\>[0]\AgdaComment{{-}{-}\textbackslash{}listLib}\<%
\end{code}
} 

\newcommand{\listLibforallInList}{
\begin{code}%
\>[0]\AgdaFunction{∀inList}\AgdaSpace{}%
\AgdaSymbol{:}\AgdaSpace{}%
\AgdaSymbol{\{}\AgdaBound{X}\AgdaSpace{}%
\AgdaSymbol{:}\AgdaSpace{}%
\AgdaPrimitiveType{Set}\AgdaSymbol{\}(}\AgdaBound{l}\AgdaSpace{}%
\AgdaSymbol{:}\AgdaSpace{}%
\AgdaDatatype{List}\AgdaSpace{}%
\AgdaBound{X}\AgdaSymbol{)(}\AgdaBound{P}\AgdaSpace{}%
\AgdaSymbol{:}\AgdaSpace{}%
\AgdaBound{X}\AgdaSpace{}%
\AgdaSymbol{→}\AgdaSpace{}%
\AgdaPrimitiveType{Set}\AgdaSymbol{)}\AgdaSpace{}%
\AgdaSymbol{→}\AgdaSpace{}%
\AgdaPrimitiveType{Set}\<%
\\
\>[0]\AgdaFunction{∀inList}\AgdaSpace{}%
\AgdaInductiveConstructor{[]}\AgdaSpace{}%
\AgdaBound{P}%
\>[20]\AgdaSymbol{=}\AgdaSpace{}%
\AgdaRecord{⊤}\<%
\\
\>[0]\AgdaFunction{∀inList}\AgdaSpace{}%
\AgdaSymbol{(}\AgdaBound{x}\AgdaSpace{}%
\AgdaInductiveConstructor{∷}\AgdaSpace{}%
\AgdaBound{l}\AgdaSymbol{)}\AgdaSpace{}%
\AgdaBound{P}%
\>[20]\AgdaSymbol{=}\AgdaSpace{}%
\AgdaBound{P}\AgdaSpace{}%
\AgdaBound{x}\AgdaSpace{}%
\AgdaFunction{×}%
\>[29]\AgdaFunction{∀inList}\AgdaSpace{}%
\AgdaBound{l}%
\>[40]\AgdaBound{P}\<%
\end{code}
} 

\AgdaHide{
\begin{code}%
\>[0]\<%
\\
\>[0]\AgdaComment{{-}{-}\textbackslash{}listLib}\<%
\end{code}
} 

\newcommand{\listLibnonNil}{
\begin{code}%
\>[0]\AgdaFunction{nonNil}\AgdaSpace{}%
\AgdaSymbol{:}\AgdaSpace{}%
\AgdaSymbol{\{}\AgdaBound{X}\AgdaSpace{}%
\AgdaSymbol{:}\AgdaSpace{}%
\AgdaPrimitiveType{Set}\AgdaSymbol{\}(}\AgdaBound{l}\AgdaSpace{}%
\AgdaSymbol{:}\AgdaSpace{}%
\AgdaDatatype{List}\AgdaSpace{}%
\AgdaBound{X}\AgdaSymbol{)}\AgdaSpace{}%
\AgdaSymbol{→}\AgdaSpace{}%
\AgdaDatatype{Bool}\<%
\\
\>[0]\AgdaFunction{nonNil}\AgdaSpace{}%
\AgdaInductiveConstructor{[]}\AgdaSpace{}%
\AgdaSymbol{=}\AgdaSpace{}%
\AgdaInductiveConstructor{true}\<%
\\
\>[0]\AgdaFunction{nonNil}\AgdaSpace{}%
\AgdaSymbol{(\_}\AgdaSpace{}%
\AgdaInductiveConstructor{∷}\AgdaSpace{}%
\AgdaSymbol{\_)}\AgdaSpace{}%
\AgdaSymbol{=}\AgdaSpace{}%
\AgdaInductiveConstructor{false}\<%
\\
\\
\>[0]\AgdaFunction{NonNil}\AgdaSpace{}%
\AgdaSymbol{:}\AgdaSpace{}%
\AgdaSymbol{\{}\AgdaBound{X}\AgdaSpace{}%
\AgdaSymbol{:}\AgdaSpace{}%
\AgdaPrimitiveType{Set}\AgdaSymbol{\}(}\AgdaBound{l}\AgdaSpace{}%
\AgdaSymbol{:}\AgdaSpace{}%
\AgdaDatatype{List}\AgdaSpace{}%
\AgdaBound{X}\AgdaSymbol{)}\AgdaSpace{}%
\AgdaSymbol{→}\AgdaSpace{}%
\AgdaPrimitiveType{Set}\<%
\\
\>[0]\AgdaFunction{NonNil}\AgdaSpace{}%
\AgdaBound{l}\AgdaSpace{}%
\AgdaSymbol{=}\AgdaSpace{}%
\AgdaFunction{T}\AgdaSpace{}%
\AgdaSymbol{(}\AgdaFunction{nonNil}\AgdaSpace{}%
\AgdaBound{l}\AgdaSymbol{)}\<%
\end{code}
} 

\AgdaHide{
\begin{code}%
\>[0]\<%
\\
\\
\>[0]\AgdaComment{{-}{-} adds the index to it but adds in addition n to the result}\<%
\\
\>[0]\AgdaComment{{-}{-} used as an auxiliary function in the next part}\<%
\\
\>[0]\<%
\end{code}
} 

\newcommand{\listLiblisttoListWithIndex}{
\begin{code}%
\>[0]\AgdaFunction{list2ListWithIndexaux}\AgdaSpace{}%
\AgdaSymbol{:}\AgdaSpace{}%
\AgdaSymbol{\{}\AgdaBound{X}\AgdaSpace{}%
\AgdaSymbol{:}\AgdaSpace{}%
\AgdaPrimitiveType{Set}\AgdaSymbol{\}(}\AgdaBound{n}\AgdaSpace{}%
\AgdaSymbol{:}\AgdaSpace{}%
\AgdaDatatype{ℕ}\AgdaSymbol{)}\AgdaSpace{}%
\AgdaSymbol{(}\AgdaBound{l}\AgdaSpace{}%
\AgdaSymbol{:}\AgdaSpace{}%
\AgdaDatatype{List}\AgdaSpace{}%
\AgdaBound{X}\AgdaSymbol{)}\AgdaSpace{}%
\AgdaSymbol{→}\AgdaSpace{}%
\AgdaDatatype{List}\AgdaSpace{}%
\AgdaSymbol{(}\AgdaBound{X}\AgdaSpace{}%
\AgdaFunction{×}\AgdaSpace{}%
\AgdaDatatype{ℕ}\AgdaSymbol{)}\<%
\\
\>[0]\AgdaFunction{list2ListWithIndexaux}\AgdaSpace{}%
\AgdaBound{n}\AgdaSpace{}%
\AgdaInductiveConstructor{[]}\AgdaSpace{}%
\AgdaSymbol{=}\AgdaSpace{}%
\AgdaInductiveConstructor{[]}\<%
\\
\>[0]\AgdaFunction{list2ListWithIndexaux}\AgdaSpace{}%
\AgdaBound{n}\AgdaSpace{}%
\AgdaSymbol{(}\AgdaBound{x}\AgdaSpace{}%
\AgdaInductiveConstructor{∷}\AgdaSpace{}%
\AgdaBound{l}\AgdaSymbol{)}\AgdaSpace{}%
\AgdaSymbol{=}\AgdaSpace{}%
\AgdaSymbol{(}\AgdaBound{x}\AgdaSpace{}%
\AgdaInductiveConstructor{,}\AgdaSpace{}%
\AgdaBound{n}\AgdaSymbol{)}\AgdaSpace{}%
\AgdaInductiveConstructor{∷}\AgdaSpace{}%
\AgdaFunction{list2ListWithIndexaux}\AgdaSpace{}%
\AgdaSymbol{(}\AgdaInductiveConstructor{suc}\AgdaSpace{}%
\AgdaBound{n}\AgdaSymbol{)}\AgdaSpace{}%
\AgdaBound{l}\<%
\\
\\
\>[0]\AgdaFunction{list2ListWithIndex}\AgdaSpace{}%
\AgdaSymbol{:}\AgdaSpace{}%
\AgdaSymbol{\{}\AgdaBound{X}\AgdaSpace{}%
\AgdaSymbol{:}\AgdaSpace{}%
\AgdaPrimitiveType{Set}\AgdaSymbol{\}(}\AgdaBound{l}\AgdaSpace{}%
\AgdaSymbol{:}\AgdaSpace{}%
\AgdaDatatype{List}\AgdaSpace{}%
\AgdaBound{X}\AgdaSymbol{)}\AgdaSpace{}%
\AgdaSymbol{→}\AgdaSpace{}%
\AgdaDatatype{List}\AgdaSpace{}%
\AgdaSymbol{(}\AgdaBound{X}\AgdaSpace{}%
\AgdaFunction{×}\AgdaSpace{}%
\AgdaDatatype{ℕ}\AgdaSymbol{)}\<%
\\
\>[0]\AgdaFunction{list2ListWithIndex}\AgdaSpace{}%
\AgdaBound{l}\AgdaSpace{}%
\AgdaSymbol{=}\AgdaSpace{}%
\AgdaFunction{list2ListWithIndexaux}\AgdaSpace{}%
\AgdaNumber{0}\AgdaSpace{}%
\AgdaBound{l}\<%
\end{code}
} 

\AgdaHide{
\begin{code}%
\>[0]\<%
\end{code}
} 


\AgdaHide{
\begin{code}%
\>[0]\AgdaKeyword{module}\AgdaSpace{}%
\AgdaModule{agdaIntro.finn}\AgdaSpace{}%
\AgdaKeyword{where}\<%
\\
\\
\>[0]\AgdaKeyword{open}\AgdaSpace{}%
\AgdaKeyword{import}\AgdaSpace{}%
\AgdaModule{Data.Nat}\<%
\\
\>[0]\AgdaKeyword{open}\AgdaSpace{}%
\AgdaKeyword{import}\AgdaSpace{}%
\AgdaModule{Data.Empty}\<%
\\
\\
\>[0]\AgdaComment{{-}{-}\textbackslash{}lagdaFinn}\<%
\end{code}
} 

\newcommand{\lagdaFinnFin}{
\begin{code}%
\>[0]\AgdaKeyword{data}\AgdaSpace{}%
\AgdaDatatype{Fin}\AgdaSpace{}%
\AgdaSymbol{:}\AgdaSpace{}%
\AgdaDatatype{ℕ}\AgdaSpace{}%
\AgdaSymbol{→}\AgdaSpace{}%
\AgdaPrimitiveType{Set}\AgdaSpace{}%
\AgdaKeyword{where}\<%
\\
\>[0][@{}l@{\AgdaIndent{0}}]%
\>[2]\AgdaInductiveConstructor{zero}\AgdaSpace{}%
\AgdaSymbol{:}\AgdaSpace{}%
\AgdaSymbol{\{}\AgdaBound{n}\AgdaSpace{}%
\AgdaSymbol{:}\AgdaSpace{}%
\AgdaDatatype{ℕ}\AgdaSymbol{\}}\AgdaSpace{}%
\AgdaSymbol{→}\AgdaSpace{}%
\AgdaDatatype{Fin}\AgdaSpace{}%
\AgdaSymbol{(}\AgdaInductiveConstructor{suc}\AgdaSpace{}%
\AgdaBound{n}\AgdaSymbol{)}\<%
\\
\>[0][@{}l@{\AgdaIndent{0}}]%
\>[2]\AgdaInductiveConstructor{suc}%
\>[7]\AgdaSymbol{:}\AgdaSpace{}%
\AgdaSymbol{\{}\AgdaBound{n}\AgdaSpace{}%
\AgdaSymbol{:}\AgdaSpace{}%
\AgdaDatatype{ℕ}\AgdaSymbol{\}}\AgdaSpace{}%
\AgdaSymbol{(}\AgdaBound{i}\AgdaSpace{}%
\AgdaSymbol{:}\AgdaSpace{}%
\AgdaDatatype{Fin}\AgdaSpace{}%
\AgdaBound{n}\AgdaSymbol{)}\AgdaSpace{}%
\AgdaSymbol{→}\AgdaSpace{}%
\AgdaDatatype{Fin}\AgdaSpace{}%
\AgdaSymbol{(}\AgdaInductiveConstructor{suc}\AgdaSpace{}%
\AgdaBound{n}\AgdaSymbol{)}\<%
\end{code}
} 

\AgdaHide{
\begin{code}%
\>[0]\<%
\\
\>[0]\AgdaComment{{-}{-}\textbackslash{}lagdaFinn}\<%
\end{code}
} 

\newcommand{\lagdaFinnEvenOdd}{
\begin{code}%
\>[0]\AgdaKeyword{mutual}\<%
\\
\>[0][@{}l@{\AgdaIndent{0}}]%
\>[2]\AgdaKeyword{data}\AgdaSpace{}%
\AgdaDatatype{Even}\AgdaSpace{}%
\AgdaSymbol{:}\AgdaSpace{}%
\AgdaDatatype{ℕ}\AgdaSpace{}%
\AgdaSymbol{→}\AgdaSpace{}%
\AgdaPrimitiveType{Set}\AgdaSpace{}%
\AgdaKeyword{where}\<%
\\
\>[2][@{}l@{\AgdaIndent{0}}]%
\>[4]\AgdaInductiveConstructor{0p}%
\>[10]\AgdaSymbol{:}\AgdaSpace{}%
\AgdaDatatype{Even}\AgdaSpace{}%
\AgdaNumber{0}\<%
\\
\>[2][@{}l@{\AgdaIndent{0}}]%
\>[4]\AgdaInductiveConstructor{sucp}%
\>[10]\AgdaSymbol{:}\AgdaSpace{}%
\AgdaSymbol{\{}\AgdaBound{n}\AgdaSpace{}%
\AgdaSymbol{:}\AgdaSpace{}%
\AgdaDatatype{ℕ}\AgdaSymbol{\}}%
\>[21]\AgdaSymbol{→}%
\>[24]\AgdaDatatype{Odd}\AgdaSpace{}%
\AgdaBound{n}%
\>[32]\AgdaSymbol{→}%
\>[35]\AgdaDatatype{Even}%
\>[41]\AgdaSymbol{(}\AgdaInductiveConstructor{suc}%
\>[47]\AgdaBound{n}\AgdaSymbol{)}\<%
\\
\\
\>[0][@{}l@{\AgdaIndent{0}}]%
\>[2]\AgdaKeyword{data}\AgdaSpace{}%
\AgdaDatatype{Odd}\AgdaSpace{}%
\AgdaSymbol{:}\AgdaSpace{}%
\AgdaDatatype{ℕ}\AgdaSpace{}%
\AgdaSymbol{→}\AgdaSpace{}%
\AgdaPrimitiveType{Set}\AgdaSpace{}%
\AgdaKeyword{where}\<%
\\
\>[2][@{}l@{\AgdaIndent{0}}]%
\>[4]\AgdaInductiveConstructor{sucp}%
\>[10]\AgdaSymbol{:}\AgdaSpace{}%
\AgdaSymbol{\{}\AgdaBound{n}\AgdaSpace{}%
\AgdaSymbol{:}\AgdaSpace{}%
\AgdaDatatype{ℕ}\AgdaSymbol{\}}%
\>[21]\AgdaSymbol{→}%
\>[24]\AgdaDatatype{Even}\AgdaSpace{}%
\AgdaBound{n}%
\>[32]\AgdaSymbol{→}%
\>[35]\AgdaDatatype{Odd}%
\>[41]\AgdaSymbol{(}\AgdaInductiveConstructor{suc}%
\>[47]\AgdaBound{n}\AgdaSymbol{)}\<%
\end{code}
} 

\AgdaHide{
\begin{code}%
\>[0]\<%
\\
\>[0]\AgdaComment{{-}{-}\textbackslash{}lagdaFinn}\<%
\end{code}
} 

\newcommand{\lagdaFinnpostulateFalsity}{
\begin{code}%
\>[0]\AgdaKeyword{postulate}\AgdaSpace{}%
\AgdaPostulate{nonExistentElement}\AgdaSpace{}%
\AgdaSymbol{:}\AgdaSpace{}%
\AgdaDatatype{Fin}\AgdaSpace{}%
\AgdaNumber{0}\<%
\end{code}
} 

\AgdaHide{
\begin{code}%
\>[0]\<%
\end{code}
} 


\AgdaHide{
\begin{code}%
\>[0]\AgdaKeyword{module}\AgdaSpace{}%
\AgdaModule{agdaIntro.student}\AgdaSpace{}%
\AgdaKeyword{where}\<%
\\
\\
\>[0]\AgdaKeyword{open}\AgdaSpace{}%
\AgdaKeyword{import}\AgdaSpace{}%
\AgdaModule{Data.Nat}\<%
\\
\>[0]\AgdaKeyword{open}\AgdaSpace{}%
\AgdaKeyword{import}\AgdaSpace{}%
\AgdaModule{Data.String}\<%
\\
\\
\>[0]\AgdaComment{{-}{-}\textbackslash{}student}\<%
\end{code}
} 

\newcommand{\studentStudentRecord}{
\begin{code}%
\>[0]\AgdaKeyword{record}\AgdaSpace{}%
\AgdaRecord{Student}\AgdaSpace{}%
\AgdaSymbol{:}\AgdaSpace{}%
\AgdaPrimitiveType{Set}\AgdaSpace{}%
\AgdaKeyword{where}\<%
\\
\>[0][@{}l@{\AgdaIndent{0}}]%
\>[2]\AgdaKeyword{constructor}\AgdaSpace{}%
\AgdaInductiveConstructor{student}\<%
\\
\>[0][@{}l@{\AgdaIndent{0}}]%
\>[2]\agdaField%
\>[9]\AgdaField{name}%
\>[17]\AgdaSymbol{:}\AgdaSpace{}%
\AgdaPostulate{String}\<%
\\
\>[2][@{}l@{\AgdaIndent{0}}]%
\>[9]\AgdaField{studnr}%
\>[17]\AgdaSymbol{:}\AgdaSpace{}%
\AgdaDatatype{ℕ}\<%
\\
\>[0]\AgdaKeyword{open}\AgdaSpace{}%
\AgdaModule{Student}\AgdaSpace{}%
\AgdaKeyword{public}\<%
\end{code}
} 

\AgdaHide{
\begin{code}%
\>[0]\<%
\\
\>[0]\AgdaFunction{exampleStudent}\AgdaSpace{}%
\AgdaSymbol{:}\AgdaSpace{}%
\AgdaRecord{Student}\<%
\\
\>[0]\AgdaComment{{-}{-}\textbackslash{}student}\<%
\end{code}
} 

\newcommand{\studentexampleStudent}{
\begin{code}%
\>[0]\AgdaFunction{exampleStudent}\AgdaSpace{}%
\AgdaSymbol{=}\AgdaSpace{}%
\AgdaInductiveConstructor{student}\AgdaSpace{}%
\AgdaString{"John"}\AgdaSpace{}%
\AgdaNumber{123456}\<%
\end{code}
} 

\AgdaHide{
\begin{code}%
\>[0]\<%
\\
\>[0]\AgdaFunction{f}\AgdaSpace{}%
\AgdaSymbol{:}\AgdaSpace{}%
\AgdaRecord{Student}\AgdaSpace{}%
\AgdaSymbol{→}\AgdaSpace{}%
\AgdaPostulate{String}\<%
\\
\>[0]\AgdaFunction{f}\AgdaSpace{}%
\AgdaSymbol{=}\AgdaSpace{}%
\AgdaField{name}\<%
\\
\\
\>[0]\AgdaFunction{john}\AgdaSpace{}%
\AgdaSymbol{:}\AgdaSpace{}%
\AgdaPostulate{String}\<%
\\
\>[0]\AgdaComment{{-}{-}\textbackslash{}student}\<%
\end{code}
} 

\newcommand{\studentjohn}{
\begin{code}%
\>[0]\AgdaFunction{john}\AgdaSpace{}%
\AgdaSymbol{=}\AgdaSpace{}%
\AgdaFunction{exampleStudent}\AgdaSpace{}%
\AgdaSymbol{.}\AgdaField{name}\<%
\end{code}
} 

\AgdaHide{
\begin{code}%
\>[0]\<%
\\
\>[0]\AgdaComment{{-}{-}\textbackslash{}student}\<%
\end{code}
} 

\newcommand{\studenthypotheticalStudent}{
\begin{code}%
\>[0]\AgdaKeyword{postulate}\AgdaSpace{}%
\AgdaPostulate{hypotheticalStudent}\AgdaSpace{}%
\AgdaSymbol{:}\AgdaSpace{}%
\AgdaRecord{Student}\<%
\end{code}
} 

\AgdaHide{
\begin{code}%
\>[0]\<%
\end{code}
} 


\AgdaHide{
\begin{code}%
\>[0]\AgdaKeyword{module}\AgdaSpace{}%
\AgdaModule{agdaIntro.exampleFinFun}\AgdaSpace{}%
\AgdaKeyword{where}\<%
\\
\\
\>[0]\AgdaKeyword{open}\AgdaSpace{}%
\AgdaKeyword{import}\AgdaSpace{}%
\AgdaModule{agdaIntro.finn}\<%
\\
\>[0]\AgdaKeyword{open}\AgdaSpace{}%
\AgdaKeyword{import}\AgdaSpace{}%
\AgdaModule{Data.Nat}\<%
\\
\\
\>[0]\<%
\end{code}
} 

\newcommand{\lagdaExampleFinFunToN}{
\begin{code}%
\>[0]\AgdaFunction{toℕ}\AgdaSpace{}%
\AgdaSymbol{:}\AgdaSpace{}%
\AgdaSymbol{∀}\AgdaSpace{}%
\AgdaSymbol{\{}\AgdaBound{n}\AgdaSymbol{\}}\AgdaSpace{}%
\AgdaSymbol{→}\AgdaSpace{}%
\AgdaDatatype{Fin}\AgdaSpace{}%
\AgdaBound{n}\AgdaSpace{}%
\AgdaSymbol{→}\AgdaSpace{}%
\AgdaDatatype{ℕ}\<%
\\
\>[0]\AgdaFunction{toℕ}%
\>[5]\AgdaInductiveConstructor{zero}%
\>[14]\AgdaSymbol{=}%
\>[17]\AgdaNumber{0}\<%
\\
\>[0]\AgdaFunction{toℕ}%
\>[5]\AgdaSymbol{(}\AgdaInductiveConstructor{suc}\AgdaSpace{}%
\AgdaBound{n}\AgdaSymbol{)}%
\>[14]\AgdaSymbol{=}%
\>[17]\AgdaInductiveConstructor{suc}\AgdaSpace{}%
\AgdaSymbol{(}\AgdaFunction{toℕ}\AgdaSpace{}%
\AgdaBound{n}\AgdaSymbol{)}\<%
\end{code}
} 

\AgdaHide{
\begin{code}%
\>[0]\<%
\\
\>[0]\<%
\end{code}
} 

\newcommand{\lagdaExampleFinFunPlus}{
\begin{code}%
\>[0]\AgdaKeyword{mutual}\<%
\\
\>[0][@{}l@{\AgdaIndent{0}}]%
\>[2]\AgdaFunction{\_+e\_}\AgdaSpace{}%
\AgdaSymbol{:}\AgdaSpace{}%
\AgdaSymbol{∀}\AgdaSpace{}%
\AgdaSymbol{\{}\AgdaBound{n}\AgdaSpace{}%
\AgdaBound{m}\AgdaSymbol{\}}\AgdaSpace{}%
\AgdaSymbol{→}%
\>[20]\AgdaDatatype{Even}\AgdaSpace{}%
\AgdaBound{n}%
\>[29]\AgdaSymbol{→}%
\>[32]\AgdaDatatype{Even}\AgdaSpace{}%
\AgdaBound{m}%
\>[40]\AgdaSymbol{→}%
\>[43]\AgdaDatatype{Even}%
\>[49]\AgdaSymbol{(}\AgdaBound{n}\AgdaSpace{}%
\AgdaPrimitive{+}\AgdaSpace{}%
\AgdaBound{m}\AgdaSymbol{)}\<%
\\
\>[0][@{}l@{\AgdaIndent{0}}]%
\>[2]\AgdaInductiveConstructor{0p}%
\>[10]\AgdaFunction{+e}%
\>[14]\AgdaBound{p}%
\>[17]\AgdaSymbol{=}%
\>[20]\AgdaBound{p}\<%
\\
\>[0][@{}l@{\AgdaIndent{0}}]%
\>[2]\AgdaInductiveConstructor{sucp}\AgdaSpace{}%
\AgdaBound{p}%
\>[10]\AgdaFunction{+e}%
\>[14]\AgdaBound{q}%
\>[17]\AgdaSymbol{=}%
\>[20]\AgdaInductiveConstructor{sucp}%
\>[26]\AgdaSymbol{(}\AgdaBound{p}\AgdaSpace{}%
\AgdaFunction{+o}\AgdaSpace{}%
\AgdaBound{q}\AgdaSymbol{)}\<%
\\
\\
\>[0][@{}l@{\AgdaIndent{0}}]%
\>[2]\AgdaFunction{\_+o\_}\AgdaSpace{}%
\AgdaSymbol{:}\AgdaSpace{}%
\AgdaSymbol{∀}\AgdaSpace{}%
\AgdaSymbol{\{}\AgdaBound{n}\AgdaSpace{}%
\AgdaBound{m}\AgdaSymbol{\}}%
\>[18]\AgdaSymbol{→}%
\>[21]\AgdaDatatype{Odd}\AgdaSpace{}%
\AgdaBound{n}%
\>[29]\AgdaSymbol{→}%
\>[32]\AgdaDatatype{Even}\AgdaSpace{}%
\AgdaBound{m}%
\>[40]\AgdaSymbol{→}%
\>[43]\AgdaDatatype{Odd}%
\>[49]\AgdaSymbol{(}\AgdaBound{n}\AgdaSpace{}%
\AgdaPrimitive{+}\AgdaSpace{}%
\AgdaBound{m}\AgdaSymbol{)}\<%
\\
\>[0][@{}l@{\AgdaIndent{0}}]%
\>[2]\AgdaInductiveConstructor{sucp}\AgdaSpace{}%
\AgdaBound{p}%
\>[10]\AgdaFunction{+o}%
\>[14]\AgdaBound{q}%
\>[17]\AgdaSymbol{=}%
\>[20]\AgdaInductiveConstructor{sucp}%
\>[26]\AgdaSymbol{(}\AgdaBound{p}\AgdaSpace{}%
\AgdaFunction{+e}\AgdaSpace{}%
\AgdaBound{q}\AgdaSymbol{)}\<%
\end{code}
} 

\AgdaHide{
\begin{code}%
\>[0]\<%
\end{code}
} 


\AgdaHide{
\begin{code}%
\>[0]\AgdaSymbol{\{{-}\#}\AgdaSpace{}%
\AgdaKeyword{OPTIONS}\AgdaSpace{}%
\AgdaOption{{-}{-}without{-}K}\AgdaSpace{}%
\AgdaSymbol{\#{-}\}}\<%
\\
\\
\>[0]\AgdaKeyword{module}\AgdaSpace{}%
\AgdaModule{agdaIntro.Equality}\AgdaSpace{}%
\AgdaKeyword{where}\<%
\\
\\
\>[0]\AgdaKeyword{infix}\AgdaSpace{}%
\AgdaNumber{4}\AgdaSpace{}%
\AgdaDatatype{\_≡\_}\<%
\\
\\
\>[0]\AgdaComment{{-}{-}\textbackslash{}EqualityModule}\<%
\end{code}
} 

\newcommand{\EqualityModuleDefEquality}{
\begin{code}%
\>[0]\AgdaKeyword{data}\AgdaSpace{}%
\AgdaDatatype{\_≡\_}\AgdaSpace{}%
\AgdaSymbol{\{}\AgdaBound{a}\AgdaSymbol{\}}\AgdaSpace{}%
\AgdaSymbol{\{}\AgdaBound{A}\AgdaSpace{}%
\AgdaSymbol{:}\AgdaSpace{}%
\AgdaPrimitiveType{Set}\AgdaSpace{}%
\AgdaBound{a}\AgdaSymbol{\}}\AgdaSpace{}%
\AgdaSymbol{(}\AgdaBound{x}\AgdaSpace{}%
\AgdaSymbol{:}\AgdaSpace{}%
\AgdaBound{A}\AgdaSymbol{)}\AgdaSpace{}%
\AgdaSymbol{:}\AgdaSpace{}%
\AgdaBound{A}\AgdaSpace{}%
\AgdaSymbol{→}\AgdaSpace{}%
\AgdaPrimitiveType{Set}\AgdaSpace{}%
\AgdaBound{a}\AgdaSpace{}%
\AgdaKeyword{where}\<%
\\
\>[0][@{}l@{\AgdaIndent{0}}]%
\>[2]\AgdaInductiveConstructor{refl}\AgdaSpace{}%
\AgdaSymbol{:}\AgdaSpace{}%
\AgdaBound{x}\AgdaSpace{}%
\AgdaDatatype{≡}\AgdaSpace{}%
\AgdaBound{x}\<%
\end{code}
} 

\AgdaHide{
\begin{code}%
\>[0]\<%
\end{code}
} 


\AgdaHide{
\begin{code}%
\>[0]\<%
\\
\\
\>[0]\AgdaKeyword{module}\AgdaSpace{}%
\AgdaModule{bitcoinLedgerModel}\AgdaSpace{}%
\AgdaKeyword{where}\<%
\\
\\
\>[0]\AgdaComment{{-}{-} open import bool}\<%
\\
\>[0]\AgdaKeyword{open}\AgdaSpace{}%
\AgdaKeyword{import}\AgdaSpace{}%
\AgdaModule{libraries.listLib}\<%
\\
\>[0]\AgdaKeyword{open}\AgdaSpace{}%
\AgdaKeyword{import}\AgdaSpace{}%
\AgdaModule{libraries.natLib}\<%
\\
\>[0]\AgdaKeyword{open}\AgdaSpace{}%
\AgdaKeyword{import}\AgdaSpace{}%
\AgdaModule{Data.Nat}\<%
\\
\>[0]\AgdaKeyword{open}\AgdaSpace{}%
\AgdaKeyword{import}\AgdaSpace{}%
\AgdaModule{Data.List}\<%
\\
\>[0]\AgdaKeyword{open}\AgdaSpace{}%
\AgdaKeyword{import}\AgdaSpace{}%
\AgdaModule{Data.Unit}\<%
\\
\>[0]\AgdaKeyword{open}\AgdaSpace{}%
\AgdaKeyword{import}\AgdaSpace{}%
\AgdaModule{Data.Bool}\<%
\\
\>[0]\AgdaKeyword{open}\AgdaSpace{}%
\AgdaKeyword{import}\AgdaSpace{}%
\AgdaModule{Data.Product}\<%
\\
\>[0]\AgdaKeyword{open}\AgdaSpace{}%
\AgdaKeyword{import}\AgdaSpace{}%
\AgdaModule{Data.Nat.Base}\<%
\\
\\
\>[0]\AgdaKeyword{infixr}\AgdaSpace{}%
\AgdaNumber{3}\AgdaSpace{}%
\AgdaInductiveConstructor{\_+msg\_}\<%
\\
\\
\>[0]\AgdaFunction{Time}\AgdaSpace{}%
\AgdaSymbol{:}\AgdaSpace{}%
\AgdaPrimitiveType{Set}\<%
\\
\>[0]\AgdaFunction{Amount}\AgdaSpace{}%
\AgdaSymbol{:}\AgdaSpace{}%
\AgdaPrimitiveType{Set}\<%
\\
\>[0]\AgdaFunction{Address}\AgdaSpace{}%
\AgdaSymbol{:}\AgdaSpace{}%
\AgdaPrimitiveType{Set}\<%
\\
\>[0]\AgdaFunction{TXID}\AgdaSpace{}%
\AgdaSymbol{:}\AgdaSpace{}%
\AgdaPrimitiveType{Set}\<%
\\
\>[0]\AgdaFunction{Signature}\AgdaSpace{}%
\AgdaSymbol{:}\AgdaSpace{}%
\AgdaPrimitiveType{Set}\<%
\\
\>[0]\AgdaFunction{PublicKey}\AgdaSpace{}%
\AgdaSymbol{:}\AgdaSpace{}%
\AgdaPrimitiveType{Set}\<%
\\
\\
\>[0]\AgdaComment{{-}{-} \textbackslash{}bitcoinVersFive}\<%
\end{code}
} 

\newcommand{\bitcoinVersFivebasicdef}{
\begin{code}%
\>[0]\AgdaFunction{Time}%
\>[11]\AgdaSymbol{=}%
\>[14]\AgdaDatatype{ℕ}\<%
\end{code}
} 

\AgdaHide{
\begin{code}%
\>[0]\<%
\\
\>[0]\AgdaFunction{Amount}%
\>[11]\AgdaSymbol{=}%
\>[14]\AgdaDatatype{ℕ}\<%
\\
\>[0]\AgdaFunction{Address}%
\>[11]\AgdaSymbol{=}%
\>[14]\AgdaDatatype{ℕ}\<%
\\
\>[0]\AgdaFunction{TXID}%
\>[11]\AgdaSymbol{=}%
\>[14]\AgdaDatatype{ℕ}\<%
\\
\>[0]\AgdaFunction{Signature}%
\>[11]\AgdaSymbol{=}%
\>[14]\AgdaDatatype{ℕ}\<%
\\
\>[0]\AgdaFunction{PublicKey}%
\>[11]\AgdaSymbol{=}%
\>[14]\AgdaDatatype{ℕ}\<%
\\
\\
\\
\>[0]\AgdaComment{{-}{-} \textbackslash{}bitcoinVersFive}\<%
\end{code}
} 

\newcommand{\bitcoinVersFiveMsg}{
\begin{code}%
\>[0]\AgdaKeyword{data}\AgdaSpace{}%
\AgdaDatatype{Msg}\AgdaSpace{}%
\AgdaSymbol{:}\AgdaSpace{}%
\AgdaPrimitiveType{Set}\AgdaSpace{}%
\AgdaKeyword{where}\<%
\\
\>[0][@{}l@{\AgdaIndent{0}}]%
\>[3]\AgdaInductiveConstructor{nat}%
\>[11]\AgdaSymbol{:}%
\>[14]\AgdaSymbol{(}\AgdaBound{n}\AgdaSpace{}%
\AgdaSymbol{:}\AgdaSpace{}%
\AgdaDatatype{ℕ}\AgdaSymbol{)}%
\>[30]\AgdaSymbol{→}\AgdaSpace{}%
\AgdaDatatype{Msg}\<%
\\
\>[0][@{}l@{\AgdaIndent{0}}]%
\>[3]\AgdaInductiveConstructor{\_+msg\_}%
\>[11]\AgdaSymbol{:}%
\>[14]\AgdaSymbol{(}\AgdaBound{m}\AgdaSpace{}%
\AgdaBound{m'}\AgdaSpace{}%
\AgdaSymbol{:}\AgdaSpace{}%
\AgdaDatatype{Msg}\AgdaSymbol{)}%
\>[31]\AgdaSymbol{→}\AgdaSpace{}%
\AgdaDatatype{Msg}\<%
\\
\>[0][@{}l@{\AgdaIndent{0}}]%
\>[3]\AgdaInductiveConstructor{list}%
\>[11]\AgdaSymbol{:}%
\>[14]\AgdaSymbol{(}\AgdaBound{l}%
\>[18]\AgdaSymbol{:}\AgdaSpace{}%
\AgdaDatatype{List}\AgdaSpace{}%
\AgdaDatatype{Msg}\AgdaSymbol{)}%
\>[31]\AgdaSymbol{→}\AgdaSpace{}%
\AgdaDatatype{Msg}\<%
\\
\\
\>[0]\AgdaKeyword{postulate}\AgdaSpace{}%
\AgdaPostulate{hashMsg}\AgdaSpace{}%
\AgdaSymbol{:}\AgdaSpace{}%
\AgdaDatatype{Msg}\AgdaSpace{}%
\AgdaSymbol{→}\AgdaSpace{}%
\AgdaDatatype{ℕ}\<%
\end{code}
} 

\AgdaHide{
\begin{code}%
\>[0]\<%
\\
\\
\\
\>[0]\AgdaComment{{-}{-} \textbackslash{}bitcoinVersFive}\<%
\end{code}
} 

\newcommand{\bitcoinVersFivePublicKey}{
\begin{code}%
\>[0]\AgdaKeyword{postulate}\AgdaSpace{}%
\AgdaPostulate{publicKey2Address}\AgdaSpace{}%
\AgdaSymbol{:}\AgdaSpace{}%
\AgdaSymbol{(}\AgdaBound{pubk}\AgdaSpace{}%
\AgdaSymbol{:}\AgdaSpace{}%
\AgdaFunction{PublicKey}\AgdaSymbol{)}\AgdaSpace{}%
\AgdaSymbol{→}\AgdaSpace{}%
\AgdaFunction{Address}\<%
\end{code}
} 

\AgdaHide{
\begin{code}%
\>[0]\<%
\\
\>[0]\AgdaComment{{-}{-} Signed means that Msg msg has been signed}\<%
\\
\>[0]\AgdaComment{{-}{-} by private key corresponding to pubk}\<%
\\
\\
\>[0]\AgdaComment{{-}{-} \textbackslash{}bitcoinVersFive}\<%
\end{code}
} 

\newcommand{\bitcoinVersFiveSignature}{
\begin{code}%
\>[0]\AgdaKeyword{postulate}\AgdaSpace{}%
\AgdaPostulate{Signed}\AgdaSpace{}%
\AgdaSymbol{:}\AgdaSpace{}%
\AgdaSymbol{(}\AgdaBound{msg}\AgdaSpace{}%
\AgdaSymbol{:}\AgdaSpace{}%
\AgdaDatatype{Msg}\AgdaSymbol{)(}\AgdaBound{publicKey}\AgdaSpace{}%
\AgdaSymbol{:}\AgdaSpace{}%
\AgdaFunction{PublicKey}\AgdaSymbol{)(}\AgdaBound{s}\AgdaSpace{}%
\AgdaSymbol{:}\AgdaSpace{}%
\AgdaFunction{Signature}\AgdaSymbol{)}\AgdaSpace{}%
\AgdaSymbol{→}\AgdaSpace{}%
\AgdaPrimitiveType{Set}\<%
\\
\\
\>[0]\AgdaKeyword{record}\AgdaSpace{}%
\AgdaRecord{SignedWithSigPbk}\AgdaSpace{}%
\AgdaSymbol{(}\AgdaBound{msg}\AgdaSpace{}%
\AgdaSymbol{:}\AgdaSpace{}%
\AgdaDatatype{Msg}\AgdaSymbol{)(}\AgdaBound{address}\AgdaSpace{}%
\AgdaSymbol{:}\AgdaSpace{}%
\AgdaFunction{Address}\AgdaSymbol{)}\AgdaSpace{}%
\AgdaSymbol{:}\AgdaSpace{}%
\AgdaPrimitiveType{Set}\AgdaSpace{}%
\AgdaKeyword{where}\<%
\\
\>[0][@{}l@{\AgdaIndent{0}}]%
\>[2]\agdaField%
\>[9]\AgdaField{publicKey}%
\>[20]\AgdaSymbol{:}%
\>[23]\AgdaFunction{PublicKey}\<%
\\
\>[2][@{}l@{\AgdaIndent{0}}]%
\>[9]\AgdaField{pbkCorrect}\AgdaSpace{}%
\AgdaSymbol{:}%
\>[23]\AgdaPostulate{publicKey2Address}\AgdaSpace{}%
\AgdaField{publicKey}\AgdaSpace{}%
\AgdaFunction{≡ℕ}%
\>[55]\AgdaBound{address}\<%
\\
\>[2][@{}l@{\AgdaIndent{0}}]%
\>[9]\AgdaField{signature}%
\>[20]\AgdaSymbol{:}%
\>[23]\AgdaFunction{Signature}\<%
\\
\>[2][@{}l@{\AgdaIndent{0}}]%
\>[9]\AgdaField{signed}%
\>[20]\AgdaSymbol{:}%
\>[23]\AgdaPostulate{Signed}\AgdaSpace{}%
\AgdaBound{msg}\AgdaSpace{}%
\AgdaField{publicKey}\AgdaSpace{}%
\AgdaField{signature}\<%
\end{code}
} 

\AgdaHide{
\begin{code}%
\>[0]\<%
\\
\>[0]\AgdaComment{\{{-}
\_≡PbkBool\_ : (pubk pubk' : PublicKey) → Bool
pubk ≡PbkBool pubk' = publicKey2Address pubk  ≡ℕb publicKey2Address pubk'
{-}\}}\<%
\\
\\
\>[0]\AgdaComment{\{{-} {-}{-} Unused but correct code

\_≡Pbk\_ : PublicKey → PublicKey → Set
key1 ≡Pbk key2 = T (key1 ≡PbkBool key2)
{-}\}}\<%
\\
\\
\>[0]\AgdaComment{{-}{-} \textbackslash{}bitcoinVersFive}\<%
\end{code}
} 

\newcommand{\bitcoinVersFiveTXField}{
\begin{code}%
\>[0]\AgdaKeyword{record}\AgdaSpace{}%
\AgdaRecord{TXField}\AgdaSpace{}%
\AgdaSymbol{:}\AgdaSpace{}%
\AgdaPrimitiveType{Set}\AgdaSpace{}%
\AgdaKeyword{where}\<%
\\
\>[0][@{}l@{\AgdaIndent{0}}]%
\>[2]\AgdaKeyword{constructor}\AgdaSpace{}%
\AgdaInductiveConstructor{txField}\<%
\\
\>[0][@{}l@{\AgdaIndent{0}}]%
\>[2]\agdaField%
\>[9]\AgdaField{amount}%
\>[20]\AgdaSymbol{:}%
\>[23]\AgdaFunction{Amount}\<%
\\
\>[2][@{}l@{\AgdaIndent{0}}]%
\>[9]\AgdaField{address}%
\>[20]\AgdaSymbol{:}%
\>[23]\AgdaFunction{Address}\<%
\end{code}
} 

\AgdaHide{
\begin{code}%
\>[0]\<%
\\
\>[0]\AgdaKeyword{open}\AgdaSpace{}%
\AgdaModule{TXField}\AgdaSpace{}%
\AgdaKeyword{public}\<%
\\
\\
\>[0]\AgdaComment{{-}{-}\textbackslash{}bitcoinVersFive}\<%
\end{code}
} 

\newcommand{\bitcoinVersFiveTXFieldToMsg}{
\begin{code}%
\>[0]\AgdaFunction{txField2Msg}\AgdaSpace{}%
\AgdaSymbol{:}\AgdaSpace{}%
\AgdaSymbol{(}\AgdaBound{inp}\AgdaSpace{}%
\AgdaSymbol{:}\AgdaSpace{}%
\AgdaRecord{TXField}\AgdaSymbol{)}\AgdaSpace{}%
\AgdaSymbol{→}\AgdaSpace{}%
\AgdaDatatype{Msg}\<%
\\
\>[0]\AgdaFunction{txField2Msg}\AgdaSpace{}%
\AgdaBound{inp}%
\>[17]\AgdaSymbol{=}%
\>[21]\AgdaInductiveConstructor{nat}\AgdaSpace{}%
\AgdaSymbol{(}\AgdaField{amount}\AgdaSpace{}%
\AgdaBound{inp}\AgdaSymbol{)}\AgdaSpace{}%
\AgdaInductiveConstructor{+msg}\AgdaSpace{}%
\AgdaInductiveConstructor{nat}\AgdaSpace{}%
\AgdaSymbol{(}\AgdaField{address}\AgdaSpace{}%
\AgdaBound{inp}\AgdaSymbol{)}\<%
\\
\\
\>[0]\AgdaFunction{txFieldList2Msg}\AgdaSpace{}%
\AgdaSymbol{:}\AgdaSpace{}%
\AgdaSymbol{(}\AgdaBound{inp}\AgdaSpace{}%
\AgdaSymbol{:}\AgdaSpace{}%
\AgdaDatatype{List}\AgdaSpace{}%
\AgdaRecord{TXField}\AgdaSymbol{)}\AgdaSpace{}%
\AgdaSymbol{→}\AgdaSpace{}%
\AgdaDatatype{Msg}\<%
\\
\>[0]\AgdaFunction{txFieldList2Msg}\AgdaSpace{}%
\AgdaBound{inp}%
\>[21]\AgdaSymbol{=}\AgdaSpace{}%
\AgdaInductiveConstructor{list}\AgdaSpace{}%
\AgdaSymbol{(}\AgdaFunction{mapL}\AgdaSpace{}%
\AgdaFunction{txField2Msg}\AgdaSpace{}%
\AgdaBound{inp}\AgdaSymbol{)}\<%
\end{code}
} 

\AgdaHide{
\begin{code}%
\>[0]\<%
\\
\\
\>[0]\AgdaComment{{-}{-} \textbackslash{}bitcoinVersFive}\<%
\end{code}
} 

\newcommand{\bitcoinVersFiveTXFieldToTotalAmount}{
\begin{code}%
\>[0]\AgdaFunction{txFieldList2TotalAmount}\AgdaSpace{}%
\AgdaSymbol{:}\AgdaSpace{}%
\AgdaSymbol{(}\AgdaBound{inp}\AgdaSpace{}%
\AgdaSymbol{:}\AgdaSpace{}%
\AgdaDatatype{List}\AgdaSpace{}%
\AgdaRecord{TXField}\AgdaSymbol{)}\AgdaSpace{}%
\AgdaSymbol{→}\AgdaSpace{}%
\AgdaFunction{Amount}\<%
\\
\>[0]\AgdaFunction{txFieldList2TotalAmount}\AgdaSpace{}%
\AgdaBound{inp}\AgdaSpace{}%
\AgdaSymbol{=}\AgdaSpace{}%
\AgdaFunction{sumListViaf}\AgdaSpace{}%
\AgdaField{amount}\AgdaSpace{}%
\AgdaBound{inp}\<%
\end{code}
} 

\AgdaHide{
\begin{code}%
\>[0]\<%
\\
\>[0]\AgdaComment{{-}{-} \textbackslash{}bitcoinVersFive}\<%
\end{code}
} 

\newcommand{\bitcoinVersFiveTXUnsigned}{
\begin{code}%
\>[0]\AgdaKeyword{record}\AgdaSpace{}%
\AgdaRecord{TXUnsigned}\AgdaSpace{}%
\AgdaSymbol{:}\AgdaSpace{}%
\AgdaPrimitiveType{Set}\AgdaSpace{}%
\AgdaKeyword{where}\<%
\\
\>[0][@{}l@{\AgdaIndent{0}}]%
\>[2]\agdaField%
\>[9]\AgdaField{inputs}%
\>[18]\AgdaSymbol{:}\AgdaSpace{}%
\AgdaDatatype{List}\AgdaSpace{}%
\AgdaRecord{TXField}\<%
\\
\>[2][@{}l@{\AgdaIndent{0}}]%
\>[9]\AgdaField{outputs}%
\>[18]\AgdaSymbol{:}\AgdaSpace{}%
\AgdaDatatype{List}\AgdaSpace{}%
\AgdaRecord{TXField}\<%
\end{code}
} 

\AgdaHide{
\begin{code}%
\>[0]\<%
\\
\>[0]\AgdaKeyword{open}\AgdaSpace{}%
\AgdaModule{TXUnsigned}\AgdaSpace{}%
\AgdaKeyword{public}\<%
\\
\\
\>[0]\AgdaComment{{-}{-} \textbackslash{}bitcoinVersFive}\<%
\end{code}
} 

\newcommand{\bitcoinVersFiveTXToMsg}{
\begin{code}%
\>[0]\AgdaFunction{txUnsigned2Msg}\AgdaSpace{}%
\AgdaSymbol{:}%
\>[18]\AgdaSymbol{(}\AgdaBound{transac}\AgdaSpace{}%
\AgdaSymbol{:}\AgdaSpace{}%
\AgdaRecord{TXUnsigned}\AgdaSymbol{)}\AgdaSpace{}%
\AgdaSymbol{→}\AgdaSpace{}%
\AgdaDatatype{Msg}\<%
\\
\>[0]\AgdaFunction{txUnsigned2Msg}\AgdaSpace{}%
\AgdaBound{transac}\AgdaSpace{}%
\AgdaSymbol{=}\AgdaSpace{}%
\AgdaFunction{txFieldList2Msg}\AgdaSpace{}%
\AgdaSymbol{(}\AgdaField{inputs}\AgdaSpace{}%
\AgdaBound{transac}\AgdaSymbol{)}%
\>[59]\AgdaInductiveConstructor{+msg}\AgdaSpace{}%
\AgdaFunction{txFieldList2Msg}\AgdaSpace{}%
\AgdaSymbol{(}\AgdaField{outputs}\AgdaSpace{}%
\AgdaBound{transac}\AgdaSymbol{)}\<%
\end{code}
} 

\AgdaHide{
\begin{code}%
\>[0]\<%
\\
\>[0]\AgdaComment{{-}{-} \textbackslash{}bitcoinVersFive}\<%
\end{code}
} 

\newcommand{\bitcoinVersFiveTXInputToMsg}{
\begin{code}%
\>[0]\AgdaFunction{txInput2Msg}\AgdaSpace{}%
\AgdaSymbol{:}\AgdaSpace{}%
\AgdaSymbol{(}\AgdaBound{inp}\AgdaSpace{}%
\AgdaSymbol{:}\AgdaSpace{}%
\AgdaRecord{TXField}\AgdaSymbol{)(}\AgdaBound{outp}\AgdaSpace{}%
\AgdaSymbol{:}\AgdaSpace{}%
\AgdaDatatype{List}\AgdaSpace{}%
\AgdaRecord{TXField}\AgdaSymbol{)}\AgdaSpace{}%
\AgdaSymbol{→}\AgdaSpace{}%
\AgdaDatatype{Msg}\<%
\\
\>[0]\AgdaFunction{txInput2Msg}\AgdaSpace{}%
\AgdaBound{inp}\AgdaSpace{}%
\AgdaBound{outp}\AgdaSpace{}%
\AgdaSymbol{=}\AgdaSpace{}%
\AgdaFunction{txField2Msg}\AgdaSpace{}%
\AgdaBound{inp}\AgdaSpace{}%
\AgdaInductiveConstructor{+msg}\AgdaSpace{}%
\AgdaFunction{txFieldList2Msg}\AgdaSpace{}%
\AgdaBound{outp}\<%
\end{code}
} 

\AgdaHide{
\begin{code}%
\>[0]\<%
\\
\>[0]\AgdaComment{{-}{-} \textbackslash{}bitcoinVersFive}\<%
\end{code}
} 

\newcommand{\bitcoinVersFiveTXSign}{
\begin{code}%
\>[0]\AgdaFunction{tx2Signaux}\AgdaSpace{}%
\AgdaSymbol{:}\AgdaSpace{}%
\AgdaSymbol{(}\AgdaBound{inp}\AgdaSpace{}%
\AgdaSymbol{:}\AgdaSpace{}%
\AgdaDatatype{List}\AgdaSpace{}%
\AgdaRecord{TXField}\AgdaSymbol{)}\AgdaSpace{}%
\AgdaSymbol{(}\AgdaBound{outp}\AgdaSpace{}%
\AgdaSymbol{:}\AgdaSpace{}%
\AgdaDatatype{List}\AgdaSpace{}%
\AgdaRecord{TXField}\AgdaSymbol{)}%
\>[57]\AgdaSymbol{→}\AgdaSpace{}%
\AgdaPrimitiveType{Set}\<%
\\
\>[0]\AgdaFunction{tx2Signaux}\AgdaSpace{}%
\AgdaInductiveConstructor{[]}%
\>[28]\AgdaBound{outp}%
\>[34]\AgdaSymbol{=}\AgdaSpace{}%
\AgdaRecord{⊤}\<%
\\
\>[0]\AgdaFunction{tx2Signaux}\AgdaSpace{}%
\AgdaSymbol{(}\AgdaBound{inp}\AgdaSpace{}%
\AgdaInductiveConstructor{∷}\AgdaSpace{}%
\AgdaBound{restinp}\AgdaSymbol{)}%
\>[28]\AgdaBound{outp}%
\>[34]\AgdaSymbol{=}\<%
\\
\>[0][@{}l@{\AgdaIndent{0}}]%
\>[4]\AgdaRecord{SignedWithSigPbk}\AgdaSpace{}%
\AgdaSymbol{(}\AgdaFunction{txInput2Msg}\AgdaSpace{}%
\AgdaBound{inp}\AgdaSpace{}%
\AgdaBound{outp}\AgdaSymbol{)}\AgdaSpace{}%
\AgdaSymbol{(}\AgdaField{address}\AgdaSpace{}%
\AgdaBound{inp}\AgdaSymbol{)}\AgdaSpace{}%
\AgdaFunction{×}%
\>[61]\AgdaFunction{tx2Signaux}\AgdaSpace{}%
\AgdaBound{restinp}\AgdaSpace{}%
\AgdaBound{outp}\<%
\\
\\
\>[0]\AgdaFunction{tx2Sign}\AgdaSpace{}%
\AgdaSymbol{:}\AgdaSpace{}%
\AgdaRecord{TXUnsigned}\AgdaSpace{}%
\AgdaSymbol{→}\AgdaSpace{}%
\AgdaPrimitiveType{Set}\<%
\\
\>[0]\AgdaFunction{tx2Sign}\AgdaSpace{}%
\AgdaBound{tr}\AgdaSpace{}%
\AgdaSymbol{=}\AgdaSpace{}%
\AgdaFunction{tx2Signaux}\AgdaSpace{}%
\AgdaSymbol{(}\AgdaField{inputs}\AgdaSpace{}%
\AgdaBound{tr}\AgdaSymbol{)}\AgdaSpace{}%
\AgdaSymbol{(}\AgdaField{outputs}\AgdaSpace{}%
\AgdaBound{tr}\AgdaSymbol{)}\<%
\end{code}
} 

\AgdaHide{
\begin{code}%
\>[0]\<%
\\
\\
\>[0]\AgdaComment{{-}{-} \textbackslash{}bitcoinVersFive}\<%
\end{code}
} 

\newcommand{\bitcoinVersFiveTX}{
\begin{code}%
\>[0]\AgdaKeyword{record}\AgdaSpace{}%
\AgdaRecord{TX}\AgdaSpace{}%
\AgdaSymbol{:}\AgdaSpace{}%
\AgdaPrimitiveType{Set}\AgdaSpace{}%
\AgdaKeyword{where}\<%
\\
\>[0][@{}l@{\AgdaIndent{0}}]%
\>[2]\agdaField%
\>[9]\AgdaField{tx}%
\>[18]\AgdaSymbol{:}%
\>[21]\AgdaRecord{TXUnsigned}\<%
\\
\>[2][@{}l@{\AgdaIndent{0}}]%
\>[9]\AgdaField{cor}%
\>[18]\AgdaSymbol{:}\AgdaSpace{}%
\AgdaFunction{txFieldList2TotalAmount}\AgdaSpace{}%
\AgdaSymbol{(}\AgdaField{inputs}\AgdaSpace{}%
\AgdaField{tx}\AgdaSymbol{)}\AgdaSpace{}%
\AgdaFunction{≥}\AgdaSpace{}%
\AgdaFunction{txFieldList2TotalAmount}\AgdaSpace{}%
\AgdaSymbol{(}\AgdaField{outputs}\AgdaSpace{}%
\AgdaField{tx}\AgdaSymbol{)}\<%
\\
\>[2][@{}l@{\AgdaIndent{0}}]%
\>[9]\AgdaField{nonEmpt}%
\>[18]\AgdaSymbol{:}\AgdaSpace{}%
\AgdaFunction{NonNil}\AgdaSpace{}%
\AgdaSymbol{(}\AgdaField{inputs}\AgdaSpace{}%
\AgdaField{tx}\AgdaSymbol{)}\AgdaSpace{}%
\AgdaFunction{×}\AgdaSpace{}%
\AgdaFunction{NonNil}\AgdaSpace{}%
\AgdaSymbol{(}\AgdaField{outputs}\AgdaSpace{}%
\AgdaField{tx}\AgdaSymbol{)}\<%
\\
\>[2][@{}l@{\AgdaIndent{0}}]%
\>[9]\AgdaField{sig}%
\>[18]\AgdaSymbol{:}\AgdaSpace{}%
\AgdaFunction{tx2Sign}\AgdaSpace{}%
\AgdaField{tx}\<%
\end{code}
} 

\AgdaHide{
\begin{code}%
\>[0]\<%
\\
\>[0]\AgdaKeyword{open}\AgdaSpace{}%
\AgdaModule{TX}\AgdaSpace{}%
\AgdaKeyword{public}\<%
\\
\\
\>[0]\AgdaFunction{Ledger}\AgdaSpace{}%
\AgdaSymbol{:}\AgdaSpace{}%
\AgdaPrimitiveType{Set}\<%
\\
\\
\>[0]\AgdaComment{{-}{-} \textbackslash{}bitcoinVersFive}\<%
\end{code}
} 

\newcommand{\bitcoinVersFiveLedger}{
\begin{code}%
\>[0]\AgdaFunction{Ledger}\AgdaSpace{}%
\AgdaSymbol{=}\AgdaSpace{}%
\AgdaSymbol{(}\AgdaBound{addr}\AgdaSpace{}%
\AgdaSymbol{:}\AgdaSpace{}%
\AgdaFunction{Address}\AgdaSymbol{)}\AgdaSpace{}%
\AgdaSymbol{→}\AgdaSpace{}%
\AgdaFunction{Amount}\<%
\\
\\
\>[0]\AgdaFunction{initialLedger}\AgdaSpace{}%
\AgdaSymbol{:}\AgdaSpace{}%
\AgdaFunction{Ledger}\<%
\\
\>[0]\AgdaFunction{initialLedger}\AgdaSpace{}%
\AgdaBound{addr}\AgdaSpace{}%
\AgdaSymbol{=}\AgdaSpace{}%
\AgdaNumber{0}\<%
\end{code}
} 

\AgdaHide{
\begin{code}%
\>[0]\<%
\\
\>[0]\AgdaComment{{-}{-} \textbackslash{}bitcoinVersFive}\<%
\end{code}
} 

\newcommand{\bitcoinVersFiveaddTXFieldToLedger}{
\begin{code}%
\>[0]\AgdaFunction{addTXFieldToLedger}\AgdaSpace{}%
\AgdaSymbol{:}%
\>[22]\AgdaSymbol{(}\AgdaBound{tr}\AgdaSpace{}%
\AgdaSymbol{:}\AgdaSpace{}%
\AgdaRecord{TXField}\AgdaSymbol{)(}\AgdaBound{oldLedger}\AgdaSpace{}%
\AgdaSymbol{:}\AgdaSpace{}%
\AgdaFunction{Ledger}\AgdaSymbol{)}\AgdaSpace{}%
\AgdaSymbol{→}\AgdaSpace{}%
\AgdaFunction{Ledger}\<%
\\
\>[0]\AgdaFunction{addTXFieldToLedger}%
\>[20]\AgdaBound{tr}\AgdaSpace{}%
\AgdaBound{oldLedger}\AgdaSpace{}%
\AgdaBound{pubk}\AgdaSpace{}%
\AgdaSymbol{=}\<%
\\
\>[0][@{}l@{\AgdaIndent{0}}]%
\>[9]\AgdaFunction{if}%
\>[14]\AgdaBound{pubk}\AgdaSpace{}%
\AgdaFunction{≡ℕb}\AgdaSpace{}%
\AgdaField{address}\AgdaSpace{}%
\AgdaBound{tr}\AgdaSpace{}%
\AgdaFunction{then}\AgdaSpace{}%
\AgdaBound{oldLedger}\AgdaSpace{}%
\AgdaBound{pubk}\AgdaSpace{}%
\AgdaPrimitive{+}%
\>[57]\AgdaField{amount}\AgdaSpace{}%
\AgdaBound{tr}\AgdaSpace{}%
\AgdaFunction{else}\AgdaSpace{}%
\AgdaBound{oldLedger}\AgdaSpace{}%
\AgdaBound{pubk}\<%
\\
\\
\>[0]\AgdaFunction{addTXFieldListToLedger}%
\>[24]\AgdaSymbol{:}%
\>[27]\AgdaSymbol{(}\AgdaBound{tr}\AgdaSpace{}%
\AgdaSymbol{:}\AgdaSpace{}%
\AgdaDatatype{List}\AgdaSpace{}%
\AgdaRecord{TXField}\AgdaSymbol{)(}\AgdaBound{oldLedger}\AgdaSpace{}%
\AgdaSymbol{:}\AgdaSpace{}%
\AgdaFunction{Ledger}\AgdaSymbol{)}\AgdaSpace{}%
\AgdaSymbol{→}\AgdaSpace{}%
\AgdaFunction{Ledger}\<%
\\
\>[0]\AgdaFunction{addTXFieldListToLedger}\AgdaSpace{}%
\AgdaInductiveConstructor{[]}%
\>[33]\AgdaBound{oldLedger}%
\>[44]\AgdaSymbol{=}%
\>[47]\AgdaBound{oldLedger}\<%
\\
\>[0]\AgdaFunction{addTXFieldListToLedger}\AgdaSpace{}%
\AgdaSymbol{(}\AgdaBound{x}\AgdaSpace{}%
\AgdaInductiveConstructor{∷}\AgdaSpace{}%
\AgdaBound{tr}\AgdaSymbol{)}%
\>[33]\AgdaBound{oldLedger}%
\>[44]\AgdaSymbol{=}\<%
\\
\>[0][@{}l@{\AgdaIndent{0}}]%
\>[6]\AgdaFunction{addTXFieldListToLedger}\AgdaSpace{}%
\AgdaBound{tr}\AgdaSpace{}%
\AgdaSymbol{(}\AgdaFunction{addTXFieldToLedger}\AgdaSpace{}%
\AgdaBound{x}\AgdaSpace{}%
\AgdaBound{oldLedger}\AgdaSymbol{)}\<%
\end{code}
} 

\AgdaHide{
\begin{code}%
\>[0]\<%
\\
\\
\>[0]\AgdaComment{{-}{-} \textbackslash{}bitcoinVersFive}\<%
\end{code}
} 

\newcommand{\bitcoinVersFivesubtrTXFieldFromLedger}{
\begin{code}%
\>[0]\AgdaFunction{subtrTXFieldFromLedger}%
\>[28]\AgdaSymbol{:}%
\>[31]\AgdaSymbol{(}\AgdaBound{tr}\AgdaSpace{}%
\AgdaSymbol{:}\AgdaSpace{}%
\AgdaRecord{TXField}\AgdaSymbol{)}%
\>[52]\AgdaSymbol{(}\AgdaBound{oldLedger}\AgdaSpace{}%
\AgdaSymbol{:}\AgdaSpace{}%
\AgdaFunction{Ledger}\AgdaSymbol{)}%
\>[74]\AgdaSymbol{→}%
\>[77]\AgdaFunction{Ledger}\<%
\\
\>[0]\AgdaFunction{subtrTXFieldListFromLedger}%
\>[28]\AgdaSymbol{:}%
\>[31]\AgdaSymbol{(}\AgdaBound{tr}\AgdaSpace{}%
\AgdaSymbol{:}\AgdaSpace{}%
\AgdaDatatype{List}\AgdaSpace{}%
\AgdaRecord{TXField}\AgdaSymbol{)}%
\>[52]\AgdaSymbol{(}\AgdaBound{oldLedger}\AgdaSpace{}%
\AgdaSymbol{:}\AgdaSpace{}%
\AgdaFunction{Ledger}\AgdaSymbol{)}%
\>[74]\AgdaSymbol{→}%
\>[77]\AgdaFunction{Ledger}\<%
\end{code}
} 

\AgdaHide{
\begin{code}%
\>[0]\AgdaFunction{subtrTXFieldFromLedger}%
\>[24]\AgdaBound{tr}\AgdaSpace{}%
\AgdaBound{oldLedger}\AgdaSpace{}%
\AgdaBound{pubk}\AgdaSpace{}%
\AgdaSymbol{=}\<%
\\
\>[0][@{}l@{\AgdaIndent{0}}]%
\>[9]\AgdaFunction{if}%
\>[14]\AgdaBound{pubk}\AgdaSpace{}%
\AgdaFunction{≡ℕb}\AgdaSpace{}%
\AgdaField{address}\AgdaSpace{}%
\AgdaBound{tr}\<%
\\
\>[0][@{}l@{\AgdaIndent{0}}]%
\>[9]\AgdaFunction{then}\AgdaSpace{}%
\AgdaBound{oldLedger}\AgdaSpace{}%
\AgdaBound{pubk}\AgdaSpace{}%
\AgdaPrimitive{∸}%
\>[32]\AgdaField{amount}\AgdaSpace{}%
\AgdaBound{tr}\<%
\\
\>[0][@{}l@{\AgdaIndent{0}}]%
\>[9]\AgdaFunction{else}\AgdaSpace{}%
\AgdaBound{oldLedger}\AgdaSpace{}%
\AgdaBound{pubk}\<%
\\
\\
\>[0]\AgdaFunction{subtrTXFieldListFromLedger}\AgdaSpace{}%
\AgdaInductiveConstructor{[]}\AgdaSpace{}%
\AgdaBound{oldLedger}\AgdaSpace{}%
\AgdaSymbol{=}\AgdaSpace{}%
\AgdaBound{oldLedger}\<%
\\
\>[0]\AgdaFunction{subtrTXFieldListFromLedger}\AgdaSpace{}%
\AgdaSymbol{(}\AgdaBound{x}\AgdaSpace{}%
\AgdaInductiveConstructor{∷}\AgdaSpace{}%
\AgdaBound{tr}\AgdaSymbol{)}\AgdaSpace{}%
\AgdaBound{oldLedger}\AgdaSpace{}%
\AgdaSymbol{=}\<%
\\
\>[0][@{}l@{\AgdaIndent{0}}]%
\>[11]\AgdaFunction{subtrTXFieldListFromLedger}\AgdaSpace{}%
\AgdaBound{tr}\AgdaSpace{}%
\AgdaSymbol{(}\AgdaFunction{subtrTXFieldFromLedger}\AgdaSpace{}%
\AgdaBound{x}\AgdaSpace{}%
\AgdaBound{oldLedger}\AgdaSymbol{)}\<%
\end{code}
} 

\AgdaHide{
\begin{code}%
\>[0]\<%
\\
\>[0]\AgdaComment{{-}{-} \textbackslash{}bitcoinVersFive}\<%
\end{code}
} 

\newcommand{\bitcoinVersFiveupdateLedgerByTX}{
\begin{code}%
\>[0]\AgdaFunction{updateLedgerByTX}\AgdaSpace{}%
\AgdaSymbol{:}%
\>[20]\AgdaSymbol{(}\AgdaBound{tr}\AgdaSpace{}%
\AgdaSymbol{:}\AgdaSpace{}%
\AgdaRecord{TX}\AgdaSymbol{)(}\AgdaBound{oldLedger}\AgdaSpace{}%
\AgdaSymbol{:}\AgdaSpace{}%
\AgdaFunction{Ledger}\AgdaSymbol{)}\AgdaSpace{}%
\AgdaSymbol{→}\AgdaSpace{}%
\AgdaFunction{Ledger}\<%
\\
\>[0]\AgdaFunction{updateLedgerByTX}\AgdaSpace{}%
\AgdaBound{tr}\AgdaSpace{}%
\AgdaBound{oldLedger}\AgdaSpace{}%
\AgdaSymbol{=}%
\>[33]\AgdaFunction{addTXFieldListToLedger}\AgdaSpace{}%
\AgdaSymbol{(}\AgdaField{outputs}\AgdaSpace{}%
\AgdaSymbol{(}\AgdaField{tx}\AgdaSpace{}%
\AgdaBound{tr}\AgdaSymbol{))}\<%
\\
\>[33][@{}l@{\AgdaIndent{0}}]%
\>[35]\AgdaSymbol{(}\AgdaFunction{subtrTXFieldListFromLedger}\AgdaSpace{}%
\AgdaSymbol{(}\AgdaField{inputs}\AgdaSpace{}%
\AgdaSymbol{(}\AgdaField{tx}\AgdaSpace{}%
\AgdaBound{tr}\AgdaSymbol{))}\AgdaSpace{}%
\AgdaBound{oldLedger}\AgdaSpace{}%
\AgdaSymbol{)}\<%
\end{code}
} 

\AgdaHide{
\begin{code}%
\>[0]\<%
\\
\\
\>[0]\AgdaComment{{-}{-} \textbackslash{}bitcoinVersFive}\<%
\end{code}
} 

\newcommand{\bitcoinVersFivecorrectInput}{
\begin{code}%
\>[0]\AgdaFunction{correctInput}\AgdaSpace{}%
\AgdaSymbol{:}\AgdaSpace{}%
\AgdaSymbol{(}\AgdaBound{tr}\AgdaSpace{}%
\AgdaSymbol{:}\AgdaSpace{}%
\AgdaRecord{TXField}\AgdaSymbol{)}\AgdaSpace{}%
\AgdaSymbol{(}\AgdaBound{ledger}\AgdaSpace{}%
\AgdaSymbol{:}\AgdaSpace{}%
\AgdaFunction{Ledger}\AgdaSymbol{)}\AgdaSpace{}%
\AgdaSymbol{→}\AgdaSpace{}%
\AgdaPrimitiveType{Set}\<%
\\
\>[0]\AgdaFunction{correctInput}\AgdaSpace{}%
\AgdaBound{tr}\AgdaSpace{}%
\AgdaBound{ledger}\AgdaSpace{}%
\AgdaSymbol{=}\AgdaSpace{}%
\AgdaBound{ledger}\AgdaSpace{}%
\AgdaSymbol{(}\AgdaField{address}\AgdaSpace{}%
\AgdaBound{tr}\AgdaSymbol{)}\AgdaSpace{}%
\AgdaFunction{≥}\AgdaSpace{}%
\AgdaField{amount}\AgdaSpace{}%
\AgdaBound{tr}\<%
\end{code}
} 

\AgdaHide{
\begin{code}%
\>[0]\<%
\\
\>[0]\AgdaComment{{-}{-} \textbackslash{}bitcoinVersFive}\<%
\end{code}
} 

\newcommand{\bitcoinVersFivecorrectInputs}{
\begin{code}%
\>[0]\AgdaFunction{correctInputs}\AgdaSpace{}%
\AgdaSymbol{:}\AgdaSpace{}%
\AgdaSymbol{(}\AgdaBound{tr}\AgdaSpace{}%
\AgdaSymbol{:}\AgdaSpace{}%
\AgdaDatatype{List}\AgdaSpace{}%
\AgdaRecord{TXField}\AgdaSymbol{)}\AgdaSpace{}%
\AgdaSymbol{(}\AgdaBound{ledger}\AgdaSpace{}%
\AgdaSymbol{:}\AgdaSpace{}%
\AgdaFunction{Ledger}\AgdaSymbol{)}\AgdaSpace{}%
\AgdaSymbol{→}\AgdaSpace{}%
\AgdaPrimitiveType{Set}\<%
\\
\>[0]\AgdaFunction{correctInputs}\AgdaSpace{}%
\AgdaInductiveConstructor{[]}%
\>[24]\AgdaBound{ledger}%
\>[32]\AgdaSymbol{=}\AgdaSpace{}%
\AgdaRecord{⊤}\<%
\\
\>[0]\AgdaFunction{correctInputs}\AgdaSpace{}%
\AgdaSymbol{(}\AgdaBound{x}\AgdaSpace{}%
\AgdaInductiveConstructor{∷}\AgdaSpace{}%
\AgdaBound{tr}\AgdaSymbol{)}%
\>[24]\AgdaBound{ledger}%
\>[32]\AgdaSymbol{=}%
\>[358I]\AgdaFunction{correctInput}\AgdaSpace{}%
\AgdaBound{x}\AgdaSpace{}%
\AgdaBound{ledger}\AgdaSpace{}%
\AgdaFunction{×}\<%
\\
\>[32][@{}l@{\AgdaIndent{0}}]\<[358I]%
\>[34]\AgdaFunction{correctInputs}\AgdaSpace{}%
\AgdaBound{tr}\AgdaSpace{}%
\AgdaSymbol{(}\AgdaFunction{subtrTXFieldFromLedger}\AgdaSpace{}%
\AgdaBound{x}\AgdaSpace{}%
\AgdaBound{ledger}\AgdaSymbol{)}\<%
\\
\\
\>[0]\AgdaFunction{correctTX}\AgdaSpace{}%
\AgdaSymbol{:}\AgdaSpace{}%
\AgdaSymbol{(}\AgdaBound{tr}\AgdaSpace{}%
\AgdaSymbol{:}\AgdaSpace{}%
\AgdaRecord{TX}\AgdaSymbol{)}\AgdaSpace{}%
\AgdaSymbol{(}\AgdaBound{ledger}\AgdaSpace{}%
\AgdaSymbol{:}\AgdaSpace{}%
\AgdaFunction{Ledger}\AgdaSymbol{)}\AgdaSpace{}%
\AgdaSymbol{→}\AgdaSpace{}%
\AgdaPrimitiveType{Set}\<%
\\
\>[0]\AgdaFunction{correctTX}\AgdaSpace{}%
\AgdaBound{tr}\AgdaSpace{}%
\AgdaBound{ledger}\AgdaSpace{}%
\AgdaSymbol{=}\AgdaSpace{}%
\AgdaFunction{correctInputs}\AgdaSpace{}%
\AgdaSymbol{(}\AgdaField{outputs}\AgdaSpace{}%
\AgdaSymbol{(}\AgdaField{tx}\AgdaSpace{}%
\AgdaBound{tr}\AgdaSymbol{))}\AgdaSpace{}%
\AgdaBound{ledger}\<%
\end{code}
} 

\AgdaHide{
\begin{code}%
\>[0]\<%
\\
\\
\\
\>[0]\AgdaFunction{UnMinedBlock}\AgdaSpace{}%
\AgdaSymbol{:}\AgdaSpace{}%
\AgdaPrimitiveType{Set}\<%
\\
\\
\>[0]\AgdaComment{{-}{-} \textbackslash{}bitcoinVersFive}\<%
\end{code}
} 

\newcommand{\bitcoinVersFiveUnminedBlock}{
\begin{code}%
\>[0]\AgdaFunction{UnMinedBlock}\AgdaSpace{}%
\AgdaSymbol{=}\AgdaSpace{}%
\AgdaDatatype{List}\AgdaSpace{}%
\AgdaRecord{TX}\<%
\end{code}
} 

\AgdaHide{
\begin{code}%
\>[0]\<%
\\
\\
\>[0]\AgdaComment{{-}{-} \textbackslash{}bitcoinVersFive}\<%
\end{code}
} 

\newcommand{\bitcoinVersFivetxtoTXfee}{
\begin{code}%
\>[0]\AgdaFunction{tx2TXFee}\AgdaSpace{}%
\AgdaSymbol{:}\AgdaSpace{}%
\AgdaRecord{TX}\AgdaSpace{}%
\AgdaSymbol{→}\AgdaSpace{}%
\AgdaFunction{Amount}\<%
\\
\>[0]\AgdaFunction{tx2TXFee}\AgdaSpace{}%
\AgdaBound{tr}\AgdaSpace{}%
\AgdaSymbol{=}\<%
\\
\>[0][@{}l@{\AgdaIndent{0}}]%
\>[4]\AgdaFunction{txFieldList2TotalAmount}\AgdaSpace{}%
\AgdaSymbol{(}\AgdaField{outputs}\AgdaSpace{}%
\AgdaSymbol{(}\AgdaField{tx}\AgdaSpace{}%
\AgdaBound{tr}\AgdaSymbol{))}\AgdaSpace{}%
\AgdaPrimitive{∸}\AgdaSpace{}%
\AgdaFunction{txFieldList2TotalAmount}\AgdaSpace{}%
\AgdaSymbol{(}\AgdaField{inputs}\AgdaSpace{}%
\AgdaSymbol{(}\AgdaField{tx}\AgdaSpace{}%
\AgdaBound{tr}\AgdaSymbol{))}\<%
\\
\\
\>[0]\AgdaFunction{unMinedBlock2TXFee}\AgdaSpace{}%
\AgdaSymbol{:}\AgdaSpace{}%
\AgdaFunction{UnMinedBlock}\AgdaSpace{}%
\AgdaSymbol{→}\AgdaSpace{}%
\AgdaFunction{Amount}\<%
\\
\>[0]\AgdaFunction{unMinedBlock2TXFee}\AgdaSpace{}%
\AgdaBound{bl}\AgdaSpace{}%
\AgdaSymbol{=}\AgdaSpace{}%
\AgdaFunction{sumListViaf}\AgdaSpace{}%
\AgdaFunction{tx2TXFee}%
\>[46]\AgdaBound{bl}\<%
\end{code}
} 

\AgdaHide{
\begin{code}%
\>[0]\<%
\\
\\
\\
\>[0]\AgdaComment{{-}{-} \textbackslash{}bitcoinVersFive}\<%
\end{code}
} 

\newcommand{\bitcoinVersFivecorrectBlock}{
\begin{code}%
\>[0]\AgdaFunction{correctUnminedBlock}\AgdaSpace{}%
\AgdaSymbol{:}\AgdaSpace{}%
\AgdaSymbol{(}\AgdaBound{block}\AgdaSpace{}%
\AgdaSymbol{:}\AgdaSpace{}%
\AgdaFunction{UnMinedBlock}\AgdaSymbol{)(}\AgdaBound{oldLedger}\AgdaSpace{}%
\AgdaSymbol{:}\AgdaSpace{}%
\AgdaFunction{Ledger}\AgdaSymbol{)→}\AgdaSpace{}%
\AgdaPrimitiveType{Set}\<%
\\
\>[0]\AgdaFunction{correctUnminedBlock}%
\>[21]\AgdaInductiveConstructor{[]}%
\>[35]\AgdaBound{oldLedger}%
\>[46]\AgdaSymbol{=}\AgdaSpace{}%
\AgdaRecord{⊤}\<%
\\
\>[0]\AgdaFunction{correctUnminedBlock}%
\>[21]\AgdaSymbol{(}\AgdaBound{tr}\AgdaSpace{}%
\AgdaInductiveConstructor{∷}\AgdaSpace{}%
\AgdaBound{block}\AgdaSymbol{)}%
\>[35]\AgdaBound{oldLedger}%
\>[46]\AgdaSymbol{=}\<%
\\
\>[0][@{}l@{\AgdaIndent{0}}]%
\>[4]\AgdaFunction{correctTX}\AgdaSpace{}%
\AgdaBound{tr}\AgdaSpace{}%
\AgdaBound{oldLedger}\AgdaSpace{}%
\AgdaFunction{×}%
\>[30]\AgdaFunction{correctUnminedBlock}%
\>[51]\AgdaBound{block}\AgdaSpace{}%
\AgdaSymbol{(}\AgdaFunction{updateLedgerByTX}\AgdaSpace{}%
\AgdaBound{tr}\AgdaSpace{}%
\AgdaBound{oldLedger}\AgdaSymbol{)}\<%
\\
\\
\>[0]\AgdaFunction{updateLedgerUnminedBlock}\AgdaSpace{}%
\AgdaSymbol{:}\AgdaSpace{}%
\AgdaSymbol{(}\AgdaBound{block}\AgdaSpace{}%
\AgdaSymbol{:}\AgdaSpace{}%
\AgdaFunction{UnMinedBlock}\AgdaSymbol{)(}\AgdaBound{oldLedger}\AgdaSpace{}%
\AgdaSymbol{:}\AgdaSpace{}%
\AgdaFunction{Ledger}\AgdaSymbol{)}\AgdaSpace{}%
\AgdaSymbol{→}\AgdaSpace{}%
\AgdaFunction{Ledger}\<%
\\
\>[0]\AgdaFunction{updateLedgerUnminedBlock}\AgdaSpace{}%
\AgdaInductiveConstructor{[]}%
\>[39]\AgdaBound{oldLedger}%
\>[50]\AgdaSymbol{=}\AgdaSpace{}%
\AgdaBound{oldLedger}\<%
\\
\>[0]\AgdaFunction{updateLedgerUnminedBlock}\AgdaSpace{}%
\AgdaSymbol{(}\AgdaBound{tr}\AgdaSpace{}%
\AgdaInductiveConstructor{∷}\AgdaSpace{}%
\AgdaBound{block}\AgdaSymbol{)}%
\>[39]\AgdaBound{oldLedger}%
\>[50]\AgdaSymbol{=}\<%
\\
\>[0][@{}l@{\AgdaIndent{0}}]%
\>[4]\AgdaFunction{updateLedgerUnminedBlock}\AgdaSpace{}%
\AgdaBound{block}\AgdaSpace{}%
\AgdaSymbol{(}\AgdaFunction{updateLedgerByTX}\AgdaSpace{}%
\AgdaBound{tr}\AgdaSpace{}%
\AgdaBound{oldLedger}\AgdaSymbol{)}\<%
\end{code}
} 

\AgdaHide{
\begin{code}%
\>[0]\<%
\\
\>[0]\AgdaFunction{BlockUnchecked}\AgdaSpace{}%
\AgdaSymbol{:}\AgdaSpace{}%
\AgdaPrimitiveType{Set}\<%
\\
\\
\>[0]\AgdaComment{{-}{-} \textbackslash{}bitcoinVersFive}\<%
\end{code}
} 

\newcommand{\bitcoinVersFiveBlockUnchecked}{
\begin{code}%
\>[0]\AgdaFunction{BlockUnchecked}\AgdaSpace{}%
\AgdaSymbol{=}\AgdaSpace{}%
\AgdaDatatype{List}\AgdaSpace{}%
\AgdaRecord{TXField}\AgdaSpace{}%
\AgdaFunction{×}\AgdaSpace{}%
\AgdaFunction{UnMinedBlock}\<%
\\
\\
\>[0]\AgdaFunction{block2TXFee}\AgdaSpace{}%
\AgdaSymbol{:}\AgdaSpace{}%
\AgdaFunction{BlockUnchecked}\AgdaSpace{}%
\AgdaSymbol{→}\AgdaSpace{}%
\AgdaFunction{Amount}\<%
\\
\>[0]\AgdaFunction{block2TXFee}\AgdaSpace{}%
\AgdaSymbol{(}\AgdaBound{outputMiner}\AgdaSpace{}%
\AgdaInductiveConstructor{,}\AgdaSpace{}%
\AgdaBound{block}\AgdaSymbol{)}\AgdaSpace{}%
\AgdaSymbol{=}\AgdaSpace{}%
\AgdaFunction{unMinedBlock2TXFee}\AgdaSpace{}%
\AgdaBound{block}\<%
\end{code}
} 

\AgdaHide{
\begin{code}%
\>[0]\<%
\\
\>[0]\AgdaComment{\{{-}
upDateLedgerMining : (minerOutput  : List TXField)
                     (oldLedger : Ledger)
                     → Ledger
upDateLedgerMining minerOutput oldLedger =
           addTXFieldListToLedger minerOutput oldLedger
{-}{-}              (txField reward miner)
{-}\}}\<%
\\
\\
\>[0]\AgdaComment{{-}{-} \textbackslash{}bitcoinVersFive}\<%
\end{code}
} 

\newcommand{\bitcoinVersFiveCorrectMinedBlock}{
\begin{code}%
\>[0]\AgdaFunction{correctMinedBlock}\AgdaSpace{}%
\AgdaSymbol{:}\AgdaSpace{}%
\AgdaSymbol{(}\AgdaBound{reward}\AgdaSpace{}%
\AgdaSymbol{:}\AgdaSpace{}%
\AgdaFunction{Amount}\AgdaSymbol{)(}\AgdaBound{block}\AgdaSpace{}%
\AgdaSymbol{:}\AgdaSpace{}%
\AgdaFunction{BlockUnchecked}\AgdaSymbol{)(}\AgdaBound{oldLedger}\AgdaSpace{}%
\AgdaSymbol{:}\AgdaSpace{}%
\AgdaFunction{Ledger}\AgdaSymbol{)}\AgdaSpace{}%
\AgdaSymbol{→}\AgdaSpace{}%
\AgdaPrimitiveType{Set}\<%
\\
\\
\>[0]\AgdaFunction{correctMinedBlock}\AgdaSpace{}%
\AgdaBound{reward}\AgdaSpace{}%
\AgdaSymbol{(}\AgdaBound{outputMiner}\AgdaSpace{}%
\AgdaInductiveConstructor{,}\AgdaSpace{}%
\AgdaBound{block}\AgdaSymbol{)}\AgdaSpace{}%
\AgdaBound{oldLedger}\AgdaSpace{}%
\AgdaSymbol{=}\<%
\\
\>[0][@{}l@{\AgdaIndent{0}}]%
\>[6]\AgdaFunction{correctUnminedBlock}\AgdaSpace{}%
\AgdaBound{block}\AgdaSpace{}%
\AgdaBound{oldLedger}\AgdaSpace{}%
\AgdaFunction{×}\<%
\\
\>[0][@{}l@{\AgdaIndent{0}}]%
\>[6]\AgdaFunction{txFieldList2TotalAmount}\AgdaSpace{}%
\AgdaBound{outputMiner}\AgdaSpace{}%
\AgdaFunction{≡ℕ}\AgdaSpace{}%
\AgdaSymbol{(}\AgdaBound{reward}\AgdaSpace{}%
\AgdaPrimitive{+}\AgdaSpace{}%
\AgdaFunction{unMinedBlock2TXFee}\AgdaSpace{}%
\AgdaBound{block}\AgdaSymbol{)}\<%
\end{code}
} 

\AgdaHide{
\begin{code}%
\>[0]\AgdaComment{{-}{-}          (upDateLedgerMining reward miner )}\<%
\\
\\
\>[0]\AgdaComment{{-}{-} \textbackslash{}bitcoinVersFive}\<%
\end{code}
} 

\newcommand{\bitcoinVersFiveupdateLedgerBlock}{
\begin{code}%
\>[0]\AgdaFunction{updateLedgerBlock}\AgdaSpace{}%
\AgdaSymbol{:}\AgdaSpace{}%
\AgdaSymbol{(}\AgdaBound{block}\AgdaSpace{}%
\AgdaSymbol{:}\AgdaSpace{}%
\AgdaFunction{BlockUnchecked}\AgdaSymbol{)(}\AgdaBound{oldLedger}\AgdaSpace{}%
\AgdaSymbol{:}\AgdaSpace{}%
\AgdaFunction{Ledger}\AgdaSymbol{)→}\AgdaSpace{}%
\AgdaFunction{Ledger}\<%
\\
\>[0]\AgdaFunction{updateLedgerBlock}\AgdaSpace{}%
\AgdaSymbol{(}\AgdaBound{outputMiner}\AgdaSpace{}%
\AgdaInductiveConstructor{,}\AgdaSpace{}%
\AgdaBound{block}\AgdaSymbol{)}\AgdaSpace{}%
\AgdaBound{oldLedger}\AgdaSpace{}%
\AgdaSymbol{=}\<%
\\
\>[0][@{}l@{\AgdaIndent{0}}]%
\>[3]\AgdaFunction{addTXFieldListToLedger}%
\>[27]\AgdaBound{outputMiner}\AgdaSpace{}%
\AgdaSymbol{(}\AgdaFunction{updateLedgerUnminedBlock}\AgdaSpace{}%
\AgdaBound{block}\AgdaSpace{}%
\AgdaBound{oldLedger}\AgdaSymbol{)}\<%
\end{code}
} 

\AgdaHide{
\begin{code}%
\>[0]\<%
\\
\>[0]\AgdaComment{{-}{-} next blockchain}\<%
\\
\>[0]\AgdaComment{{-}{-} reward can be a function f : Time → Amount}\<%
\\
\\
\>[0]\AgdaFunction{BlockChainUnchecked}\AgdaSpace{}%
\AgdaSymbol{:}\AgdaSpace{}%
\AgdaPrimitiveType{Set}\<%
\\
\\
\>[0]\AgdaComment{{-}{-} \textbackslash{}bitcoinVersFive}\<%
\end{code}
} 

\newcommand{\bitcoinVersFiveBlockChainUnchecked}{
\begin{code}%
\>[0]\AgdaFunction{BlockChainUnchecked}\AgdaSpace{}%
\AgdaSymbol{=}\AgdaSpace{}%
\AgdaDatatype{List}\AgdaSpace{}%
\AgdaFunction{BlockUnchecked}\<%
\end{code}
} 

\AgdaHide{
\begin{code}%
\>[0]\<%
\\
\>[0]\AgdaComment{{-}{-} \textbackslash{}bitcoinVersFive}\<%
\end{code}
} 

\newcommand{\bitcoinVersFiveCorrectBlockChain}{
\begin{code}%
\>[0]\AgdaFunction{CorrectBlockChain}%
\>[505I]\AgdaSymbol{:}%
\>[506I]\AgdaSymbol{(}\AgdaBound{blockReward}\AgdaSpace{}%
\AgdaSymbol{:}\AgdaSpace{}%
\AgdaFunction{Time}\AgdaSpace{}%
\AgdaSymbol{→}\AgdaSpace{}%
\AgdaFunction{Amount}\AgdaSymbol{)}\<%
\\
\>[505I][@{}l@{\AgdaIndent{0}}]\<[506I]%
\>[20]\AgdaSymbol{(}\AgdaBound{startTime}\AgdaSpace{}%
\AgdaSymbol{:}\AgdaSpace{}%
\AgdaFunction{Time}\AgdaSymbol{)}\<%
\\
\>[505I][@{}l@{\AgdaIndent{0}}]%
\>[20]\AgdaSymbol{(}\AgdaBound{sartLedger}\AgdaSpace{}%
\AgdaSymbol{:}\AgdaSpace{}%
\AgdaFunction{Ledger}\AgdaSymbol{)}\<%
\\
\>[505I][@{}l@{\AgdaIndent{0}}]%
\>[20]\AgdaSymbol{(}\AgdaBound{bc}%
\>[25]\AgdaSymbol{:}\AgdaSpace{}%
\AgdaFunction{BlockChainUnchecked}\AgdaSymbol{)}\<%
\\
\>[505I][@{}l@{\AgdaIndent{0}}]%
\>[20]\AgdaSymbol{→}\AgdaSpace{}%
\AgdaPrimitiveType{Set}\<%
\\
\>[0]\AgdaFunction{CorrectBlockChain}\AgdaSpace{}%
\AgdaBound{blockReward}\AgdaSpace{}%
\AgdaBound{startTime}\AgdaSpace{}%
\AgdaBound{startLedger}\AgdaSpace{}%
\AgdaInductiveConstructor{[]}\AgdaSpace{}%
\AgdaSymbol{=}\AgdaSpace{}%
\AgdaRecord{⊤}\<%
\\
\>[0]\AgdaFunction{CorrectBlockChain}\AgdaSpace{}%
\AgdaBound{blockReward}\AgdaSpace{}%
\AgdaBound{startTime}\AgdaSpace{}%
\AgdaBound{startLedger}\AgdaSpace{}%
\AgdaSymbol{(}\AgdaBound{block}\AgdaSpace{}%
\AgdaInductiveConstructor{∷}\AgdaSpace{}%
\AgdaBound{restbc}\AgdaSymbol{)}\<%
\\
\>[0][@{}l@{\AgdaIndent{0}}]%
\>[3]\AgdaSymbol{=}%
\>[529I]\AgdaFunction{correctMinedBlock}\AgdaSpace{}%
\AgdaSymbol{(}\AgdaBound{blockReward}\AgdaSpace{}%
\AgdaBound{startTime}\AgdaSymbol{)}\AgdaSpace{}%
\AgdaBound{block}\AgdaSpace{}%
\AgdaBound{startLedger}\<%
\\
\>[3][@{}l@{\AgdaIndent{0}}]\<[529I]%
\>[5]\AgdaFunction{×}%
\>[8]\AgdaFunction{CorrectBlockChain}\AgdaSpace{}%
\AgdaBound{blockReward}\AgdaSpace{}%
\AgdaSymbol{(}\AgdaInductiveConstructor{suc}\AgdaSpace{}%
\AgdaBound{startTime}\AgdaSymbol{)}\<%
\\
\>[5][@{}l@{\AgdaIndent{0}}]%
\>[8]\AgdaSymbol{(}\AgdaFunction{updateLedgerBlock}\AgdaSpace{}%
\AgdaBound{block}\AgdaSpace{}%
\AgdaBound{startLedger}\AgdaSymbol{)}%
\>[47]\AgdaBound{restbc}\<%
\end{code}
} 

\AgdaHide{
\begin{code}%
\>[0]\<%
\\
\>[0]\AgdaComment{{-}{-} \textbackslash{}bitcoinVersFive}\<%
\end{code}
} 

\newcommand{\bitcoinVersFiveFinalLedger}{
\begin{code}%
\>[0]\AgdaFunction{FinalLedger}\AgdaSpace{}%
\AgdaSymbol{:}%
\>[15]\AgdaSymbol{(}\AgdaBound{txFeePrevious}\AgdaSpace{}%
\AgdaSymbol{:}\AgdaSpace{}%
\AgdaFunction{Amount}\AgdaSymbol{)}%
\>[44]\AgdaSymbol{(}\AgdaBound{blockReward}\AgdaSpace{}%
\AgdaSymbol{:}\AgdaSpace{}%
\AgdaFunction{Time}\AgdaSpace{}%
\AgdaSymbol{→}\AgdaSpace{}%
\AgdaFunction{Amount}\AgdaSymbol{)}\<%
\\
\>[15][@{}l@{\AgdaIndent{0}}]%
\>[16]\AgdaSymbol{(}\AgdaBound{startTime}\AgdaSpace{}%
\AgdaSymbol{:}\AgdaSpace{}%
\AgdaFunction{Time}\AgdaSymbol{)}%
\>[45]\AgdaSymbol{(}\AgdaBound{startLedger}\AgdaSpace{}%
\AgdaSymbol{:}\AgdaSpace{}%
\AgdaFunction{Ledger}\AgdaSymbol{)}\<%
\\
\>[15][@{}l@{\AgdaIndent{0}}]%
\>[16]\AgdaSymbol{(}\AgdaBound{bc}%
\>[21]\AgdaSymbol{:}\AgdaSpace{}%
\AgdaFunction{BlockChainUnchecked}\AgdaSymbol{)}%
\>[45]\AgdaSymbol{→}\AgdaSpace{}%
\AgdaFunction{Ledger}\<%
\\
\>[0]\AgdaFunction{FinalLedger}\AgdaSpace{}%
\AgdaBound{trfee}\AgdaSpace{}%
\AgdaBound{blockReward}\AgdaSpace{}%
\AgdaBound{startTime}\AgdaSpace{}%
\AgdaBound{startLedger}\AgdaSpace{}%
\AgdaInductiveConstructor{[]}\AgdaSpace{}%
\AgdaSymbol{=}\AgdaSpace{}%
\AgdaBound{startLedger}\<%
\\
\>[0]\AgdaFunction{FinalLedger}\AgdaSpace{}%
\AgdaBound{trfee}\AgdaSpace{}%
\AgdaBound{blockReward}\AgdaSpace{}%
\AgdaBound{startTime}\AgdaSpace{}%
\AgdaBound{startLedger}\AgdaSpace{}%
\AgdaSymbol{(}\AgdaBound{block}\AgdaSpace{}%
\AgdaInductiveConstructor{∷}\AgdaSpace{}%
\AgdaBound{restbc}\AgdaSymbol{)}%
\>[70]\AgdaSymbol{=}\<%
\\
\>[0][@{}l@{\AgdaIndent{0}}]%
\>[2]\AgdaFunction{FinalLedger}\AgdaSpace{}%
\AgdaSymbol{(}\AgdaFunction{block2TXFee}\AgdaSpace{}%
\AgdaBound{block}\AgdaSymbol{)}\AgdaSpace{}%
\AgdaBound{blockReward}\AgdaSpace{}%
\AgdaSymbol{(}\AgdaInductiveConstructor{suc}\AgdaSpace{}%
\AgdaBound{startTime}\AgdaSymbol{)}\<%
\\
\>[2][@{}l@{\AgdaIndent{0}}]%
\>[5]\AgdaSymbol{(}\AgdaFunction{updateLedgerBlock}\AgdaSpace{}%
\AgdaBound{block}\AgdaSpace{}%
\AgdaBound{startLedger}\AgdaSymbol{)}\AgdaSpace{}%
\AgdaBound{restbc}\<%
\end{code}
} 

\AgdaHide{
\begin{code}%
\>[0]\<%
\\
\>[0]\AgdaComment{{-}{-} \textbackslash{}bitcoinVersFive}\<%
\end{code}
} 

\newcommand{\bitcoinVersFiveBlockchain}{
\begin{code}%
\>[0]\AgdaKeyword{record}\AgdaSpace{}%
\AgdaRecord{BlockChain}\AgdaSpace{}%
\AgdaSymbol{(}\AgdaBound{blockReward}\AgdaSpace{}%
\AgdaSymbol{:}\AgdaSpace{}%
\AgdaFunction{Time}\AgdaSpace{}%
\AgdaSymbol{→}\AgdaSpace{}%
\AgdaFunction{Amount}\AgdaSymbol{)}\AgdaSpace{}%
\AgdaSymbol{:}\AgdaSpace{}%
\AgdaPrimitiveType{Set}\AgdaSpace{}%
\AgdaKeyword{where}\<%
\\
\>[0][@{}l@{\AgdaIndent{0}}]%
\>[2]\agdaField%
\>[9]\AgdaField{blockchain}%
\>[21]\AgdaSymbol{:}\AgdaSpace{}%
\AgdaFunction{BlockChainUnchecked}\<%
\\
\>[2][@{}l@{\AgdaIndent{0}}]%
\>[9]\AgdaField{correct}%
\>[21]\AgdaSymbol{:}\AgdaSpace{}%
\AgdaFunction{CorrectBlockChain}\AgdaSpace{}%
\AgdaBound{blockReward}\AgdaSpace{}%
\AgdaNumber{0}\AgdaSpace{}%
\AgdaFunction{initialLedger}\AgdaSpace{}%
\AgdaField{blockchain}\<%
\end{code}
} 

\AgdaHide{
\begin{code}%
\>[0]\<%
\\
\>[0]\AgdaKeyword{open}\AgdaSpace{}%
\AgdaModule{BlockChain}\AgdaSpace{}%
\AgdaKeyword{public}\<%
\\
\>[0]\<%
\end{code}
} 

\newcommand{\bitcoinVersFiveblockChaintoFinalLedger}{
\begin{code}%
\>[0]\AgdaFunction{blockChain2FinalLedger}\AgdaSpace{}%
\AgdaSymbol{:}%
\>[26]\AgdaSymbol{(}\AgdaBound{blockReward}\AgdaSpace{}%
\AgdaSymbol{:}\AgdaSpace{}%
\AgdaFunction{Time}\AgdaSpace{}%
\AgdaSymbol{→}\AgdaSpace{}%
\AgdaFunction{Amount}\AgdaSymbol{)(}\AgdaBound{bc}\AgdaSpace{}%
\AgdaSymbol{:}\AgdaSpace{}%
\AgdaRecord{BlockChain}\AgdaSpace{}%
\AgdaBound{blockReward}\AgdaSymbol{)}\<%
\\
\>[26][@{}l@{\AgdaIndent{0}}]%
\>[27]\AgdaSymbol{→}\AgdaSpace{}%
\AgdaFunction{Ledger}\<%
\\
\>[0]\AgdaFunction{blockChain2FinalLedger}\AgdaSpace{}%
\AgdaBound{blockReward}\AgdaSpace{}%
\AgdaBound{bc}\AgdaSpace{}%
\AgdaSymbol{=}\<%
\\
\>[0][@{}l@{\AgdaIndent{0}}]%
\>[3]\AgdaFunction{FinalLedger}\AgdaSpace{}%
\AgdaNumber{0}\AgdaSpace{}%
\AgdaBound{blockReward}\AgdaSpace{}%
\AgdaNumber{0}\AgdaSpace{}%
\AgdaFunction{initialLedger}\AgdaSpace{}%
\AgdaSymbol{(}\AgdaField{blockchain}\AgdaSpace{}%
\AgdaBound{bc}\AgdaSymbol{)}\<%
\end{code}
} 

\AgdaHide{
\begin{code}%
\>[0]\<%
\end{code}
} 


\AgdaHide{
\begin{code}%
\>[0]\<%
\\
\>[0]\AgdaKeyword{module}\AgdaSpace{}%
\AgdaModule{bitcoinTreeModel}\AgdaSpace{}%
\AgdaKeyword{where}\<%
\\
\\
\>[0]\AgdaComment{{-}{-} open import bool}\<%
\\
\>[0]\AgdaKeyword{open}\AgdaSpace{}%
\AgdaKeyword{import}\AgdaSpace{}%
\AgdaModule{libraries.listLib}\<%
\\
\>[0]\AgdaKeyword{open}\AgdaSpace{}%
\AgdaKeyword{import}\AgdaSpace{}%
\AgdaModule{libraries.natLib}\<%
\\
\>[0]\AgdaKeyword{open}\AgdaSpace{}%
\AgdaKeyword{import}\AgdaSpace{}%
\AgdaModule{libraries.finLib}\<%
\\
\>[0]\AgdaKeyword{open}\AgdaSpace{}%
\AgdaKeyword{import}\AgdaSpace{}%
\AgdaModule{Data.Nat}\<%
\\
\>[0]\AgdaKeyword{open}\AgdaSpace{}%
\AgdaKeyword{import}\AgdaSpace{}%
\AgdaModule{Data.Empty}\<%
\\
\>[0]\AgdaKeyword{open}\AgdaSpace{}%
\AgdaKeyword{import}\AgdaSpace{}%
\AgdaModule{Data.Fin}\AgdaSpace{}%
\AgdaKeyword{hiding}\AgdaSpace{}%
\AgdaSymbol{(}\AgdaFunction{\_+\_}\AgdaSpace{}%
\AgdaSymbol{;}\AgdaSpace{}%
\AgdaFunction{\_≤\_}\AgdaSpace{}%
\AgdaSymbol{)}\<%
\\
\>[0]\AgdaKeyword{open}\AgdaSpace{}%
\AgdaKeyword{import}\AgdaSpace{}%
\AgdaModule{Data.List}\<%
\\
\>[0]\AgdaKeyword{open}\AgdaSpace{}%
\AgdaKeyword{import}\AgdaSpace{}%
\AgdaModule{Data.Unit}\AgdaSpace{}%
\AgdaKeyword{hiding}\AgdaSpace{}%
\AgdaSymbol{(}\AgdaRecord{\_≤\_}\AgdaSpace{}%
\AgdaSymbol{)}\<%
\\
\>[0]\AgdaKeyword{open}\AgdaSpace{}%
\AgdaKeyword{import}\AgdaSpace{}%
\AgdaModule{Data.Bool}\<%
\\
\>[0]\AgdaKeyword{open}\AgdaSpace{}%
\AgdaKeyword{import}\AgdaSpace{}%
\AgdaModule{Data.Product}\<%
\\
\>[0]\AgdaKeyword{open}\AgdaSpace{}%
\AgdaKeyword{import}\AgdaSpace{}%
\AgdaModule{Data.Nat.Base}\<%
\\
\>[0]\AgdaKeyword{open}\AgdaSpace{}%
\AgdaKeyword{import}\AgdaSpace{}%
\AgdaModule{Data.Maybe}\<%
\\
\>[0]\AgdaKeyword{open}\AgdaSpace{}%
\AgdaKeyword{import}\AgdaSpace{}%
\AgdaModule{Relation.Binary.PropositionalEquality}\<%
\\
\>[0]\AgdaKeyword{open}\AgdaSpace{}%
\AgdaKeyword{import}\AgdaSpace{}%
\AgdaModule{Relation.Nullary}\<%
\\
\\
\\
\>[0]\AgdaComment{{-}{-} \#\#\#\#\#\#\#\#\#\#\#\#\#\#\#\#\#\#\#\#\#\#\#\#\#\#\#\#\#\#\#\#\#\#\#\#\#}\<%
\\
\>[0]\AgdaComment{{-}{-} \#  Preliminaries                    \#}\<%
\\
\>[0]\AgdaComment{{-}{-} \#\#\#\#\#\#\#\#\#\#\#\#\#\#\#\#\#\#\#\#\#\#\#\#\#\#\#\#\#\#\#\#\#\#\#\#\#}\<%
\\
\\
\\
\>[0]\AgdaKeyword{infixr}\AgdaSpace{}%
\AgdaNumber{3}\AgdaSpace{}%
\AgdaInductiveConstructor{\_+msg\_}\<%
\\
\\
\>[0]\AgdaFunction{Time}\AgdaSpace{}%
\AgdaSymbol{:}\AgdaSpace{}%
\AgdaPrimitiveType{Set}\<%
\\
\>[0]\AgdaFunction{Time}\AgdaSpace{}%
\AgdaSymbol{=}\AgdaSpace{}%
\AgdaDatatype{ℕ}\<%
\\
\\
\>[0]\AgdaFunction{Amount}\AgdaSpace{}%
\AgdaSymbol{:}\AgdaSpace{}%
\AgdaPrimitiveType{Set}\<%
\\
\>[0]\AgdaFunction{Amount}\AgdaSpace{}%
\AgdaSymbol{=}\AgdaSpace{}%
\AgdaDatatype{ℕ}\<%
\\
\\
\>[0]\AgdaFunction{Address}\AgdaSpace{}%
\AgdaSymbol{:}\AgdaSpace{}%
\AgdaPrimitiveType{Set}\<%
\\
\>[0]\AgdaFunction{Address}\AgdaSpace{}%
\AgdaSymbol{=}\AgdaSpace{}%
\AgdaDatatype{ℕ}\<%
\\
\\
\>[0]\AgdaFunction{TXID}\AgdaSpace{}%
\AgdaSymbol{:}\AgdaSpace{}%
\AgdaPrimitiveType{Set}\<%
\\
\>[0]\AgdaFunction{TXID}\AgdaSpace{}%
\AgdaSymbol{=}\AgdaSpace{}%
\AgdaDatatype{ℕ}\<%
\\
\\
\>[0]\AgdaFunction{Signature}\AgdaSpace{}%
\AgdaSymbol{:}\AgdaSpace{}%
\AgdaPrimitiveType{Set}\<%
\\
\>[0]\AgdaFunction{Signature}%
\>[11]\AgdaSymbol{=}%
\>[14]\AgdaDatatype{ℕ}\<%
\\
\\
\>[0]\AgdaFunction{PublicKey}\AgdaSpace{}%
\AgdaSymbol{:}\AgdaSpace{}%
\AgdaPrimitiveType{Set}\<%
\\
\>[0]\AgdaFunction{PublicKey}%
\>[11]\AgdaSymbol{=}%
\>[14]\AgdaDatatype{ℕ}\<%
\\
\\
\>[0]\AgdaComment{{-}{-} \textbackslash{}bitcoinVersSix}\<%
\end{code}
} 

\newcommand{\bitcoinVersSixMsg}{
\begin{code}%
\>[0]\AgdaKeyword{data}\AgdaSpace{}%
\AgdaDatatype{Msg}\AgdaSpace{}%
\AgdaSymbol{:}\AgdaSpace{}%
\AgdaPrimitiveType{Set}\AgdaSpace{}%
\AgdaKeyword{where}\<%
\\
\>[0][@{}l@{\AgdaIndent{0}}]%
\>[3]\AgdaInductiveConstructor{nat}\AgdaSpace{}%
\AgdaSymbol{:}\AgdaSpace{}%
\AgdaSymbol{(}\AgdaBound{n}\AgdaSpace{}%
\AgdaSymbol{:}\AgdaSpace{}%
\AgdaDatatype{ℕ}\AgdaSymbol{)}\AgdaSpace{}%
\AgdaSymbol{→}\AgdaSpace{}%
\AgdaDatatype{Msg}\<%
\\
\>[0][@{}l@{\AgdaIndent{0}}]%
\>[3]\AgdaInductiveConstructor{\_+msg\_}\AgdaSpace{}%
\AgdaSymbol{:}\AgdaSpace{}%
\AgdaSymbol{(}\AgdaBound{m}\AgdaSpace{}%
\AgdaBound{m'}\AgdaSpace{}%
\AgdaSymbol{:}\AgdaSpace{}%
\AgdaDatatype{Msg}\AgdaSymbol{)}\AgdaSpace{}%
\AgdaSymbol{→}\AgdaSpace{}%
\AgdaDatatype{Msg}\<%
\\
\>[0][@{}l@{\AgdaIndent{0}}]%
\>[3]\AgdaInductiveConstructor{list}\AgdaSpace{}%
\AgdaSymbol{:}\AgdaSpace{}%
\AgdaSymbol{(}\AgdaBound{l}%
\>[14]\AgdaSymbol{:}\AgdaSpace{}%
\AgdaDatatype{List}\AgdaSpace{}%
\AgdaDatatype{Msg}\AgdaSymbol{)}\AgdaSpace{}%
\AgdaSymbol{→}\AgdaSpace{}%
\AgdaDatatype{Msg}\<%
\end{code}
} 

\AgdaHide{
\begin{code}%
\>[0]\<%
\\
\>[0]\AgdaKeyword{postulate}\AgdaSpace{}%
\AgdaPostulate{hashMsg}\AgdaSpace{}%
\AgdaSymbol{:}\AgdaSpace{}%
\AgdaDatatype{Msg}\AgdaSpace{}%
\AgdaSymbol{→}\AgdaSpace{}%
\AgdaDatatype{ℕ}\<%
\\
\\
\>[0]\AgdaComment{{-}{-} +msg could be replaced by a list with two elements; we use +msg}\<%
\\
\>[0]\AgdaComment{{-}{-} since it is closer to how messages are serialised where}\<%
\\
\>[0]\AgdaComment{{-}{-} a list requires an extra length argument.x}\<%
\\
\\
\>[0]\AgdaComment{{-}{-} \textbackslash{}bitcoinVersSix}\<%
\\
\>[0]\AgdaComment{{-}{-} already in \textbackslash{}bitcoinVersFive}\<%
\end{code}
} 

\newcommand{\bitcoinVersSixPublicKey}{
\begin{code}%
\>[0]\AgdaKeyword{postulate}\AgdaSpace{}%
\AgdaPostulate{publicKey2Address}\AgdaSpace{}%
\AgdaSymbol{:}\AgdaSpace{}%
\AgdaSymbol{(}\AgdaBound{pubk}\AgdaSpace{}%
\AgdaSymbol{:}\AgdaSpace{}%
\AgdaFunction{PublicKey}\AgdaSymbol{)}\AgdaSpace{}%
\AgdaSymbol{→}\AgdaSpace{}%
\AgdaFunction{Address}\<%
\end{code}
} 

\AgdaHide{
\begin{code}%
\>[0]\<%
\\
\>[0]\AgdaComment{{-}{-} Signed means that Msg msg has been signed}\<%
\\
\>[0]\AgdaComment{{-}{-} by private key corresponding to pubk}\<%
\\
\\
\>[0]\AgdaComment{{-}{-} \textbackslash{}bitcoinVersSix}\<%
\\
\>[0]\AgdaComment{{-}{-} already in \textbackslash{}bitcoinVersFive}\<%
\end{code}
} 

\newcommand{\bitcoinVersSixSigned}{
\begin{code}%
\>[0]\AgdaKeyword{postulate}\AgdaSpace{}%
\AgdaPostulate{Signed}\AgdaSpace{}%
\AgdaSymbol{:}\AgdaSpace{}%
\AgdaSymbol{(}\AgdaBound{msg}\AgdaSpace{}%
\AgdaSymbol{:}\AgdaSpace{}%
\AgdaDatatype{Msg}\AgdaSymbol{)(}\AgdaBound{publicKey}\AgdaSpace{}%
\AgdaSymbol{:}\AgdaSpace{}%
\AgdaFunction{PublicKey}\AgdaSymbol{)(}\AgdaBound{s}\AgdaSpace{}%
\AgdaSymbol{:}\AgdaSpace{}%
\AgdaFunction{Signature}\AgdaSymbol{)}\AgdaSpace{}%
\AgdaSymbol{→}\AgdaSpace{}%
\AgdaPrimitiveType{Set}\<%
\\
\>[0]\AgdaComment{{-}{-} postulate sign2ℕ : (msg : Msg)(addr : Address)(s : Signed msg addr)}\<%
\\
\>[0]\AgdaComment{{-}{-}                    → ℕ}\<%
\end{code}
} 

\AgdaHide{
\begin{code}%
\>[0]\<%
\\
\>[0]\AgdaComment{{-}{-} \textbackslash{}bitcoinVersSix}\<%
\end{code}
} 

\newcommand{\bitcoinVersSixminerReward}{
\begin{code}%
\>[0]\AgdaKeyword{postulate}\AgdaSpace{}%
\AgdaPostulate{blockReward}\AgdaSpace{}%
\AgdaSymbol{:}\AgdaSpace{}%
\AgdaSymbol{(}\AgdaBound{t}\AgdaSpace{}%
\AgdaSymbol{:}\AgdaSpace{}%
\AgdaFunction{Time}\AgdaSymbol{)}\AgdaSpace{}%
\AgdaSymbol{→}\AgdaSpace{}%
\AgdaFunction{Amount}\<%
\\
\>[0]\AgdaKeyword{postulate}\AgdaSpace{}%
\AgdaPostulate{blockMaturationTime}\AgdaSpace{}%
\AgdaSymbol{:}\AgdaSpace{}%
\AgdaFunction{Time}\<%
\end{code}
} 

\AgdaHide{
\begin{code}%
\>[0]\AgdaComment{{-}{-} reward for miner at time t}\<%
\\
\>[0]\AgdaComment{{-}{-} how long the miner needs to wait before his reward can be spent}\<%
\\
\>[0]\AgdaComment{{-}{-} should be 101}\<%
\\
\\
\\
\>[0]\AgdaComment{{-}{-} \#\#\#\#\#\#\#\#\#\#\#\#\#\#\#\#\#\#\#\#\#\#\#\#\#\#\#\#\#\#\#\#\#\#\#\#\#\#\#\#\#\#\#\#\#\#\#\#\#\#\#\#\#\#\#\#\#\#\#\#\#\#\#\#\#\#\#\#\#\#\#\#\#\#\#\#\#}\<%
\\
\>[0]\AgdaComment{{-}{-} \#  Inductive{-}Recursive Definition of                                        \#}\<%
\\
\>[0]\AgdaComment{{-}{-} \#       Transaction Tree, Transactions, and                                 \#}\<%
\\
\>[0]\AgdaComment{{-}{-} \#  Unspent Transaction Outputs                                              \#}\<%
\\
\>[0]\AgdaComment{{-}{-} \#\#\#\#\#\#\#\#\#\#\#\#\#\#\#\#\#\#\#\#\#\#\#\#\#\#\#\#\#\#\#\#\#\#\#\#\#\#\#\#\#\#\#\#\#\#\#\#\#\#\#\#\#\#\#\#\#\#\#\#\#\#\#\#\#\#\#\#\#\#\#\#\#\#\#\#\#}\<%
\\
\\
\>[0]\AgdaComment{{-}{-} \textbackslash{}bitcoinVersSix}\<%
\end{code}
} 

\newcommand{\bitcoinVersSixTXOutputfield}{
\begin{code}%
\>[0]\AgdaKeyword{record}\AgdaSpace{}%
\AgdaRecord{TXOutputfield}\AgdaSpace{}%
\AgdaSymbol{:}\AgdaSpace{}%
\AgdaPrimitiveType{Set}\AgdaSpace{}%
\AgdaKeyword{where}\<%
\\
\>[0][@{}l@{\AgdaIndent{0}}]%
\>[2]\AgdaKeyword{constructor}\AgdaSpace{}%
\AgdaInductiveConstructor{txOutputfield}\<%
\\
\>[0][@{}l@{\AgdaIndent{0}}]%
\>[2]\agdaField%
\>[9]\AgdaField{amount}\AgdaSpace{}%
\AgdaSymbol{:}\AgdaSpace{}%
\AgdaFunction{Amount}\<%
\\
\>[2][@{}l@{\AgdaIndent{0}}]%
\>[9]\AgdaField{address}\AgdaSpace{}%
\AgdaSymbol{:}\AgdaSpace{}%
\AgdaFunction{Address}\<%
\end{code}
} 

\AgdaHide{
\begin{code}%
\>[0]\AgdaKeyword{open}\AgdaSpace{}%
\AgdaModule{TXOutputfield}\AgdaSpace{}%
\AgdaKeyword{public}\<%
\\
\\
\\
\>[0]\AgdaKeyword{mutual}\<%
\\
\>[0]\AgdaComment{{-}{-} \textbackslash{}bitcoinVersSix}\<%
\end{code}
} 

\newcommand{\bitcoinVersSixTXTree}{
\begin{code}%
\>[0][@{}l@{\AgdaIndent{1}}]%
\>[2]\AgdaKeyword{data}\AgdaSpace{}%
\AgdaDatatype{TXTree}%
\>[15]\AgdaSymbol{:}\AgdaSpace{}%
\AgdaPrimitiveType{Set}\AgdaSpace{}%
\AgdaKeyword{where}\<%
\\
\>[2][@{}l@{\AgdaIndent{0}}]%
\>[4]\AgdaInductiveConstructor{genesisTree}%
\>[17]\AgdaSymbol{:}%
\>[20]\AgdaDatatype{TXTree}\<%
\\
\>[2][@{}l@{\AgdaIndent{0}}]%
\>[4]\AgdaInductiveConstructor{txtree}%
\>[17]\AgdaSymbol{:}%
\>[20]\AgdaSymbol{(}\AgdaBound{tree}\AgdaSpace{}%
\AgdaSymbol{:}\AgdaSpace{}%
\AgdaDatatype{TXTree}\AgdaSymbol{)(}\AgdaBound{tx}\AgdaSpace{}%
\AgdaSymbol{:}\AgdaSpace{}%
\AgdaDatatype{TX}\AgdaSpace{}%
\AgdaBound{tree}\AgdaSymbol{)}\AgdaSpace{}%
\AgdaSymbol{→}\AgdaSpace{}%
\AgdaDatatype{TXTree}\<%
\end{code}
} 

\AgdaHide{
\begin{code}%
\>[0]\<%
\\
\>[0]\AgdaComment{{-}{-} \textbackslash{}bitcoinVersSix}\<%
\end{code}
} 

\newcommand{\bitcoinVersSixTX}{
\begin{code}%
\>[0][@{}l@{\AgdaIndent{1}}]%
\>[2]\AgdaKeyword{data}\AgdaSpace{}%
\AgdaDatatype{TX}\AgdaSpace{}%
\AgdaSymbol{(}\AgdaBound{tr}\AgdaSpace{}%
\AgdaSymbol{:}\AgdaSpace{}%
\AgdaDatatype{TXTree}\AgdaSymbol{)}\AgdaSpace{}%
\AgdaSymbol{:}%
\>[27]\AgdaPrimitiveType{Set}\AgdaSpace{}%
\AgdaKeyword{where}\<%
\\
\>[2][@{}l@{\AgdaIndent{0}}]%
\>[5]\AgdaInductiveConstructor{normalTX}%
\>[15]\AgdaSymbol{:}%
\>[18]\AgdaSymbol{(}\AgdaBound{inputs}%
\>[28]\AgdaSymbol{:}%
\>[31]\AgdaFunction{TxInputs}\AgdaSpace{}%
\AgdaBound{tr}\AgdaSymbol{)}%
\>[45]\AgdaSymbol{(}\AgdaBound{outputs}%
\>[55]\AgdaSymbol{:}%
\>[58]\AgdaDatatype{List}\AgdaSpace{}%
\AgdaRecord{TXOutputfield}\AgdaSymbol{)}%
\>[79]\AgdaSymbol{→}\AgdaSpace{}%
\AgdaDatatype{TX}\AgdaSpace{}%
\AgdaBound{tr}\<%
\\
\>[2][@{}l@{\AgdaIndent{0}}]%
\>[5]\AgdaInductiveConstructor{coinbase}%
\>[15]\AgdaSymbol{:}%
\>[18]\AgdaSymbol{(}\AgdaBound{time}\AgdaSpace{}%
\AgdaSymbol{:}\AgdaSpace{}%
\AgdaFunction{Time}\AgdaSymbol{)}%
\>[45]\AgdaSymbol{(}\AgdaBound{outputs}%
\>[55]\AgdaSymbol{:}%
\>[58]\AgdaDatatype{List}\AgdaSpace{}%
\AgdaRecord{TXOutputfield}\AgdaSymbol{)}%
\>[79]\AgdaSymbol{→}\AgdaSpace{}%
\AgdaDatatype{TX}\AgdaSpace{}%
\AgdaBound{tr}\<%
\end{code}
} 

\AgdaHide{
\begin{code}%
\>[0]\<%
\\
\>[0]\AgdaComment{{-}{-} \textbackslash{}bitcoinVersSix}\<%
\end{code}
} 

\newcommand{\bitcoinVersSixTXOutput}{
\begin{code}%
\>[0][@{}l@{\AgdaIndent{1}}]%
\>[2]\AgdaKeyword{record}\AgdaSpace{}%
\AgdaRecord{TXOutput}\AgdaSpace{}%
\AgdaSymbol{:}\AgdaSpace{}%
\AgdaPrimitiveType{Set}\AgdaSpace{}%
\AgdaKeyword{where}\<%
\\
\>[2][@{}l@{\AgdaIndent{0}}]%
\>[4]\AgdaKeyword{inductive}\<%
\\
\>[2][@{}l@{\AgdaIndent{0}}]%
\>[4]\AgdaKeyword{constructor}\AgdaSpace{}%
\AgdaInductiveConstructor{txOutput}\<%
\\
\>[2][@{}l@{\AgdaIndent{0}}]%
\>[4]\agdaField%
\>[11]\AgdaField{trTree}%
\>[20]\AgdaSymbol{:}\AgdaSpace{}%
\AgdaDatatype{TXTree}\<%
\\
\>[4][@{}l@{\AgdaIndent{0}}]%
\>[11]\AgdaField{tx}%
\>[20]\AgdaSymbol{:}\AgdaSpace{}%
\AgdaDatatype{TX}\AgdaSpace{}%
\AgdaField{trTree}\<%
\\
\>[4][@{}l@{\AgdaIndent{0}}]%
\>[11]\AgdaField{output}%
\>[20]\AgdaSymbol{:}\AgdaSpace{}%
\AgdaDatatype{Fin}\AgdaSpace{}%
\AgdaSymbol{(}\AgdaFunction{nrOutputs}\AgdaSpace{}%
\AgdaField{trTree}%
\>[45]\AgdaField{tx}\AgdaSymbol{)}\<%
\end{code}
} 

\AgdaHide{
\begin{code}%
\>[0]\<%
\\
\\
\>[0]\AgdaComment{{-}{-} \textbackslash{}bitcoinVersSix}\<%
\end{code}
} 

\newcommand{\bitcoinVersSixutxo}{
\begin{code}%
\>[0][@{}l@{\AgdaIndent{1}}]%
\>[2]\AgdaFunction{utxoMinusNewInputs}\AgdaSpace{}%
\AgdaSymbol{:}\AgdaSpace{}%
\AgdaSymbol{(}\AgdaBound{tr}\AgdaSpace{}%
\AgdaSymbol{:}\AgdaSpace{}%
\AgdaDatatype{TXTree}\AgdaSymbol{)(}\AgdaBound{tx}\AgdaSpace{}%
\AgdaSymbol{:}\AgdaSpace{}%
\AgdaDatatype{TX}\AgdaSpace{}%
\AgdaBound{tr}\AgdaSymbol{)}\AgdaSpace{}%
\AgdaSymbol{→}\AgdaSpace{}%
\AgdaDatatype{List}\AgdaSpace{}%
\AgdaRecord{TXOutput}\<%
\\
\>[0][@{}l@{\AgdaIndent{1}}]%
\>[2]\AgdaFunction{utxo}\AgdaSpace{}%
\AgdaSymbol{:}\AgdaSpace{}%
\AgdaSymbol{(}\AgdaBound{tr}\AgdaSpace{}%
\AgdaSymbol{:}\AgdaSpace{}%
\AgdaDatatype{TXTree}\AgdaSymbol{)}\AgdaSpace{}%
\AgdaSymbol{→}\AgdaSpace{}%
\AgdaDatatype{List}\AgdaSpace{}%
\AgdaRecord{TXOutput}\<%
\end{code}
} 

\AgdaHide{
\begin{code}%
\>[0]\<%
\\
\>[0]\AgdaComment{{-}{-} \textbackslash{}bitcoinVersSix}\<%
\end{code}
} 

\newcommand{\bitcoinVersSixTxInputs}{
\begin{code}%
\>[0][@{}l@{\AgdaIndent{1}}]%
\>[2]\AgdaFunction{TxInputs}\AgdaSpace{}%
\AgdaSymbol{:}\AgdaSpace{}%
\AgdaSymbol{(}\AgdaBound{tr}\AgdaSpace{}%
\AgdaSymbol{:}\AgdaSpace{}%
\AgdaDatatype{TXTree}\AgdaSymbol{)}\AgdaSpace{}%
\AgdaSymbol{→}\AgdaSpace{}%
\AgdaPrimitiveType{Set}\<%
\\
\>[0][@{}l@{\AgdaIndent{1}}]%
\>[2]\AgdaFunction{TxInputs}\AgdaSpace{}%
\AgdaBound{tr}\AgdaSpace{}%
\AgdaSymbol{=}\AgdaSpace{}%
\AgdaDatatype{SubList+}\AgdaSpace{}%
\AgdaSymbol{(}\AgdaFunction{PublicKey}\AgdaSpace{}%
\AgdaFunction{×}%
\>[39]\AgdaFunction{Signature}\AgdaSymbol{)}\AgdaSpace{}%
\AgdaSymbol{(}\AgdaFunction{utxo}\AgdaSpace{}%
\AgdaBound{tr}\AgdaSymbol{)}\<%
\end{code}
} 

\AgdaHide{
\begin{code}%
\>[0]\<%
\\
\>[0]\AgdaComment{{-}{-} \textbackslash{}bitcoinVersSix}\<%
\end{code}
} 

\newcommand{\bitcoinVersSixutxoMinusNewInputs}{
\begin{code}%
\>[0][@{}l@{\AgdaIndent{1}}]%
\>[2]\AgdaFunction{utxoMinusNewInputs}\AgdaSpace{}%
\AgdaBound{tr}\AgdaSpace{}%
\AgdaSymbol{(}\AgdaInductiveConstructor{normalTX}\AgdaSpace{}%
\AgdaBound{inputs}\AgdaSpace{}%
\AgdaBound{outputs}\AgdaSymbol{)}%
\>[51]\AgdaSymbol{=}%
\>[54]\AgdaFunction{listMinusSubList+}\AgdaSpace{}%
\AgdaSymbol{(}\AgdaFunction{utxo}\AgdaSpace{}%
\AgdaBound{tr}\AgdaSymbol{)}\AgdaSpace{}%
\AgdaBound{inputs}\<%
\\
\>[0][@{}l@{\AgdaIndent{1}}]%
\>[2]\AgdaFunction{utxoMinusNewInputs}\AgdaSpace{}%
\AgdaBound{tr}\AgdaSpace{}%
\AgdaSymbol{(}\AgdaInductiveConstructor{coinbase}\AgdaSpace{}%
\AgdaBound{time}\AgdaSpace{}%
\AgdaBound{outputs}\AgdaSymbol{)}%
\>[51]\AgdaSymbol{=}%
\>[54]\AgdaFunction{utxo}\AgdaSpace{}%
\AgdaBound{tr}\<%
\end{code}
} 

\AgdaHide{
\begin{code}%
\>[0]\<%
\\
\>[0]\<%
\end{code}
} 

\newcommand{\bitcoinVersSixutxodef}{
\begin{code}%
\>[0][@{}l@{\AgdaIndent{1}}]%
\>[2]\AgdaFunction{utxo}\AgdaSpace{}%
\AgdaInductiveConstructor{genesisTree}%
\>[23]\AgdaSymbol{=}%
\>[26]\AgdaInductiveConstructor{[]}\<%
\\
\>[0][@{}l@{\AgdaIndent{1}}]%
\>[2]\AgdaFunction{utxo}\AgdaSpace{}%
\AgdaSymbol{(}\AgdaInductiveConstructor{txtree}\AgdaSpace{}%
\AgdaBound{tr}\AgdaSpace{}%
\AgdaBound{tx}\AgdaSymbol{)}%
\>[23]\AgdaSymbol{=}%
\>[26]\AgdaFunction{utxoMinusNewInputs}\AgdaSpace{}%
\AgdaBound{tr}\AgdaSpace{}%
\AgdaBound{tx}\AgdaSpace{}%
\AgdaFunction{\ensuremath{+\!\!+}}\AgdaSpace{}%
\AgdaFunction{tx2TXOutputs}\AgdaSpace{}%
\AgdaBound{tr}\AgdaSpace{}%
\AgdaBound{tx}\<%
\end{code}
} 

\AgdaHide{
\begin{code}%
\>[0]\<%
\\
\>[0]\AgdaComment{{-}{-}(normalTX inp outp)) =}\<%
\\
\>[0]\AgdaComment{{-}{-}           listMinusSubList+ (utxo tr) inp \ensuremath{+\!\!+} tx2TXOutputs tr  (normalTX inp outp)}\<%
\\
\>[0]\AgdaComment{{-}{-}  utxo (txtree tr (coinbase t outp)) = utxo tr \ensuremath{+\!\!+} tx2TXOutputs tr  (coinbase t outp)}\<%
\\
\\
\\
\>[0]\AgdaComment{{-}{-} \textbackslash{}bitcoinVersSix}\<%
\end{code}
} 

\newcommand{\bitcoinVersSixTXOuptuts}{
\begin{code}%
\>[0][@{}l@{\AgdaIndent{1}}]%
\>[2]\AgdaFunction{tx2TXOutputs}\AgdaSpace{}%
\AgdaSymbol{:}\AgdaSpace{}%
\AgdaSymbol{(}\AgdaBound{tr}\AgdaSpace{}%
\AgdaSymbol{:}\AgdaSpace{}%
\AgdaDatatype{TXTree}\AgdaSymbol{)(}\AgdaBound{tx}\AgdaSpace{}%
\AgdaSymbol{:}\AgdaSpace{}%
\AgdaDatatype{TX}\AgdaSpace{}%
\AgdaBound{tr}\AgdaSymbol{)}\AgdaSpace{}%
\AgdaSymbol{→}\AgdaSpace{}%
\AgdaDatatype{List}\AgdaSpace{}%
\AgdaRecord{TXOutput}\<%
\\
\>[0][@{}l@{\AgdaIndent{1}}]%
\>[2]\AgdaFunction{tx2TXOutputs}\AgdaSpace{}%
\AgdaBound{tr}\AgdaSpace{}%
\AgdaBound{tx}%
\>[22]\AgdaSymbol{=}\AgdaSpace{}%
\AgdaFunction{mapL}\AgdaSpace{}%
\AgdaSymbol{(λ}\AgdaSpace{}%
\AgdaBound{i}\AgdaSpace{}%
\AgdaSymbol{→}\AgdaSpace{}%
\AgdaInductiveConstructor{txOutput}\AgdaSpace{}%
\AgdaBound{tr}\AgdaSpace{}%
\AgdaBound{tx}\AgdaSpace{}%
\AgdaBound{i}\AgdaSymbol{)(}\AgdaFunction{listOfElementsOfFin}\AgdaSpace{}%
\AgdaSymbol{(}\AgdaFunction{nrOutputs}\AgdaSpace{}%
\AgdaBound{tr}\AgdaSpace{}%
\AgdaBound{tx}\AgdaSymbol{))}\<%
\end{code}
} 

\AgdaHide{
\begin{code}%
\>[0]\<%
\\
\>[0]\AgdaComment{{-}{-} \textbackslash{}bitcoinVersSix}\<%
\end{code}
} 

\newcommand{\bitcoinVersSixnrOutputs}{
\begin{code}%
\>[0][@{}l@{\AgdaIndent{1}}]%
\>[2]\AgdaFunction{nrOutputs}\AgdaSpace{}%
\AgdaSymbol{:}\AgdaSpace{}%
\AgdaSymbol{(}\AgdaBound{tr}\AgdaSpace{}%
\AgdaSymbol{:}\AgdaSpace{}%
\AgdaDatatype{TXTree}\AgdaSymbol{)}\AgdaSpace{}%
\AgdaSymbol{(}\AgdaBound{tx}\AgdaSpace{}%
\AgdaSymbol{:}\AgdaSpace{}%
\AgdaDatatype{TX}\AgdaSpace{}%
\AgdaBound{tr}\AgdaSymbol{)}\AgdaSpace{}%
\AgdaSymbol{→}\AgdaSpace{}%
\AgdaDatatype{ℕ}\<%
\end{code}
} 

\AgdaHide{
\begin{code}%
\>[0]\<%
\\
\>[0][@{}l@{\AgdaIndent{2}}]%
\>[2]\AgdaFunction{nrOutputs}\AgdaSpace{}%
\AgdaBound{tr}\AgdaSpace{}%
\AgdaSymbol{(}\AgdaInductiveConstructor{normalTX}\AgdaSpace{}%
\AgdaBound{inp}\AgdaSpace{}%
\AgdaBound{outp}\AgdaSymbol{)}\AgdaSpace{}%
\AgdaSymbol{=}\AgdaSpace{}%
\AgdaFunction{length}\AgdaSpace{}%
\AgdaBound{outp}\<%
\\
\>[0][@{}l@{\AgdaIndent{2}}]%
\>[2]\AgdaFunction{nrOutputs}\AgdaSpace{}%
\AgdaBound{tr}\AgdaSpace{}%
\AgdaSymbol{(}\AgdaInductiveConstructor{coinbase}\AgdaSpace{}%
\AgdaBound{t}\AgdaSpace{}%
\AgdaBound{outp}\AgdaSymbol{)}\AgdaSpace{}%
\AgdaSymbol{=}\AgdaSpace{}%
\AgdaFunction{length}\AgdaSpace{}%
\AgdaBound{outp}\<%
\\
\\
\\
\\
\>[0]\AgdaKeyword{open}\AgdaSpace{}%
\AgdaModule{TXOutput}\AgdaSpace{}%
\AgdaKeyword{public}\<%
\\
\>[0]\AgdaKeyword{open}\AgdaSpace{}%
\AgdaModule{TXTree}\AgdaSpace{}%
\AgdaKeyword{public}\<%
\\
\>[0]\AgdaKeyword{open}\AgdaSpace{}%
\AgdaModule{TX}\AgdaSpace{}%
\AgdaKeyword{public}\<%
\\
\\
\\
\>[0]\AgdaComment{{-}{-} \#\#\#\#\#\#\#\#\#\#\#\#\#\#\#\#\#\#\#\#\#\#\#\#\#\#\#\#\#\#\#\#\#\#\#\#\#\#\#\#\#\#\#\#\#\#\#\#\#\#\#\#\#\#\#\#\#\#\#\#\#\#\#\#\#\#\#\#\#\#\#\#\#\#\#\#\#}\<%
\\
\>[0]\AgdaComment{{-}{-} \#  Computing Sum of Inputs and Outputs of Transactions and Transaction Fees \#}\<%
\\
\>[0]\AgdaComment{{-}{-} \#\#\#\#\#\#\#\#\#\#\#\#\#\#\#\#\#\#\#\#\#\#\#\#\#\#\#\#\#\#\#\#\#\#\#\#\#\#\#\#\#\#\#\#\#\#\#\#\#\#\#\#\#\#\#\#\#\#\#\#\#\#\#\#\#\#\#\#\#\#\#\#\#\#\#\#\#}\<%
\\
\\
\>[0]\AgdaComment{{-}{-} \textbackslash{}bitcoinVersSix}\<%
\end{code}
} 

\newcommand{\bitcoinVersSixtxtoSumOutputs}{
\begin{code}%
\>[0]\AgdaFunction{outputs2SumAmount}\AgdaSpace{}%
\AgdaSymbol{:}\AgdaSpace{}%
\AgdaDatatype{List}\AgdaSpace{}%
\AgdaRecord{TXOutputfield}\AgdaSpace{}%
\AgdaSymbol{→}\AgdaSpace{}%
\AgdaFunction{Amount}\<%
\\
\>[0]\AgdaFunction{outputs2SumAmount}\AgdaSpace{}%
\AgdaBound{l}\AgdaSpace{}%
\AgdaSymbol{=}\AgdaSpace{}%
\AgdaFunction{sumListViaf}\AgdaSpace{}%
\AgdaField{amount}\AgdaSpace{}%
\AgdaBound{l}\<%
\\
\\
\>[0]\AgdaFunction{tx2SumOutputs}\AgdaSpace{}%
\AgdaSymbol{:}\AgdaSpace{}%
\AgdaSymbol{\{}\AgdaBound{tr}\AgdaSpace{}%
\AgdaSymbol{:}\AgdaSpace{}%
\AgdaDatatype{TXTree}\AgdaSymbol{\}(}\AgdaBound{tx}\AgdaSpace{}%
\AgdaSymbol{:}\AgdaSpace{}%
\AgdaDatatype{TX}\AgdaSpace{}%
\AgdaBound{tr}\AgdaSymbol{)}\AgdaSpace{}%
\AgdaSymbol{→}\AgdaSpace{}%
\AgdaFunction{Amount}\<%
\\
\>[0]\AgdaFunction{tx2SumOutputs}\AgdaSpace{}%
\AgdaSymbol{(}\AgdaInductiveConstructor{normalTX}\AgdaSpace{}%
\AgdaBound{inputs}\AgdaSpace{}%
\AgdaBound{outputs}\AgdaSymbol{)}%
\>[41]\AgdaSymbol{=}%
\>[44]\AgdaFunction{outputs2SumAmount}\AgdaSpace{}%
\AgdaBound{outputs}\<%
\\
\>[0]\AgdaFunction{tx2SumOutputs}\AgdaSpace{}%
\AgdaSymbol{(}\AgdaInductiveConstructor{coinbase}\AgdaSpace{}%
\AgdaBound{time}\AgdaSpace{}%
\AgdaBound{outputs}\AgdaSymbol{)}%
\>[41]\AgdaSymbol{=}%
\>[44]\AgdaFunction{outputs2SumAmount}\AgdaSpace{}%
\AgdaBound{outputs}\<%
\end{code}
} 

\AgdaHide{
\begin{code}%
\>[0]\<%
\\
\\
\>[0]\AgdaComment{{-}{-} \textbackslash{}bitcoinVersSix}\<%
\end{code}
} 

\newcommand{\bitcoinVersSixtxOutputtoOutputfield}{
\begin{code}%
\>[0]\AgdaFunction{txOutput2Outputfield}\AgdaSpace{}%
\AgdaSymbol{:}\AgdaSpace{}%
\AgdaRecord{TXOutput}\AgdaSpace{}%
\AgdaSymbol{→}\AgdaSpace{}%
\AgdaRecord{TXOutputfield}\<%
\\
\>[0]\AgdaFunction{txOutput2Outputfield}\AgdaSpace{}%
\AgdaSymbol{(}\AgdaInductiveConstructor{txOutput}\AgdaSpace{}%
\AgdaBound{trTree}\AgdaSpace{}%
\AgdaSymbol{(}\AgdaInductiveConstructor{normalTX}\AgdaSpace{}%
\AgdaBound{inputs}\AgdaSpace{}%
\AgdaBound{outputs}\AgdaSymbol{)}\AgdaSpace{}%
\AgdaBound{i}\AgdaSymbol{)}%
\>[68]\AgdaSymbol{=}\AgdaSpace{}%
\AgdaFunction{nth}\AgdaSpace{}%
\AgdaBound{outputs}\AgdaSpace{}%
\AgdaBound{i}\<%
\\
\>[0]\AgdaFunction{txOutput2Outputfield}\AgdaSpace{}%
\AgdaSymbol{(}\AgdaInductiveConstructor{txOutput}\AgdaSpace{}%
\AgdaBound{trTree}\AgdaSpace{}%
\AgdaSymbol{(}\AgdaInductiveConstructor{coinbase}\AgdaSpace{}%
\AgdaBound{time}\AgdaSpace{}%
\AgdaBound{outputs}\AgdaSymbol{)}\AgdaSpace{}%
\AgdaBound{i}\AgdaSymbol{)}%
\>[68]\AgdaSymbol{=}\AgdaSpace{}%
\AgdaFunction{nth}\AgdaSpace{}%
\AgdaBound{outputs}\AgdaSpace{}%
\AgdaBound{i}\<%
\\
\\
\>[0]\AgdaFunction{txOutput2Amount}\AgdaSpace{}%
\AgdaSymbol{:}\AgdaSpace{}%
\AgdaRecord{TXOutput}\AgdaSpace{}%
\AgdaSymbol{→}\AgdaSpace{}%
\AgdaFunction{Amount}\<%
\\
\>[0]\AgdaFunction{txOutput2Amount}\AgdaSpace{}%
\AgdaBound{output}\AgdaSpace{}%
\AgdaSymbol{=}\AgdaSpace{}%
\AgdaFunction{txOutput2Outputfield}\AgdaSpace{}%
\AgdaBound{output}\AgdaSpace{}%
\AgdaSymbol{.}\AgdaField{amount}\<%
\\
\\
\>[0]\AgdaFunction{txOutput2Address}\AgdaSpace{}%
\AgdaSymbol{:}\AgdaSpace{}%
\AgdaRecord{TXOutput}\AgdaSpace{}%
\AgdaSymbol{→}\AgdaSpace{}%
\AgdaFunction{Address}\<%
\\
\>[0]\AgdaFunction{txOutput2Address}\AgdaSpace{}%
\AgdaBound{output}\AgdaSpace{}%
\AgdaSymbol{=}\AgdaSpace{}%
\AgdaFunction{txOutput2Outputfield}\AgdaSpace{}%
\AgdaBound{output}\AgdaSpace{}%
\AgdaSymbol{.}\AgdaField{address}\<%
\end{code}
} 

\AgdaHide{
\begin{code}%
\>[0]\<%
\\
\\
\>[0]\AgdaComment{{-}{-} \textbackslash{}bitcoinVersSix}\<%
\end{code}
} 

\newcommand{\bitcoinVersSixinputstoPrevOutputsSigPbk}{
\begin{code}%
\>[0]\AgdaFunction{inputs2PrevOutputsSigPbk}%
\>[335I]\AgdaSymbol{:}%
\>[28]\AgdaSymbol{(}\AgdaBound{tr}\AgdaSpace{}%
\AgdaSymbol{:}\AgdaSpace{}%
\AgdaDatatype{TXTree}\AgdaSymbol{)(}\AgdaBound{inputs}\AgdaSpace{}%
\AgdaSymbol{:}\AgdaSpace{}%
\AgdaFunction{TxInputs}\AgdaSpace{}%
\AgdaBound{tr}\AgdaSymbol{)}\<%
\\
\>[335I][@{}l@{\AgdaIndent{0}}]%
\>[28]\AgdaSymbol{→}\AgdaSpace{}%
\AgdaDatatype{List}\AgdaSpace{}%
\AgdaSymbol{(}\AgdaRecord{TXOutput}\AgdaSpace{}%
\AgdaFunction{×}\AgdaSpace{}%
\AgdaFunction{PublicKey}\AgdaSpace{}%
\AgdaFunction{×}\AgdaSpace{}%
\AgdaFunction{Signature}\AgdaSymbol{)}\<%
\\
\>[0]\AgdaFunction{inputs2PrevOutputsSigPbk}\AgdaSpace{}%
\AgdaBound{tr}\AgdaSpace{}%
\AgdaBound{inputs}\AgdaSpace{}%
\AgdaSymbol{=}\AgdaSpace{}%
\AgdaFunction{subList+2List}\AgdaSpace{}%
\AgdaBound{inputs}\<%
\\
\\
\>[0]\AgdaFunction{inputs2PrevOutputs}\AgdaSpace{}%
\AgdaSymbol{:}\AgdaSpace{}%
\AgdaSymbol{(}\AgdaBound{tr}\AgdaSpace{}%
\AgdaSymbol{:}\AgdaSpace{}%
\AgdaDatatype{TXTree}\AgdaSymbol{)(}\AgdaBound{inputs}\AgdaSpace{}%
\AgdaSymbol{:}\AgdaSpace{}%
\AgdaFunction{TxInputs}\AgdaSpace{}%
\AgdaBound{tr}\AgdaSymbol{)}\AgdaSpace{}%
\AgdaSymbol{→}\AgdaSpace{}%
\AgdaDatatype{List}\AgdaSpace{}%
\AgdaRecord{TXOutput}\<%
\\
\>[0]\AgdaFunction{inputs2PrevOutputs}\AgdaSpace{}%
\AgdaBound{tr}\AgdaSpace{}%
\AgdaBound{inputs}\AgdaSpace{}%
\AgdaSymbol{=}\AgdaSpace{}%
\AgdaFunction{mapL}\AgdaSpace{}%
\AgdaField{proj₁}\AgdaSpace{}%
\AgdaSymbol{(}\AgdaFunction{inputs2PrevOutputsSigPbk}\AgdaSpace{}%
\AgdaBound{tr}\AgdaSpace{}%
\AgdaBound{inputs}\AgdaSymbol{)}\<%
\\
\\
\>[0]\AgdaFunction{inputs2Sum}\AgdaSpace{}%
\AgdaSymbol{:}\AgdaSpace{}%
\AgdaSymbol{(}\AgdaBound{tr}\AgdaSpace{}%
\AgdaSymbol{:}\AgdaSpace{}%
\AgdaDatatype{TXTree}\AgdaSymbol{)(}\AgdaBound{inputs}\AgdaSpace{}%
\AgdaSymbol{:}\AgdaSpace{}%
\AgdaFunction{TxInputs}\AgdaSpace{}%
\AgdaBound{tr}\AgdaSymbol{)}\AgdaSpace{}%
\AgdaSymbol{→}\AgdaSpace{}%
\AgdaFunction{Amount}\<%
\\
\>[0]\AgdaFunction{inputs2Sum}%
\>[12]\AgdaBound{tr}\AgdaSpace{}%
\AgdaBound{inputs}\AgdaSpace{}%
\AgdaSymbol{=}\AgdaSpace{}%
\AgdaFunction{sumListViaf}\AgdaSpace{}%
\AgdaFunction{txOutput2Amount}\AgdaSpace{}%
\AgdaSymbol{(}\AgdaFunction{inputs2PrevOutputs}\AgdaSpace{}%
\AgdaBound{tr}\AgdaSpace{}%
\AgdaBound{inputs}\AgdaSymbol{)}\<%
\end{code}
} 

\AgdaHide{
\begin{code}%
\>[0]\<%
\\
\>[0]\AgdaComment{{-}{-} \textbackslash{}bitcoinVersSix}\<%
\end{code}
} 

\newcommand{\bitcoinVersSixtxTreetoTimeNextTobeMinedBlock}{
\begin{code}%
\>[0]\AgdaFunction{txTree2TimeNextTobeMinedBlock}\AgdaSpace{}%
\AgdaSymbol{:}\AgdaSpace{}%
\AgdaSymbol{(}\AgdaBound{tr}\AgdaSpace{}%
\AgdaSymbol{:}\AgdaSpace{}%
\AgdaDatatype{TXTree}\AgdaSymbol{)}\AgdaSpace{}%
\AgdaSymbol{→}\AgdaSpace{}%
\AgdaFunction{Time}\<%
\\
\>[0]\AgdaFunction{txTree2TimeNextTobeMinedBlock}\AgdaSpace{}%
\AgdaInductiveConstructor{genesisTree}%
\>[71]\AgdaSymbol{=}%
\>[74]\AgdaNumber{0}\<%
\\
\>[0]\AgdaFunction{txTree2TimeNextTobeMinedBlock}\AgdaSpace{}%
\AgdaSymbol{(}\AgdaInductiveConstructor{txtree}\AgdaSpace{}%
\AgdaBound{tree}\AgdaSpace{}%
\AgdaSymbol{(}\AgdaInductiveConstructor{normalTX}\AgdaSpace{}%
\AgdaBound{inputs}\AgdaSpace{}%
\AgdaBound{outputs}\AgdaSymbol{))}%
\>[71]\AgdaSymbol{=}\<%
\\
\>[0][@{}l@{\AgdaIndent{0}}]%
\>[12]\AgdaFunction{txTree2TimeNextTobeMinedBlock}\AgdaSpace{}%
\AgdaBound{tree}\<%
\\
\>[0]\AgdaFunction{txTree2TimeNextTobeMinedBlock}\AgdaSpace{}%
\AgdaSymbol{(}\AgdaInductiveConstructor{txtree}\AgdaSpace{}%
\AgdaBound{tree}\AgdaSpace{}%
\AgdaSymbol{(}\AgdaInductiveConstructor{coinbase}\AgdaSpace{}%
\AgdaBound{time}\AgdaSpace{}%
\AgdaBound{outputs}\AgdaSymbol{))}%
\>[71]\AgdaSymbol{=}%
\>[74]\AgdaInductiveConstructor{suc}\AgdaSpace{}%
\AgdaBound{time}\<%
\end{code}
} 

\AgdaHide{
\begin{code}%
\>[0]\<%
\\
\>[0]\AgdaComment{{-}{-} \textbackslash{}bitcoinVersSix}\<%
\end{code}
} 

\newcommand{\bitcoinVersSixtxtoSumInputs}{
\begin{code}%
\>[0]\AgdaKeyword{mutual}\<%
\\
\>[0][@{}l@{\AgdaIndent{0}}]%
\>[2]\AgdaFunction{tx2SumInputs}\AgdaSpace{}%
\AgdaSymbol{:}\AgdaSpace{}%
\AgdaSymbol{(}\AgdaBound{tr}\AgdaSpace{}%
\AgdaSymbol{:}\AgdaSpace{}%
\AgdaDatatype{TXTree}\AgdaSymbol{)(}\AgdaBound{tx}\AgdaSpace{}%
\AgdaSymbol{:}\AgdaSpace{}%
\AgdaDatatype{TX}\AgdaSpace{}%
\AgdaBound{tr}\AgdaSymbol{)}\AgdaSpace{}%
\AgdaSymbol{→}\AgdaSpace{}%
\AgdaFunction{Amount}\<%
\\
\>[0][@{}l@{\AgdaIndent{0}}]%
\>[2]\AgdaFunction{tx2SumInputs}\AgdaSpace{}%
\AgdaBound{tr}\AgdaSpace{}%
\AgdaSymbol{(}\AgdaInductiveConstructor{normalTX}\AgdaSpace{}%
\AgdaBound{inputs}\AgdaSpace{}%
\AgdaBound{outputs}\AgdaSymbol{)}%
\>[45]\AgdaSymbol{=}\AgdaSpace{}%
\AgdaFunction{inputs2Sum}\AgdaSpace{}%
\AgdaBound{tr}\AgdaSpace{}%
\AgdaBound{inputs}\<%
\\
\>[0][@{}l@{\AgdaIndent{0}}]%
\>[2]\AgdaFunction{tx2SumInputs}\AgdaSpace{}%
\AgdaBound{tr}\AgdaSpace{}%
\AgdaSymbol{(}\AgdaInductiveConstructor{coinbase}\AgdaSpace{}%
\AgdaBound{time}\AgdaSpace{}%
\AgdaBound{outputs}\AgdaSymbol{)}%
\>[45]\AgdaSymbol{=}\<%
\\
\>[2][@{}l@{\AgdaIndent{0}}]%
\>[5]\AgdaFunction{txTree2RecentTXFees}\AgdaSpace{}%
\AgdaBound{tr}\AgdaSpace{}%
\AgdaPrimitive{+}\AgdaSpace{}%
\AgdaPostulate{blockReward}\AgdaSpace{}%
\AgdaSymbol{(}\AgdaFunction{txTree2TimeNextTobeMinedBlock}\AgdaSpace{}%
\AgdaBound{tr}\AgdaSymbol{)}\<%
\\
\\
\>[0][@{}l@{\AgdaIndent{0}}]%
\>[2]\AgdaFunction{txTree2RecentTXFees}\AgdaSpace{}%
\AgdaSymbol{:}\AgdaSpace{}%
\AgdaSymbol{(}\AgdaBound{tr}\AgdaSpace{}%
\AgdaSymbol{:}\AgdaSpace{}%
\AgdaDatatype{TXTree}\AgdaSymbol{)}\AgdaSpace{}%
\AgdaSymbol{→}\AgdaSpace{}%
\AgdaFunction{Amount}\<%
\\
\>[0][@{}l@{\AgdaIndent{0}}]%
\>[2]\AgdaFunction{txTree2RecentTXFees}\AgdaSpace{}%
\AgdaInductiveConstructor{genesisTree}%
\>[61]\AgdaSymbol{=}\AgdaSpace{}%
\AgdaNumber{0}\<%
\\
\>[0][@{}l@{\AgdaIndent{0}}]%
\>[2]\AgdaFunction{txTree2RecentTXFees}\AgdaSpace{}%
\AgdaSymbol{(}\AgdaInductiveConstructor{txtree}\AgdaSpace{}%
\AgdaBound{tr}\AgdaSpace{}%
\AgdaSymbol{(}\AgdaInductiveConstructor{normalTX}\AgdaSpace{}%
\AgdaBound{inputs}\AgdaSpace{}%
\AgdaBound{outputs}\AgdaSymbol{))}%
\>[61]\AgdaSymbol{=}\<%
\\
\>[2][@{}l@{\AgdaIndent{0}}]%
\>[5]\AgdaFunction{txTree2RecentTXFees}\AgdaSpace{}%
\AgdaBound{tr}\AgdaSpace{}%
\AgdaPrimitive{+}\AgdaSpace{}%
\AgdaSymbol{(}\AgdaFunction{inputs2Sum}\AgdaSpace{}%
\AgdaBound{tr}\AgdaSpace{}%
\AgdaBound{inputs}\AgdaSpace{}%
\AgdaPrimitive{∸}\AgdaSpace{}%
\AgdaFunction{outputs2SumAmount}\AgdaSpace{}%
\AgdaBound{outputs}\AgdaSymbol{)}\<%
\\
\>[0][@{}l@{\AgdaIndent{0}}]%
\>[2]\AgdaFunction{txTree2RecentTXFees}\AgdaSpace{}%
\AgdaSymbol{(}\AgdaInductiveConstructor{txtree}\AgdaSpace{}%
\AgdaBound{tr}\AgdaSpace{}%
\AgdaSymbol{(}\AgdaInductiveConstructor{coinbase}\AgdaSpace{}%
\AgdaBound{time}\AgdaSpace{}%
\AgdaBound{outputs}\AgdaSymbol{))}%
\>[61]\AgdaSymbol{=}\AgdaSpace{}%
\AgdaNumber{0}\<%
\end{code}
} 

\AgdaHide{
\begin{code}%
\>[0]\<%
\\
\>[0]\AgdaComment{{-}{-} \textbackslash{}bitcoinVersSix}\<%
\end{code}
} 

\newcommand{\bitcoinVersSixoutputToMaturationTime}{
\begin{code}%
\>[0]\AgdaFunction{output2MaturationTime}\AgdaSpace{}%
\AgdaSymbol{:}\AgdaSpace{}%
\AgdaRecord{TXOutput}\AgdaSpace{}%
\AgdaSymbol{→}\AgdaSpace{}%
\AgdaFunction{Time}\<%
\\
\>[0]\AgdaFunction{output2MaturationTime}\AgdaSpace{}%
\AgdaSymbol{(}\AgdaInductiveConstructor{txOutput}\AgdaSpace{}%
\AgdaBound{trTree}\AgdaSpace{}%
\AgdaSymbol{(}\AgdaInductiveConstructor{normalTX}\AgdaSpace{}%
\AgdaBound{inputs}\AgdaSpace{}%
\AgdaBound{outputs}\AgdaSymbol{)}\AgdaSpace{}%
\AgdaBound{i}\AgdaSymbol{)}\AgdaSpace{}%
\AgdaSymbol{=}\AgdaSpace{}%
\AgdaNumber{0}\<%
\\
\>[0]\AgdaFunction{output2MaturationTime}\AgdaSpace{}%
\AgdaSymbol{(}\AgdaInductiveConstructor{txOutput}\AgdaSpace{}%
\AgdaBound{trTree}\AgdaSpace{}%
\AgdaSymbol{(}\AgdaInductiveConstructor{coinbase}\AgdaSpace{}%
\AgdaBound{time}\AgdaSpace{}%
\AgdaBound{outputs}\AgdaSymbol{)}\AgdaSpace{}%
\AgdaBound{i}\AgdaSymbol{)}\AgdaSpace{}%
\AgdaSymbol{=}\AgdaSpace{}%
\AgdaBound{time}\AgdaSpace{}%
\AgdaPrimitive{+}\AgdaSpace{}%
\AgdaPostulate{blockMaturationTime}\<%
\end{code}
} 

\AgdaHide{
\begin{code}%
\>[0]\<%
\\
\\
\>[0]\AgdaComment{{-}{-} \#\#\#\#\#\#\#\#\#\#\#\#\#\#\#\#\#\#\#\#\#\#\#\#\#\#\#\#\#\#\#\#\#\#\#\#\#\#\#\#\#\#\#\#\#\#\#\#\#\#\#\#\#\#\#\#\#\#\#\#\#\#}\<%
\\
\>[0]\AgdaComment{{-}{-} \#  Computing Messages to be Signed and Transaction IDs       \#}\<%
\\
\>[0]\AgdaComment{{-}{-} \#\#\#\#\#\#\#\#\#\#\#\#\#\#\#\#\#\#\#\#\#\#\#\#\#\#\#\#\#\#\#\#\#\#\#\#\#\#\#\#\#\#\#\#\#\#\#\#\#\#\#\#\#\#\#\#\#\#\#\#\#\#}\<%
\\
\\
\>[0]\AgdaComment{{-}{-} \textbackslash{}bitcoinVersSix}\<%
\end{code}
} 

\newcommand{\bitcoinVersSixtxOutputfieldtoMsg}{
\begin{code}%
\>[0]\AgdaFunction{txOutputfield2Msg}\AgdaSpace{}%
\AgdaSymbol{:}\AgdaSpace{}%
\AgdaRecord{TXOutputfield}\AgdaSpace{}%
\AgdaSymbol{→}\AgdaSpace{}%
\AgdaDatatype{Msg}\<%
\\
\>[0]\AgdaFunction{txOutputfield2Msg}\AgdaSpace{}%
\AgdaSymbol{(}\AgdaInductiveConstructor{txOutputfield}\AgdaSpace{}%
\AgdaBound{amount₁}\AgdaSpace{}%
\AgdaBound{address₁}\AgdaSymbol{)}\AgdaSpace{}%
\AgdaSymbol{=}\AgdaSpace{}%
\AgdaInductiveConstructor{nat}\AgdaSpace{}%
\AgdaBound{amount₁}\AgdaSpace{}%
\AgdaInductiveConstructor{+msg}\AgdaSpace{}%
\AgdaInductiveConstructor{nat}\AgdaSpace{}%
\AgdaBound{address₁}\<%
\\
\\
\>[0]\AgdaFunction{outputFields2Msg}\AgdaSpace{}%
\AgdaSymbol{:}\AgdaSpace{}%
\AgdaSymbol{(}\AgdaBound{outp}\AgdaSpace{}%
\AgdaSymbol{:}\AgdaSpace{}%
\AgdaDatatype{List}\AgdaSpace{}%
\AgdaRecord{TXOutputfield}\AgdaSymbol{)}\AgdaSpace{}%
\AgdaSymbol{→}\AgdaSpace{}%
\AgdaDatatype{Msg}\<%
\\
\>[0]\AgdaFunction{outputFields2Msg}\AgdaSpace{}%
\AgdaBound{outp}\AgdaSpace{}%
\AgdaSymbol{=}\AgdaSpace{}%
\AgdaInductiveConstructor{list}\AgdaSpace{}%
\AgdaSymbol{(}\AgdaFunction{mapL}\AgdaSpace{}%
\AgdaFunction{txOutputfield2Msg}\AgdaSpace{}%
\AgdaBound{outp}\AgdaSymbol{)}\<%
\end{code}
} 

\AgdaHide{
\begin{code}%
\>[0]\<%
\\
\>[0]\AgdaKeyword{mutual}\<%
\\
\\
\>[0]\AgdaComment{{-}{-} \textbackslash{}bitcoinVersSix}\<%
\end{code}
} 

\newcommand{\bitcoinVersSixtxtoMsg}{
\begin{code}%
\>[0][@{}l@{\AgdaIndent{1}}]%
\>[2]\AgdaFunction{tx2Msg}\AgdaSpace{}%
\AgdaSymbol{:}\AgdaSpace{}%
\AgdaSymbol{(}\AgdaBound{tr}\AgdaSpace{}%
\AgdaSymbol{:}\AgdaSpace{}%
\AgdaDatatype{TXTree}\AgdaSymbol{)(}\AgdaBound{tx}\AgdaSpace{}%
\AgdaSymbol{:}\AgdaSpace{}%
\AgdaDatatype{TX}\AgdaSpace{}%
\AgdaBound{tr}\AgdaSymbol{)}\AgdaSpace{}%
\AgdaSymbol{→}\AgdaSpace{}%
\AgdaDatatype{Msg}\<%
\\
\>[0][@{}l@{\AgdaIndent{1}}]%
\>[2]\AgdaFunction{tx2Msg}\AgdaSpace{}%
\AgdaBound{tr}\AgdaSpace{}%
\AgdaSymbol{(}\AgdaInductiveConstructor{normalTX}\AgdaSpace{}%
\AgdaBound{inputs₁}\AgdaSpace{}%
\AgdaBound{outputs₁}\AgdaSymbol{)}%
\>[518I]\AgdaSymbol{=}%
\>[43]\AgdaInductiveConstructor{list}\AgdaSpace{}%
\AgdaSymbol{(}\AgdaFunction{mapL}\AgdaSpace{}%
\AgdaFunction{utxoIndexSig2Msg}\AgdaSpace{}%
\AgdaFunction{inputIndices}\AgdaSymbol{)}\<%
\\
\>[518I][@{}l@{\AgdaIndent{0}}]%
\>[43]\AgdaInductiveConstructor{+msg}\AgdaSpace{}%
\AgdaFunction{outputFields2Msg}\AgdaSpace{}%
\AgdaBound{outputs₁}\<%
\\
\>[2][@{}l@{\AgdaIndent{0}}]%
\>[6]\AgdaKeyword{where}\<%
\\
\>[6][@{}l@{\AgdaIndent{0}}]%
\>[9]\AgdaFunction{inputIndices}\AgdaSpace{}%
\AgdaSymbol{:}\AgdaSpace{}%
\AgdaDatatype{List}\AgdaSpace{}%
\AgdaSymbol{(}\AgdaDatatype{Fin}\AgdaSpace{}%
\AgdaSymbol{(}\AgdaFunction{length}\AgdaSpace{}%
\AgdaSymbol{(}\AgdaFunction{utxo}\AgdaSpace{}%
\AgdaBound{tr}\AgdaSymbol{))}\AgdaSpace{}%
\AgdaFunction{×}\AgdaSpace{}%
\AgdaFunction{PublicKey}\AgdaSpace{}%
\AgdaFunction{×}\AgdaSpace{}%
\AgdaFunction{Signature}\AgdaSymbol{)}\<%
\\
\>[6][@{}l@{\AgdaIndent{0}}]%
\>[9]\AgdaFunction{inputIndices}\AgdaSpace{}%
\AgdaSymbol{=}\AgdaSpace{}%
\AgdaFunction{subList+2IndicesOriginalList}\AgdaSpace{}%
\AgdaSymbol{(}\AgdaFunction{utxo}\AgdaSpace{}%
\AgdaBound{tr}\AgdaSymbol{)}\AgdaSpace{}%
\AgdaBound{inputs₁}\<%
\\
\\
\>[6][@{}l@{\AgdaIndent{0}}]%
\>[9]\AgdaFunction{utxoIndexSig2Msg}\AgdaSpace{}%
\AgdaSymbol{:}\AgdaSpace{}%
\AgdaDatatype{Fin}\AgdaSpace{}%
\AgdaSymbol{(}\AgdaFunction{length}\AgdaSpace{}%
\AgdaSymbol{(}\AgdaFunction{utxo}\AgdaSpace{}%
\AgdaBound{tr}\AgdaSymbol{))}\AgdaSpace{}%
\AgdaFunction{×}\AgdaSpace{}%
\AgdaFunction{PublicKey}\AgdaSpace{}%
\AgdaFunction{×}\AgdaSpace{}%
\AgdaFunction{Signature}\AgdaSpace{}%
\AgdaSymbol{→}\AgdaSpace{}%
\AgdaDatatype{Msg}\<%
\\
\>[6][@{}l@{\AgdaIndent{0}}]%
\>[9]\AgdaFunction{utxoIndexSig2Msg}\AgdaSpace{}%
\AgdaSymbol{(}\AgdaBound{i}\AgdaSpace{}%
\AgdaInductiveConstructor{,}\AgdaSpace{}%
\AgdaSymbol{(}\AgdaBound{pbk}\AgdaSpace{}%
\AgdaInductiveConstructor{,}\AgdaSpace{}%
\AgdaBound{sig}\AgdaSymbol{))}\AgdaSpace{}%
\AgdaSymbol{=}\AgdaSpace{}%
\AgdaFunction{utxo2Msg}\AgdaSpace{}%
\AgdaBound{tr}\AgdaSpace{}%
\AgdaBound{i}\AgdaSpace{}%
\AgdaInductiveConstructor{+msg}\AgdaSpace{}%
\AgdaInductiveConstructor{nat}\AgdaSpace{}%
\AgdaBound{pbk}\AgdaSpace{}%
\AgdaInductiveConstructor{+msg}\AgdaSpace{}%
\AgdaInductiveConstructor{nat}\AgdaSpace{}%
\AgdaBound{sig}\<%
\\
\\
\>[0][@{}l@{\AgdaIndent{1}}]%
\>[2]\AgdaFunction{tx2Msg}\AgdaSpace{}%
\AgdaBound{tr}\AgdaSpace{}%
\AgdaSymbol{(}\AgdaInductiveConstructor{coinbase}\AgdaSpace{}%
\AgdaBound{time}\AgdaSpace{}%
\AgdaBound{outputs₁}\AgdaSymbol{)}\AgdaSpace{}%
\AgdaSymbol{=}\AgdaSpace{}%
\AgdaInductiveConstructor{nat}\AgdaSpace{}%
\AgdaBound{time}\AgdaSpace{}%
\AgdaInductiveConstructor{+msg}\AgdaSpace{}%
\AgdaFunction{outputFields2Msg}\AgdaSpace{}%
\AgdaBound{outputs₁}\<%
\end{code}
} 

\AgdaHide{
\begin{code}%
\>[0]\<%
\\
\>[0]\AgdaComment{{-}{-} \textbackslash{}bitcoinVersSix}\<%
\end{code}
} 

\newcommand{\bitcoinVersSixtxtoid}{
\begin{code}%
\>[0][@{}l@{\AgdaIndent{1}}]%
\>[2]\AgdaFunction{tx2id}\AgdaSpace{}%
\AgdaSymbol{:}\AgdaSpace{}%
\AgdaSymbol{(}\AgdaBound{tr}\AgdaSpace{}%
\AgdaSymbol{:}\AgdaSpace{}%
\AgdaDatatype{TXTree}\AgdaSymbol{)(}\AgdaBound{tx}\AgdaSpace{}%
\AgdaSymbol{:}\AgdaSpace{}%
\AgdaDatatype{TX}\AgdaSpace{}%
\AgdaBound{tr}\AgdaSymbol{)}\AgdaSpace{}%
\AgdaSymbol{→}\AgdaSpace{}%
\AgdaDatatype{ℕ}\<%
\\
\>[0][@{}l@{\AgdaIndent{1}}]%
\>[2]\AgdaFunction{tx2id}\AgdaSpace{}%
\AgdaBound{tr}\AgdaSpace{}%
\AgdaBound{tx}\AgdaSpace{}%
\AgdaSymbol{=}\AgdaSpace{}%
\AgdaPostulate{hashMsg}\AgdaSpace{}%
\AgdaSymbol{(}\AgdaFunction{tx2Msg}\AgdaSpace{}%
\AgdaBound{tr}\AgdaSpace{}%
\AgdaBound{tx}\AgdaSymbol{)}\<%
\end{code}
} 

\AgdaHide{
\begin{code}%
\>[0]\<%
\\
\>[0][@{}l@{\AgdaIndent{2}}]%
\>[2]\AgdaComment{{-}{-} note time included following the fix for nonuniqueness of ids}\<%
\\
\>[0][@{}l@{\AgdaIndent{2}}]%
\>[2]\AgdaComment{{-}{-} in \textbackslash{}cite\{github:bip{-}0034.mediawiki\}}\<%
\\
\\
\>[0]\AgdaComment{{-}{-} \textbackslash{}bitcoinVersSix}\<%
\end{code}
} 

\newcommand{\bitcoinVersSixutxoMinusNewInputstoMsg}{
\begin{code}%
\>[0][@{}l@{\AgdaIndent{1}}]%
\>[2]\AgdaFunction{utxoMinusNewInputs2Msg}%
\>[591I]\AgdaSymbol{:}%
\>[28]\AgdaSymbol{(}\AgdaBound{tr}\AgdaSpace{}%
\AgdaSymbol{:}\AgdaSpace{}%
\AgdaDatatype{TXTree}\AgdaSymbol{)(}\AgdaBound{tx}\AgdaSpace{}%
\AgdaSymbol{:}\AgdaSpace{}%
\AgdaDatatype{TX}\AgdaSpace{}%
\AgdaBound{tr}\AgdaSymbol{)(}\AgdaBound{i}\AgdaSpace{}%
\AgdaSymbol{:}\AgdaSpace{}%
\AgdaDatatype{Fin}\AgdaSpace{}%
\AgdaSymbol{(}\AgdaFunction{length}\AgdaSpace{}%
\AgdaSymbol{(}\AgdaFunction{utxoMinusNewInputs}\AgdaSpace{}%
\AgdaBound{tr}\AgdaSpace{}%
\AgdaBound{tx}\AgdaSymbol{)))}\<%
\\
\>[591I][@{}l@{\AgdaIndent{0}}]%
\>[28]\AgdaSymbol{→}\AgdaSpace{}%
\AgdaDatatype{Msg}\<%
\\
\>[0][@{}l@{\AgdaIndent{1}}]%
\>[2]\AgdaFunction{utxoMinusNewInputs2Msg}\AgdaSpace{}%
\AgdaBound{tr}\AgdaSpace{}%
\AgdaSymbol{(}\AgdaInductiveConstructor{normalTX}\AgdaSpace{}%
\AgdaBound{inputs}\AgdaSpace{}%
\AgdaBound{outputs}\AgdaSymbol{)}\AgdaSpace{}%
\AgdaBound{i}\AgdaSpace{}%
\AgdaSymbol{=}\<%
\\
\>[2][@{}l@{\AgdaIndent{0}}]%
\>[10]\AgdaFunction{utxo2Msg}\AgdaSpace{}%
\AgdaBound{tr}\AgdaSpace{}%
\AgdaSymbol{(}\AgdaFunction{listMinusSubList+Index2OrgIndex}\AgdaSpace{}%
\AgdaSymbol{(}\AgdaFunction{utxo}\AgdaSpace{}%
\AgdaBound{tr}\AgdaSymbol{)}\AgdaSpace{}%
\AgdaBound{inputs}\AgdaSpace{}%
\AgdaBound{i}\AgdaSymbol{)}\<%
\\
\>[0][@{}l@{\AgdaIndent{1}}]%
\>[2]\AgdaFunction{utxoMinusNewInputs2Msg}\AgdaSpace{}%
\AgdaBound{tr}\AgdaSpace{}%
\AgdaSymbol{(}\AgdaInductiveConstructor{coinbase}\AgdaSpace{}%
\AgdaBound{time}\AgdaSpace{}%
\AgdaBound{outputs}\AgdaSymbol{)}\AgdaSpace{}%
\AgdaBound{i}\AgdaSpace{}%
\AgdaSymbol{=}\AgdaSpace{}%
\AgdaFunction{utxo2Msg}\AgdaSpace{}%
\AgdaBound{tr}\AgdaSpace{}%
\AgdaBound{i}\<%
\end{code}
} 

\AgdaHide{
\begin{code}%
\>[0]\<%
\\
\>[2][@{}l@{\AgdaIndent{1}}]%
\>[4]\AgdaComment{{-}{-} listMinusSubList+Index2OrgIndex}\<%
\\
\>[2][@{}l@{\AgdaIndent{1}}]%
\>[4]\AgdaComment{{-}{-} will map the result of deleting entries indices to the original}\<%
\\
\>[2][@{}l@{\AgdaIndent{1}}]%
\>[4]\AgdaComment{{-}{-} entries}\<%
\\
\>[2][@{}l@{\AgdaIndent{1}}]%
\>[4]\AgdaComment{{-}{-} we now need to add as well the new indices}\<%
\\
\\
\>[0]\AgdaComment{{-}{-} \textbackslash{}bitcoinVersSix}\<%
\end{code}
} 

\newcommand{\bitcoinVersSixutxotoMsg}{
\begin{code}%
\>[0][@{}l@{\AgdaIndent{1}}]%
\>[2]\AgdaFunction{utxo2Msg}\AgdaSpace{}%
\AgdaSymbol{:}\AgdaSpace{}%
\AgdaSymbol{(}\AgdaBound{tr}\AgdaSpace{}%
\AgdaSymbol{:}\AgdaSpace{}%
\AgdaDatatype{TXTree}\AgdaSymbol{)(}\AgdaBound{i}\AgdaSpace{}%
\AgdaSymbol{:}\AgdaSpace{}%
\AgdaDatatype{Fin}\AgdaSpace{}%
\AgdaSymbol{(}\AgdaFunction{length}\AgdaSpace{}%
\AgdaSymbol{(}\AgdaFunction{utxo}\AgdaSpace{}%
\AgdaBound{tr}\AgdaSymbol{)))}\AgdaSpace{}%
\AgdaSymbol{→}\AgdaSpace{}%
\AgdaDatatype{Msg}\<%
\\
\>[0][@{}l@{\AgdaIndent{1}}]%
\>[2]\AgdaFunction{utxo2Msg}\AgdaSpace{}%
\AgdaInductiveConstructor{genesisTree}\AgdaSpace{}%
\AgdaSymbol{()}\<%
\\
\>[0][@{}l@{\AgdaIndent{1}}]%
\>[2]\AgdaFunction{utxo2Msg}\AgdaSpace{}%
\AgdaSymbol{(}\AgdaInductiveConstructor{txtree}\AgdaSpace{}%
\AgdaBound{tr}\AgdaSpace{}%
\AgdaBound{tx}\AgdaSymbol{)}\AgdaSpace{}%
\AgdaSymbol{=}\AgdaSpace{}%
\AgdaFunction{concatListIndex2OriginIndices}\AgdaSpace{}%
\AgdaFunction{l₀}\AgdaSpace{}%
\AgdaFunction{l₁}\AgdaSpace{}%
\AgdaFunction{f₀}\AgdaSpace{}%
\AgdaFunction{f₁}\<%
\\
\>[2][@{}l@{\AgdaIndent{0}}]%
\>[4]\AgdaKeyword{module}\AgdaSpace{}%
\AgdaModule{utxo2Msgaux}\AgdaSpace{}%
\AgdaKeyword{where}\<%
\\
\>[4][@{}l@{\AgdaIndent{0}}]%
\>[8]\AgdaFunction{l₀}\AgdaSpace{}%
\AgdaSymbol{:}\AgdaSpace{}%
\AgdaDatatype{List}\AgdaSpace{}%
\AgdaRecord{TXOutput}\<%
\\
\>[4][@{}l@{\AgdaIndent{0}}]%
\>[8]\AgdaFunction{l₀}\AgdaSpace{}%
\AgdaSymbol{=}\AgdaSpace{}%
\AgdaFunction{utxoMinusNewInputs}\AgdaSpace{}%
\AgdaBound{tr}\AgdaSpace{}%
\AgdaBound{tx}\<%
\\
\\
\>[4][@{}l@{\AgdaIndent{0}}]%
\>[8]\AgdaFunction{l₁}\AgdaSpace{}%
\AgdaSymbol{:}\AgdaSpace{}%
\AgdaDatatype{List}\AgdaSpace{}%
\AgdaRecord{TXOutput}\<%
\\
\>[4][@{}l@{\AgdaIndent{0}}]%
\>[8]\AgdaFunction{l₁}\AgdaSpace{}%
\AgdaSymbol{=}\AgdaSpace{}%
\AgdaFunction{tx2TXOutputs}\AgdaSpace{}%
\AgdaBound{tr}\AgdaSpace{}%
\AgdaBound{tx}\<%
\\
\\
\>[4][@{}l@{\AgdaIndent{0}}]%
\>[8]\AgdaFunction{f₀}\AgdaSpace{}%
\AgdaSymbol{:}\AgdaSpace{}%
\AgdaDatatype{Fin}\AgdaSpace{}%
\AgdaSymbol{(}\AgdaFunction{length}\AgdaSpace{}%
\AgdaFunction{l₀}\AgdaSymbol{)}%
\>[30]\AgdaSymbol{→}\AgdaSpace{}%
\AgdaDatatype{Msg}\<%
\\
\>[4][@{}l@{\AgdaIndent{0}}]%
\>[8]\AgdaFunction{f₀}\AgdaSpace{}%
\AgdaBound{i}\AgdaSpace{}%
\AgdaSymbol{=}\AgdaSpace{}%
\AgdaFunction{utxoMinusNewInputs2Msg}\AgdaSpace{}%
\AgdaBound{tr}\AgdaSpace{}%
\AgdaBound{tx}\AgdaSpace{}%
\AgdaBound{i}\<%
\\
\\
\>[4][@{}l@{\AgdaIndent{0}}]%
\>[8]\AgdaFunction{f₁}\AgdaSpace{}%
\AgdaSymbol{:}\AgdaSpace{}%
\AgdaDatatype{Fin}\AgdaSpace{}%
\AgdaSymbol{(}\AgdaFunction{length}\AgdaSpace{}%
\AgdaFunction{l₁}\AgdaSymbol{)}\AgdaSpace{}%
\AgdaSymbol{→}\AgdaSpace{}%
\AgdaDatatype{Msg}\<%
\\
\>[4][@{}l@{\AgdaIndent{0}}]%
\>[8]\AgdaFunction{f₁}\AgdaSpace{}%
\AgdaBound{i}\AgdaSpace{}%
\AgdaSymbol{=}\AgdaSpace{}%
\AgdaInductiveConstructor{nat}\AgdaSpace{}%
\AgdaSymbol{(}\AgdaFunction{tx2id}\AgdaSpace{}%
\AgdaBound{tr}\AgdaSpace{}%
\AgdaBound{tx}\AgdaSymbol{)}\AgdaSpace{}%
\AgdaInductiveConstructor{+msg}\AgdaSpace{}%
\AgdaInductiveConstructor{nat}\AgdaSpace{}%
\AgdaSymbol{(}\AgdaFunction{toℕ}\AgdaSpace{}%
\AgdaBound{i}\AgdaSymbol{)}\<%
\end{code}
} 

\AgdaHide{
\begin{code}%
\>[0]\<%
\\
\\
\\
\>[0]\AgdaComment{{-}{-} msg to be signed is what needs to be signed by the user}\<%
\\
\>[0]\AgdaComment{{-}{-} in \textbackslash{}cite\{derosa:theFirstTransacationPt1\} he writes:\textbackslash{}\textbackslash{}}\<%
\\
\>[0]\AgdaComment{{-}{-} In practice, for each input I, the message to be signed is a slightly modified version of the transaction where:}\<%
\\
\>[0]\AgdaComment{{-}{-} {-}  The I script is set to the output script of the UTXO it refers to.}\<%
\\
\>[0]\AgdaComment{{-}{-} {-} Input scripts other than I are truncated to zero{-}length.}\<%
\\
\>[0]\AgdaComment{{-}{-} {-} A SIGHASH flag is appended.}\<%
\\
\>[0]\AgdaComment{{-}{-}}\<%
\\
\>[0]\AgdaComment{{-}{-} since in our case the output scrpt is epresented by by the address in the}\<%
\\
\>[0]\AgdaComment{{-}{-}   txoutput field, we get the following:}\<%
\\
\\
\>[0]\AgdaComment{{-}{-} \textbackslash{}bitcoinVersSix}\<%
\\
\>[0]\AgdaComment{{-}{-} unused}\<%
\\
\>[0]\AgdaFunction{txOutput2Msg}\AgdaSpace{}%
\AgdaSymbol{:}\AgdaSpace{}%
\AgdaSymbol{(}\AgdaBound{txout}%
\>[23]\AgdaSymbol{:}\AgdaSpace{}%
\AgdaRecord{TXOutput}\AgdaSymbol{)}\AgdaSpace{}%
\AgdaSymbol{→}\AgdaSpace{}%
\AgdaDatatype{Msg}\<%
\\
\>[0]\AgdaFunction{txOutput2Msg}\AgdaSpace{}%
\AgdaSymbol{(}\AgdaInductiveConstructor{txOutput}\AgdaSpace{}%
\AgdaBound{tr}\AgdaSpace{}%
\AgdaBound{tx}\AgdaSpace{}%
\AgdaBound{i}\AgdaSymbol{)}\AgdaSpace{}%
\AgdaSymbol{=}\AgdaSpace{}%
\AgdaInductiveConstructor{nat}\AgdaSpace{}%
\AgdaSymbol{(}\AgdaFunction{tx2id}\AgdaSpace{}%
\AgdaBound{tr}\AgdaSpace{}%
\AgdaBound{tx}\AgdaSymbol{)}\AgdaSpace{}%
\AgdaInductiveConstructor{+msg}\AgdaSpace{}%
\AgdaInductiveConstructor{nat}\AgdaSpace{}%
\AgdaSymbol{(}\AgdaFunction{toℕ}\AgdaSpace{}%
\AgdaBound{i}\AgdaSymbol{)}\<%
\\
\\
\\
\\
\>[0]\AgdaComment{{-}{-} \textbackslash{}bitcoinVersSix}\<%
\end{code}
} 

\newcommand{\bitcoinVersSixmsgToBeSignedByInput}{
\begin{code}%
\>[0]\AgdaFunction{msgToBeSignedByInput}\AgdaSpace{}%
\AgdaSymbol{:}\AgdaSpace{}%
\AgdaSymbol{(}\AgdaBound{txoutput}\AgdaSpace{}%
\AgdaSymbol{:}\AgdaSpace{}%
\AgdaRecord{TXOutput}\AgdaSymbol{)(}\AgdaBound{outputs}%
\>[56]\AgdaSymbol{:}\AgdaSpace{}%
\AgdaDatatype{List}\AgdaSpace{}%
\AgdaRecord{TXOutputfield}\AgdaSymbol{)}\AgdaSpace{}%
\AgdaSymbol{→}\AgdaSpace{}%
\AgdaDatatype{Msg}\<%
\\
\>[0]\AgdaFunction{msgToBeSignedByInput}\AgdaSpace{}%
\AgdaBound{txoutput}\AgdaSpace{}%
\AgdaBound{outputs}\AgdaSpace{}%
\AgdaSymbol{=}\<%
\\
\>[0][@{}l@{\AgdaIndent{0}}]%
\>[19]\AgdaSymbol{(}\AgdaInductiveConstructor{nat}\AgdaSpace{}%
\AgdaSymbol{(}\AgdaFunction{tx2id}\AgdaSpace{}%
\AgdaSymbol{(}\AgdaField{trTree}\AgdaSpace{}%
\AgdaBound{txoutput}\AgdaSymbol{)}\AgdaSpace{}%
\AgdaSymbol{(}\AgdaField{tx}\AgdaSpace{}%
\AgdaBound{txoutput}\AgdaSymbol{))}\AgdaSpace{}%
\AgdaInductiveConstructor{+msg}\<%
\\
\>[19][@{}l@{\AgdaIndent{0}}]%
\>[20]\AgdaInductiveConstructor{nat}\AgdaSpace{}%
\AgdaSymbol{(}\AgdaFunction{toℕ}\AgdaSpace{}%
\AgdaSymbol{(}\AgdaField{output}\AgdaSpace{}%
\AgdaBound{txoutput}\AgdaSymbol{))}\AgdaSpace{}%
\AgdaInductiveConstructor{+msg}\<%
\\
\>[19][@{}l@{\AgdaIndent{0}}]%
\>[20]\AgdaInductiveConstructor{nat}\AgdaSpace{}%
\AgdaSymbol{(}\AgdaFunction{txOutput2Address}\AgdaSpace{}%
\AgdaBound{txoutput}\AgdaSymbol{))}\AgdaSpace{}%
\AgdaInductiveConstructor{+msg}\<%
\\
\>[0][@{}l@{\AgdaIndent{0}}]%
\>[19]\AgdaFunction{outputFields2Msg}\AgdaSpace{}%
\AgdaBound{outputs}\<%
\end{code}
} 

\AgdaHide{
\begin{code}%
\>[0]\<%
\\
\>[0][@{}l@{\AgdaIndent{1}}]%
\>[1]\AgdaComment{{-}{-} first item is the id of previous transaction}\<%
\\
\>[0][@{}l@{\AgdaIndent{1}}]%
\>[1]\AgdaComment{{-}{-} second item is the nr of the output within this transaction}\<%
\\
\>[0][@{}l@{\AgdaIndent{1}}]%
\>[1]\AgdaComment{{-}{-} thrid item is the output address of this transcation}\<%
\\
\>[0][@{}l@{\AgdaIndent{1}}]%
\>[1]\AgdaComment{{-}{-} last item is the essage for the outputs}\<%
\\
\\
\>[1][@{}l@{\AgdaIndent{0}}]%
\>[27]\AgdaComment{{-}{-} address}\<%
\\
\>[1][@{}l@{\AgdaIndent{0}}]%
\>[27]\AgdaComment{{-}{-} with script it would be}\<%
\\
\>[1][@{}l@{\AgdaIndent{0}}]%
\>[27]\AgdaComment{{-}{-}   the scriptPubKey of the output}\<%
\\
\>[1][@{}l@{\AgdaIndent{0}}]%
\>[27]\AgdaComment{{-}{-} here it is the address}\<%
\\
\\
\>[0]\AgdaComment{{-}{-} \textbackslash{}bitcoinVersSix}\<%
\end{code}
} 

\newcommand{\bitcoinVersSixCorrectTX}{
\begin{code}%
\>[0]\AgdaFunction{CorrectTX}\AgdaSpace{}%
\AgdaSymbol{:}\AgdaSpace{}%
\AgdaSymbol{(}\AgdaBound{tr}\AgdaSpace{}%
\AgdaSymbol{:}\AgdaSpace{}%
\AgdaDatatype{TXTree}\AgdaSymbol{)(}\AgdaBound{tx}\AgdaSpace{}%
\AgdaSymbol{:}\AgdaSpace{}%
\AgdaDatatype{TX}\AgdaSpace{}%
\AgdaBound{tr}\AgdaSymbol{)}\AgdaSpace{}%
\AgdaSymbol{→}\AgdaSpace{}%
\AgdaPrimitiveType{Set}\<%
\\
\>[0]\AgdaFunction{CorrectTX}\AgdaSpace{}%
\AgdaBound{tr}\AgdaSpace{}%
\AgdaSymbol{(}\AgdaInductiveConstructor{normalTX}\AgdaSpace{}%
\AgdaBound{inputs}\AgdaSpace{}%
\AgdaBound{outputs}\AgdaSymbol{)}\AgdaSpace{}%
\AgdaSymbol{=}\<%
\\
\>[0][@{}l@{\AgdaIndent{0}}]%
\>[9]\AgdaFunction{NonNil}\AgdaSpace{}%
\AgdaSymbol{(}\AgdaFunction{inputs2PrevOutputs}\AgdaSpace{}%
\AgdaBound{tr}\AgdaSpace{}%
\AgdaBound{inputs}\AgdaSymbol{)}\AgdaSpace{}%
\AgdaFunction{×}\<%
\\
\>[0][@{}l@{\AgdaIndent{0}}]%
\>[9]\AgdaFunction{NonNil}\AgdaSpace{}%
\AgdaBound{outputs}\AgdaSpace{}%
\AgdaFunction{×}\<%
\\
\>[0][@{}l@{\AgdaIndent{0}}]%
\>[9]\AgdaSymbol{(}\AgdaFunction{inputs2Sum}\AgdaSpace{}%
\AgdaBound{tr}\AgdaSpace{}%
\AgdaBound{inputs}\AgdaSpace{}%
\AgdaFunction{≥}%
\>[34]\AgdaFunction{outputs2SumAmount}\AgdaSpace{}%
\AgdaBound{outputs}\AgdaSymbol{)}\AgdaSpace{}%
\AgdaFunction{×}\<%
\\
\>[0][@{}l@{\AgdaIndent{0}}]%
\>[9]\AgdaSymbol{(}\AgdaFunction{∀inList}%
\>[758I]\AgdaSymbol{(}\AgdaFunction{inputs2PrevOutputs}\AgdaSpace{}%
\AgdaBound{tr}\AgdaSpace{}%
\AgdaBound{inputs}\AgdaSymbol{)}\<%
\\
\>[9][@{}l@{\AgdaIndent{0}}]\<[758I]%
\>[18]\AgdaSymbol{(λ}\AgdaSpace{}%
\AgdaBound{o}\AgdaSpace{}%
\AgdaSymbol{→}\AgdaSpace{}%
\AgdaFunction{output2MaturationTime}\AgdaSpace{}%
\AgdaBound{o}\AgdaSpace{}%
\AgdaDatatype{≤}\AgdaSpace{}%
\AgdaFunction{txTree2TimeNextTobeMinedBlock}\AgdaSpace{}%
\AgdaBound{tr}\AgdaSymbol{))}\AgdaSpace{}%
\AgdaFunction{×}\<%
\\
\>[0][@{}l@{\AgdaIndent{0}}]%
\>[9]\AgdaSymbol{(}\AgdaFunction{∀inList}\AgdaSpace{}%
\AgdaSymbol{(}\AgdaFunction{inputs2PrevOutputsSigPbk}\AgdaSpace{}%
\AgdaBound{tr}\AgdaSpace{}%
\AgdaBound{inputs}\AgdaSymbol{)}\<%
\\
\>[9][@{}l@{\AgdaIndent{0}}]%
\>[12]\AgdaSymbol{(λ}%
\>[772I]\AgdaSymbol{\{(}\AgdaBound{outp}\AgdaSpace{}%
\AgdaInductiveConstructor{,}\AgdaSpace{}%
\AgdaBound{pbk}\AgdaSpace{}%
\AgdaInductiveConstructor{,}\AgdaSpace{}%
\AgdaBound{sign}\AgdaSymbol{)}\AgdaSpace{}%
\AgdaSymbol{→}\<%
\\
\>[772I][@{}l@{\AgdaIndent{0}}]%
\>[16]\AgdaPostulate{publicKey2Address}\AgdaSpace{}%
\AgdaBound{pbk}\AgdaSpace{}%
\AgdaDatatype{≡}\AgdaSpace{}%
\AgdaFunction{txOutput2Address}\AgdaSpace{}%
\AgdaBound{outp}\AgdaSpace{}%
\AgdaFunction{×}\<%
\\
\>[772I][@{}l@{\AgdaIndent{0}}]%
\>[16]\AgdaPostulate{Signed}\AgdaSpace{}%
\AgdaSymbol{(}\AgdaFunction{msgToBeSignedByInput}\AgdaSpace{}%
\AgdaBound{outp}\AgdaSpace{}%
\AgdaBound{outputs}\AgdaSymbol{)}\AgdaSpace{}%
\AgdaBound{pbk}\AgdaSpace{}%
\AgdaBound{sign}%
\>[69]\AgdaSymbol{\}))}\<%
\\
\\
\>[0]\AgdaFunction{CorrectTX}\AgdaSpace{}%
\AgdaBound{tr}\AgdaSpace{}%
\AgdaSymbol{(}\AgdaInductiveConstructor{coinbase}\AgdaSpace{}%
\AgdaBound{time}\AgdaSpace{}%
\AgdaBound{outputs}\AgdaSymbol{)}\AgdaSpace{}%
\AgdaSymbol{=}\<%
\\
\>[0][@{}l@{\AgdaIndent{0}}]%
\>[9]\AgdaFunction{NonNil}\AgdaSpace{}%
\AgdaBound{outputs}\AgdaSpace{}%
\AgdaFunction{×}\<%
\\
\>[0][@{}l@{\AgdaIndent{0}}]%
\>[9]\AgdaFunction{outputs2SumAmount}\AgdaSpace{}%
\AgdaBound{outputs}\AgdaSpace{}%
\AgdaDatatype{≡}\AgdaSpace{}%
\AgdaFunction{txTree2RecentTXFees}\AgdaSpace{}%
\AgdaBound{tr}\AgdaSpace{}%
\AgdaPrimitive{+}\AgdaSpace{}%
\AgdaPostulate{blockReward}\AgdaSpace{}%
\AgdaBound{time}\AgdaSpace{}%
\AgdaFunction{×}\<%
\\
\>[0][@{}l@{\AgdaIndent{0}}]%
\>[9]\AgdaBound{time}\AgdaSpace{}%
\AgdaDatatype{≡}\AgdaSpace{}%
\AgdaFunction{txTree2TimeNextTobeMinedBlock}\AgdaSpace{}%
\AgdaBound{tr}\<%
\end{code}
} 

\AgdaHide{
\begin{code}%
\>[0]\AgdaComment{{-}{-}         (∀inList (inputs2PrevOutputs tr inputs)}\<%
\\
\>[0]\AgdaComment{{-}{-}                  (λ o → output2MaturationTime o ≤ txTree2TimeNextTobeMinedBlock tr)) ×}\<%
\\
\>[0][@{}l@{\AgdaIndent{0}}]%
\>[9]\AgdaComment{{-}{-} expresses that inputs are spendable after maturation time}\<%
\\
\\
\>[0]\AgdaComment{{-}{-}         (∀inList (inputs2PrevOutputsSigPbk tr inputs)}\<%
\\
\>[0]\AgdaComment{{-}{-}            (λ \{(outp , pbk , sign) →}\<%
\\
\>[0]\AgdaComment{{-}{-}                publicKey2Address pbk ≡ txOutput2Address outp ×}\<%
\\
\>[0]\AgdaComment{{-}{-}                Signed (msgToBeSignedByInput outp}\<%
\\
\>[0]\AgdaComment{{-}{-}                                             outputs) pbk sign  \}))}\<%
\\
\\
\>[0][@{}l@{\AgdaIndent{0}}]%
\>[9]\AgdaComment{{-}{-} expresses that public keys in the input have address of the input}\<%
\\
\>[0][@{}l@{\AgdaIndent{0}}]%
\>[9]\AgdaComment{{-}{-}   and the input msg is signed by the private key of this public key}\<%
\\
\\
\\
\\
\\
\\
\>[0]\AgdaComment{{-}{-} \textbackslash{}bitcoinVersSix}\<%
\end{code}
} 

\newcommand{\bitcoinVersSixCorrectTxTree}{
\begin{code}%
\>[0]\AgdaFunction{CorrectTxTree}\AgdaSpace{}%
\AgdaSymbol{:}\AgdaSpace{}%
\AgdaSymbol{(}\AgdaBound{tr}\AgdaSpace{}%
\AgdaSymbol{:}\AgdaSpace{}%
\AgdaDatatype{TXTree}\AgdaSymbol{)}\AgdaSpace{}%
\AgdaSymbol{→}\AgdaSpace{}%
\AgdaPrimitiveType{Set}\<%
\\
\>[0]\AgdaFunction{CorrectTxTree}\AgdaSpace{}%
\AgdaInductiveConstructor{genesisTree}\AgdaSpace{}%
\AgdaSymbol{=}\AgdaSpace{}%
\AgdaRecord{⊤}\<%
\\
\>[0]\AgdaFunction{CorrectTxTree}\AgdaSpace{}%
\AgdaSymbol{(}\AgdaInductiveConstructor{txtree}\AgdaSpace{}%
\AgdaBound{tr}\AgdaSpace{}%
\AgdaBound{tx}\AgdaSymbol{)}\AgdaSpace{}%
\AgdaSymbol{=}\AgdaSpace{}%
\AgdaFunction{CorrectTxTree}\AgdaSpace{}%
\AgdaBound{tr}\AgdaSpace{}%
\AgdaFunction{×}\AgdaSpace{}%
\AgdaFunction{CorrectTX}\AgdaSpace{}%
\AgdaBound{tr}\AgdaSpace{}%
\AgdaBound{tx}\<%
\\
\\
\>[0]\AgdaKeyword{record}\AgdaSpace{}%
\AgdaRecord{TXTreeCorrect}\AgdaSpace{}%
\AgdaSymbol{:}\AgdaSpace{}%
\AgdaPrimitiveType{Set}\AgdaSpace{}%
\AgdaKeyword{where}\<%
\\
\>[0][@{}l@{\AgdaIndent{0}}]%
\>[2]\AgdaKeyword{constructor}\AgdaSpace{}%
\AgdaInductiveConstructor{txTreeCorrect}\<%
\\
\>[0][@{}l@{\AgdaIndent{0}}]%
\>[2]\agdaField%
\>[9]\AgdaField{txtr}%
\>[18]\AgdaSymbol{:}%
\>[21]\AgdaDatatype{TXTree}\<%
\\
\>[2][@{}l@{\AgdaIndent{0}}]%
\>[9]\AgdaField{corTxtr}%
\>[18]\AgdaSymbol{:}%
\>[21]\AgdaFunction{CorrectTxTree}\AgdaSpace{}%
\AgdaField{txtr}\<%
\\
\>[0]\<%
\end{code}
} 

\AgdaHide{
\begin{code}%
\>[0]\<%
\\
\>[0]\AgdaKeyword{open}\AgdaSpace{}%
\AgdaModule{TXTreeCorrect}\AgdaSpace{}%
\AgdaKeyword{public}\<%
\\
\\
\>[0]\AgdaComment{{-}{-} \textbackslash{}bitcoinVersSix}\<%
\end{code}
} 

\newcommand{\bitcoinVersSixTXCorrect}{
\begin{code}%
\>[0]\AgdaKeyword{record}\AgdaSpace{}%
\AgdaRecord{TXcorrect}\AgdaSpace{}%
\AgdaSymbol{(}\AgdaBound{tr}\AgdaSpace{}%
\AgdaSymbol{:}\AgdaSpace{}%
\AgdaRecord{TXTreeCorrect}\AgdaSymbol{)}\AgdaSpace{}%
\AgdaSymbol{:}\AgdaSpace{}%
\AgdaPrimitiveType{Set}\AgdaSpace{}%
\AgdaKeyword{where}\<%
\\
\>[0][@{}l@{\AgdaIndent{0}}]%
\>[2]\agdaField%
\>[10]\AgdaField{theTx}%
\>[17]\AgdaSymbol{:}%
\>[20]\AgdaDatatype{TX}\AgdaSpace{}%
\AgdaSymbol{(}\AgdaBound{tr}\AgdaSpace{}%
\AgdaSymbol{.}\AgdaField{txtr}\AgdaSymbol{)}\<%
\\
\>[2][@{}l@{\AgdaIndent{0}}]%
\>[10]\AgdaField{corTx}%
\>[17]\AgdaSymbol{:}%
\>[20]\AgdaFunction{CorrectTX}\AgdaSpace{}%
\AgdaSymbol{(}\AgdaBound{tr}\AgdaSpace{}%
\AgdaSymbol{.}\AgdaField{txtr}\AgdaSymbol{)}\AgdaSpace{}%
\AgdaField{theTx}\<%
\end{code}
} 

\AgdaHide{
\begin{code}%
\>[0]\<%
\\
\>[0]\AgdaKeyword{open}\AgdaSpace{}%
\AgdaModule{TXcorrect}\AgdaSpace{}%
\AgdaKeyword{public}\<%
\\
\\
\>[0]\AgdaComment{{-}{-} \textbackslash{}bitcoinVersSix}\<%
\end{code}
} 

\newcommand{\bitcoinVersSixaddTxtreeCorrect}{
\begin{code}%
\>[0]\AgdaFunction{initialTxTreeCorrect}\AgdaSpace{}%
\AgdaSymbol{:}\AgdaSpace{}%
\AgdaRecord{TXTreeCorrect}\<%
\\
\>[0]\AgdaFunction{initialTxTreeCorrect}\AgdaSpace{}%
\AgdaSymbol{=}\AgdaSpace{}%
\AgdaInductiveConstructor{txTreeCorrect}\AgdaSpace{}%
\AgdaInductiveConstructor{genesisTree}\AgdaSpace{}%
\AgdaInductiveConstructor{tt}\<%
\\
\\
\>[0]\AgdaFunction{addTxtreeCorrect}\AgdaSpace{}%
\AgdaSymbol{:}\AgdaSpace{}%
\AgdaSymbol{(}\AgdaBound{tr}\AgdaSpace{}%
\AgdaSymbol{:}\AgdaSpace{}%
\AgdaRecord{TXTreeCorrect}\AgdaSymbol{)(}\AgdaBound{tx}\AgdaSpace{}%
\AgdaSymbol{:}\AgdaSpace{}%
\AgdaRecord{TXcorrect}\AgdaSpace{}%
\AgdaBound{tr}\AgdaSymbol{)}\AgdaSpace{}%
\AgdaSymbol{→}\AgdaSpace{}%
\AgdaRecord{TXTreeCorrect}\<%
\\
\>[0]\AgdaFunction{addTxtreeCorrect}\AgdaSpace{}%
\AgdaBound{tr}\AgdaSpace{}%
\AgdaBound{tx}\AgdaSpace{}%
\AgdaSymbol{=}\<%
\\
\>[0][@{}l@{\AgdaIndent{0}}]%
\>[4]\AgdaInductiveConstructor{txTreeCorrect}\AgdaSpace{}%
\AgdaSymbol{(}\AgdaInductiveConstructor{txtree}\AgdaSpace{}%
\AgdaSymbol{(}\AgdaBound{tr}\AgdaSpace{}%
\AgdaSymbol{.}\AgdaField{txtr}\AgdaSymbol{)}\AgdaSpace{}%
\AgdaSymbol{(}\AgdaBound{tx}\AgdaSpace{}%
\AgdaSymbol{.}\AgdaField{theTx}\AgdaSymbol{))}\AgdaSpace{}%
\AgdaSymbol{(}\AgdaBound{tr}\AgdaSpace{}%
\AgdaSymbol{.}\AgdaField{corTxtr}\AgdaSpace{}%
\AgdaInductiveConstructor{,}\AgdaSpace{}%
\AgdaBound{tx}\AgdaSpace{}%
\AgdaSymbol{.}\AgdaField{corTx}\AgdaSymbol{)}\<%
\end{code}
} 

\AgdaHide{
\begin{code}%
\>[0]\<%
\\
\>[0]\AgdaComment{{-}{-} \#\#\#\#\#\#\#\#\#\#\#\#\#\#\#\#\#\#\#\#\#\#\#\#\#\#\#\#\#\#\#\#\#\#\#\#\#\#\#\#\#\#\#\#\#\#\#\#\#\#\#\#\#\#\#\#\#\#\#\#\#\#}\<%
\\
\>[0]\AgdaComment{{-}{-} \#  Merkle Trees                                              \#}\<%
\\
\>[0]\AgdaComment{{-}{-} \#\#\#\#\#\#\#\#\#\#\#\#\#\#\#\#\#\#\#\#\#\#\#\#\#\#\#\#\#\#\#\#\#\#\#\#\#\#\#\#\#\#\#\#\#\#\#\#\#\#\#\#\#\#\#\#\#\#\#\#\#\#}\<%
\\
\\
\\
\>[0]\AgdaComment{{-}{-} \textbackslash{}bitcoinVersSix}\<%
\end{code}
} 

\newcommand{\bitcoinVersSixMerkleInputField}{
\begin{code}%
\>[0]\AgdaKeyword{record}\AgdaSpace{}%
\AgdaRecord{MerkleInputField}\AgdaSpace{}%
\AgdaSymbol{:}\AgdaSpace{}%
\AgdaPrimitiveType{Set}\AgdaSpace{}%
\AgdaKeyword{where}\<%
\\
\>[0][@{}l@{\AgdaIndent{0}}]%
\>[2]\AgdaKeyword{constructor}\AgdaSpace{}%
\AgdaInductiveConstructor{inputField}\<%
\\
\>[0][@{}l@{\AgdaIndent{0}}]%
\>[2]\agdaField%
\>[9]\AgdaField{txid}%
\>[23]\AgdaSymbol{:}\AgdaSpace{}%
\AgdaFunction{TXID}\<%
\\
\>[2][@{}l@{\AgdaIndent{0}}]%
\>[9]\AgdaField{outputNr}%
\>[23]\AgdaSymbol{:}\AgdaSpace{}%
\AgdaDatatype{ℕ}\<%
\\
\>[2][@{}l@{\AgdaIndent{0}}]%
\>[9]\AgdaField{publicKey}%
\>[23]\AgdaSymbol{:}\AgdaSpace{}%
\AgdaFunction{PublicKey}\<%
\\
\>[2][@{}l@{\AgdaIndent{0}}]%
\>[9]\AgdaField{signature}%
\>[23]\AgdaSymbol{:}\AgdaSpace{}%
\AgdaFunction{Signature}\<%
\end{code}
} 

\AgdaHide{
\begin{code}%
\>[0]\<%
\\
\>[0]\AgdaComment{{-}{-} \textbackslash{}bitcoinVersSix}\<%
\end{code}
} 

\newcommand{\bitcoinVersSixMerkleTX}{
\begin{code}%
\>[0]\AgdaKeyword{record}\AgdaSpace{}%
\AgdaRecord{MerkleTXcoinbase}\AgdaSpace{}%
\AgdaSymbol{:}\AgdaSpace{}%
\AgdaPrimitiveType{Set}\AgdaSpace{}%
\AgdaKeyword{where}\<%
\\
\>[0][@{}l@{\AgdaIndent{0}}]%
\>[2]\AgdaKeyword{constructor}\AgdaSpace{}%
\AgdaInductiveConstructor{merkleCoinbase}\<%
\\
\>[0][@{}l@{\AgdaIndent{0}}]%
\>[2]\agdaField%
\>[9]\AgdaField{time}%
\>[18]\AgdaSymbol{:}%
\>[21]\AgdaFunction{Time}\<%
\\
\>[2][@{}l@{\AgdaIndent{0}}]%
\>[9]\AgdaField{outputs}%
\>[18]\AgdaSymbol{:}%
\>[21]\AgdaDatatype{List}\AgdaSpace{}%
\AgdaRecord{TXOutputfield}\<%
\\
\\
\>[0]\AgdaKeyword{record}\AgdaSpace{}%
\AgdaRecord{MerkleTXstandard}\AgdaSpace{}%
\AgdaSymbol{:}\AgdaSpace{}%
\AgdaPrimitiveType{Set}\AgdaSpace{}%
\AgdaKeyword{where}\<%
\\
\>[0][@{}l@{\AgdaIndent{0}}]%
\>[2]\AgdaKeyword{constructor}\AgdaSpace{}%
\AgdaInductiveConstructor{merkleTXstd}\<%
\\
\>[0][@{}l@{\AgdaIndent{0}}]%
\>[2]\agdaField%
\>[9]\AgdaField{inputs}%
\>[18]\AgdaSymbol{:}%
\>[21]\AgdaDatatype{List}\AgdaSpace{}%
\AgdaRecord{MerkleInputField}\<%
\\
\>[2][@{}l@{\AgdaIndent{0}}]%
\>[9]\AgdaField{outputs}%
\>[18]\AgdaSymbol{:}%
\>[21]\AgdaDatatype{List}\AgdaSpace{}%
\AgdaRecord{TXOutputfield}\<%
\\
\\
\>[0]\AgdaKeyword{data}\AgdaSpace{}%
\AgdaDatatype{MerkleTX}\AgdaSpace{}%
\AgdaSymbol{:}\AgdaSpace{}%
\AgdaPrimitiveType{Set}\AgdaSpace{}%
\AgdaKeyword{where}\<%
\\
\>[0][@{}l@{\AgdaIndent{0}}]%
\>[2]\AgdaInductiveConstructor{txcoinbase}%
\>[14]\AgdaSymbol{:}%
\>[17]\AgdaRecord{MerkleTXcoinbase}%
\>[35]\AgdaSymbol{→}%
\>[38]\AgdaDatatype{MerkleTX}\<%
\\
\>[0][@{}l@{\AgdaIndent{0}}]%
\>[2]\AgdaInductiveConstructor{txstd}%
\>[14]\AgdaSymbol{:}%
\>[17]\AgdaRecord{MerkleTXstandard}%
\>[35]\AgdaSymbol{→}%
\>[38]\AgdaDatatype{MerkleTX}\<%
\end{code}
} 

\AgdaHide{
\begin{code}%
\>[0]\<%
\\
\\
\\
\\
\\
\>[0]\AgdaKeyword{open}\AgdaSpace{}%
\AgdaModule{MerkleInputField}\AgdaSpace{}%
\AgdaKeyword{public}\<%
\\
\>[0]\AgdaKeyword{open}\AgdaSpace{}%
\AgdaModule{MerkleTXstandard}\AgdaSpace{}%
\AgdaKeyword{public}\<%
\\
\>[0]\AgdaKeyword{open}\AgdaSpace{}%
\AgdaModule{MerkleTXcoinbase}\AgdaSpace{}%
\AgdaKeyword{public}\<%
\\
\\
\>[0]\AgdaComment{{-}{-} \textbackslash{}bitcoinVersSix}\<%
\end{code}
} 

\newcommand{\bitcoinVersSixtxOutputsSignedtoMerkleInput}{
\begin{code}%
\>[0]\AgdaFunction{txOutpSign2MerkleInput}\AgdaSpace{}%
\AgdaSymbol{:}\AgdaSpace{}%
\AgdaRecord{TXOutput}\AgdaSpace{}%
\AgdaFunction{×}\AgdaSpace{}%
\AgdaFunction{PublicKey}\AgdaSpace{}%
\AgdaFunction{×}\AgdaSpace{}%
\AgdaFunction{Signature}\AgdaSpace{}%
\AgdaSymbol{→}\AgdaSpace{}%
\AgdaRecord{MerkleInputField}\<%
\\
\>[0]\AgdaFunction{txOutpSign2MerkleInput}\AgdaSpace{}%
\AgdaSymbol{(}\AgdaInductiveConstructor{txOutput}\AgdaSpace{}%
\AgdaBound{tr}\AgdaSpace{}%
\AgdaBound{tx₁}\AgdaSpace{}%
\AgdaBound{i}\AgdaSpace{}%
\AgdaInductiveConstructor{,}\AgdaSpace{}%
\AgdaBound{pbk}\AgdaSpace{}%
\AgdaInductiveConstructor{,}\AgdaSpace{}%
\AgdaBound{sign}\AgdaSymbol{)}\AgdaSpace{}%
\AgdaSymbol{=}\AgdaSpace{}%
\AgdaInductiveConstructor{inputField}\AgdaSpace{}%
\AgdaSymbol{(}\AgdaFunction{tx2id}\AgdaSpace{}%
\AgdaBound{tr}\AgdaSpace{}%
\AgdaBound{tx₁}\AgdaSymbol{)}\AgdaSpace{}%
\AgdaSymbol{(}\AgdaFunction{toℕ}\AgdaSpace{}%
\AgdaBound{i}\AgdaSymbol{)}\AgdaSpace{}%
\AgdaBound{pbk}\AgdaSpace{}%
\AgdaBound{sign}\<%
\\
\\
\>[0]\AgdaFunction{txInputs2MerkleInputs}\AgdaSpace{}%
\AgdaSymbol{:}\AgdaSpace{}%
\AgdaSymbol{(}\AgdaBound{tr}\AgdaSpace{}%
\AgdaSymbol{:}\AgdaSpace{}%
\AgdaDatatype{TXTree}\AgdaSymbol{)(}\AgdaBound{txInputs}\AgdaSpace{}%
\AgdaSymbol{:}\AgdaSpace{}%
\AgdaFunction{TxInputs}\AgdaSpace{}%
\AgdaBound{tr}\AgdaSymbol{)}\AgdaSpace{}%
\AgdaSymbol{→}\AgdaSpace{}%
\AgdaDatatype{List}\AgdaSpace{}%
\AgdaRecord{MerkleInputField}\<%
\\
\>[0]\AgdaFunction{txInputs2MerkleInputs}\AgdaSpace{}%
\AgdaBound{tr}\AgdaSpace{}%
\AgdaBound{txin}\AgdaSpace{}%
\AgdaSymbol{=}%
\>[33]\AgdaFunction{mapL}\AgdaSpace{}%
\AgdaFunction{txOutpSign2MerkleInput}\AgdaSpace{}%
\AgdaSymbol{(}\AgdaFunction{inputs2PrevOutputsSigPbk}\AgdaSpace{}%
\AgdaBound{tr}\AgdaSpace{}%
\AgdaBound{txin}\AgdaSymbol{)}\<%
\end{code}
} 

\AgdaHide{
\begin{code}%
\>[0]\<%
\\
\>[0]\AgdaComment{{-}{-} \textbackslash{}bitcoinVersSix}\<%
\end{code}
} 

\newcommand{\bitcoinVersSixMerkleTXCorrespondstoTX}{
\begin{code}%
\>[0]\AgdaFunction{MerkleTXCorresponds2TX}\AgdaSpace{}%
\AgdaSymbol{:}\AgdaSpace{}%
\AgdaSymbol{(}\AgdaBound{tx}\AgdaSpace{}%
\AgdaSymbol{:}\AgdaSpace{}%
\AgdaDatatype{MerkleTX}\AgdaSymbol{)(}\AgdaBound{tr}\AgdaSpace{}%
\AgdaSymbol{:}\AgdaSpace{}%
\AgdaDatatype{TXTree}\AgdaSymbol{)(}\AgdaBound{tx}\AgdaSpace{}%
\AgdaSymbol{:}\AgdaSpace{}%
\AgdaDatatype{TX}\AgdaSpace{}%
\AgdaBound{tr}\AgdaSymbol{)}\AgdaSpace{}%
\AgdaSymbol{→}\AgdaSpace{}%
\AgdaPrimitiveType{Set}\<%
\\
\>[0]\AgdaFunction{MerkleTXCorresponds2TX}\AgdaSpace{}%
\AgdaSymbol{(}\AgdaInductiveConstructor{txcoinbase}\AgdaSpace{}%
\AgdaSymbol{(}\AgdaInductiveConstructor{merkleCoinbase}\AgdaSpace{}%
\AgdaBound{time₁}\AgdaSpace{}%
\AgdaBound{outputs₁}\AgdaSymbol{))}\AgdaSpace{}%
\AgdaBound{tr}\AgdaSpace{}%
\AgdaSymbol{(}\AgdaInductiveConstructor{coinbase}\AgdaSpace{}%
\AgdaBound{time₂}\AgdaSpace{}%
\AgdaBound{outputs₂}\AgdaSymbol{)}\<%
\\
\>[0][@{}l@{\AgdaIndent{0}}]%
\>[7]\AgdaSymbol{=}\AgdaSpace{}%
\AgdaBound{time₁}\AgdaSpace{}%
\AgdaDatatype{≡}\AgdaSpace{}%
\AgdaBound{time₂}\AgdaSpace{}%
\AgdaFunction{×}\AgdaSpace{}%
\AgdaBound{outputs₁}\AgdaSpace{}%
\AgdaDatatype{≡}\AgdaSpace{}%
\AgdaBound{outputs₂}\<%
\\
\>[0]\AgdaFunction{MerkleTXCorresponds2TX}\AgdaSpace{}%
\AgdaSymbol{(}\AgdaInductiveConstructor{txstd}\AgdaSpace{}%
\AgdaSymbol{(}\AgdaInductiveConstructor{merkleTXstd}\AgdaSpace{}%
\AgdaBound{inputs₁}\AgdaSpace{}%
\AgdaBound{outputs₁}\AgdaSymbol{))}\AgdaSpace{}%
\AgdaBound{tr}\AgdaSpace{}%
\AgdaSymbol{(}\AgdaInductiveConstructor{normalTX}\AgdaSpace{}%
\AgdaBound{inputs₂}\AgdaSpace{}%
\AgdaBound{outputs₂}\AgdaSymbol{)}\<%
\\
\>[0][@{}l@{\AgdaIndent{0}}]%
\>[7]\AgdaSymbol{=}\AgdaSpace{}%
\AgdaBound{inputs₁}\AgdaSpace{}%
\AgdaDatatype{≡}\AgdaSpace{}%
\AgdaFunction{txInputs2MerkleInputs}\AgdaSpace{}%
\AgdaBound{tr}\AgdaSpace{}%
\AgdaBound{inputs₂}\AgdaSpace{}%
\AgdaFunction{×}\AgdaSpace{}%
\AgdaBound{outputs₁}\AgdaSpace{}%
\AgdaDatatype{≡}\AgdaSpace{}%
\AgdaBound{outputs₂}\<%
\\
\>[0]\AgdaFunction{MerkleTXCorresponds2TX}\AgdaSpace{}%
\AgdaSymbol{(}\AgdaInductiveConstructor{txcoinbase}\AgdaSpace{}%
\AgdaBound{x}\AgdaSymbol{)}\AgdaSpace{}%
\AgdaBound{tr}\AgdaSpace{}%
\AgdaSymbol{(}\AgdaInductiveConstructor{normalTX}\AgdaSpace{}%
\AgdaBound{inputs₁}\AgdaSpace{}%
\AgdaBound{outputs₁}\AgdaSymbol{)}%
\>[70]\AgdaSymbol{=}%
\>[73]\AgdaDatatype{⊥}\<%
\\
\>[0]\AgdaFunction{MerkleTXCorresponds2TX}\AgdaSpace{}%
\AgdaSymbol{(}\AgdaInductiveConstructor{txstd}\AgdaSpace{}%
\AgdaBound{x}\AgdaSymbol{)}\AgdaSpace{}%
\AgdaBound{tr}\AgdaSpace{}%
\AgdaSymbol{(}\AgdaInductiveConstructor{coinbase}\AgdaSpace{}%
\AgdaBound{time₁}\AgdaSpace{}%
\AgdaBound{outputs₁}\AgdaSymbol{)}%
\>[70]\AgdaSymbol{=}%
\>[73]\AgdaDatatype{⊥}\<%
\end{code}
} 

\AgdaHide{
\begin{code}%
\>[0]\<%
\\
\>[0]\AgdaComment{{-}{-} \textbackslash{}bitcoinVersSix}\<%
\end{code}
} 

\newcommand{\bitcoinVersSixMerkleTXCor}{
\begin{code}%
\>[0]\AgdaKeyword{record}\AgdaSpace{}%
\AgdaRecord{MerkleTXCor}\AgdaSpace{}%
\AgdaSymbol{(}\AgdaBound{tr}\AgdaSpace{}%
\AgdaSymbol{:}\AgdaSpace{}%
\AgdaRecord{TXTreeCorrect}\AgdaSymbol{)}\AgdaSpace{}%
\AgdaSymbol{:}\AgdaSpace{}%
\AgdaPrimitiveType{Set}\AgdaSpace{}%
\AgdaKeyword{where}\<%
\\
\>[0][@{}l@{\AgdaIndent{0}}]%
\>[2]\agdaField%
\>[9]\AgdaField{mk}%
\>[14]\AgdaSymbol{:}\AgdaSpace{}%
\AgdaDatatype{MerkleTX}\<%
\\
\>[2][@{}l@{\AgdaIndent{0}}]%
\>[9]\AgdaField{mtx}%
\>[14]\AgdaSymbol{:}\AgdaSpace{}%
\AgdaRecord{TXcorrect}\AgdaSpace{}%
\AgdaBound{tr}\<%
\\
\>[2][@{}l@{\AgdaIndent{0}}]%
\>[9]\AgdaField{cor}%
\>[14]\AgdaSymbol{:}\AgdaSpace{}%
\AgdaFunction{MerkleTXCorresponds2TX}\AgdaSpace{}%
\AgdaField{mk}\AgdaSpace{}%
\AgdaSymbol{(}\AgdaBound{tr}\AgdaSpace{}%
\AgdaSymbol{.}\AgdaField{txtr}\AgdaSymbol{)}\AgdaSpace{}%
\AgdaSymbol{(}\AgdaField{mtx}\AgdaSpace{}%
\AgdaSymbol{.}\AgdaField{theTx}\AgdaSymbol{)}\<%
\end{code}
} 

\AgdaHide{
\begin{code}%
\>[0]\<%
\\
\>[0]\AgdaKeyword{open}\AgdaSpace{}%
\AgdaModule{MerkleTXCor}\AgdaSpace{}%
\AgdaKeyword{public}\<%
\\
\\
\>[0]\AgdaComment{{-}{-} \textbackslash{}bitcoinVersSix}\<%
\end{code}
} 

\newcommand{\bitcoinVersSixupdateTXTree}{
\begin{code}%
\>[0]\AgdaFunction{updateTXTree}\AgdaSpace{}%
\AgdaSymbol{:}\AgdaSpace{}%
\AgdaSymbol{(}\AgdaBound{tr}\AgdaSpace{}%
\AgdaSymbol{:}\AgdaSpace{}%
\AgdaRecord{TXTreeCorrect}\AgdaSymbol{)(}\AgdaBound{m}\AgdaSpace{}%
\AgdaSymbol{:}\AgdaSpace{}%
\AgdaRecord{MerkleTXCor}\AgdaSpace{}%
\AgdaBound{tr}\AgdaSymbol{)}\AgdaSpace{}%
\AgdaSymbol{→}\AgdaSpace{}%
\AgdaRecord{TXTreeCorrect}\<%
\\
\>[0]\AgdaFunction{updateTXTree}\AgdaSpace{}%
\AgdaBound{tr}\AgdaSpace{}%
\AgdaBound{m}\AgdaSpace{}%
\AgdaSymbol{=}\AgdaSpace{}%
\AgdaFunction{addTxtreeCorrect}\AgdaSpace{}%
\AgdaBound{tr}\AgdaSpace{}%
\AgdaSymbol{(}\AgdaBound{m}\AgdaSpace{}%
\AgdaSymbol{.}\AgdaField{mtx}\AgdaSymbol{)}\<%
\end{code}
} 

\AgdaHide{
\begin{code}%
\>[0]\<%
\end{code}
} 

\section{Introduction}
\label{sectIntroduction}
Since its introduction in November 2008, the market capitalisation
of bitcoins has risen to over 161 Billion US-\$
(as of 27 November 17) \cite{coinmarketcap}. 
In 2017 the price for one bitcoin has risen from 
1001 US\$ on January 1 to 9665 US\$  on 27 Nov 2017
\cite{coinmarketcap}. Other cryptocurrencies
have been introduced as well, the main contender being
currently Ethereum with a market capitalisation of almost
46 Billion US-\$. Cryptocurrencies have become a major financial
instrument and might even become major currencies in the near future.

Cryptocurrencies have as well been used  for the use in 
smart contracts \cite{szabo:SmartContracts:1996,Wiki:2017:SmartContract}. The simplest form of a smart contract is where 
the buyer reserves money for the seller on the blockchain.
In order to unlock the money, the seller needs to receive
a second signature from the buyer, which is given once the 
goods are received. If the goods don't arrive on time, the money is returned
to the buyer. Smart contracts are now used in large funds, where
all contractual relationships are entirely governed by algorithms with
no legal framework being used.


Since cryptocurrencies are entirely governed by algorithms, one 
needs to have a very high guarantee that these algorithms are correct --
in case of failure there is no legal framework available to remedy
the problem. The only way to fix problems is to find a consensus amongst
users and amongst miners\footnote{In case of soft forks there are 
user activated (UASF) vs miner activated (MASF) soft forks.}
to change the protocol  and create a soft (i.e.~backward compatible)
or hard fork.
A possible mistake in the algorithms is one of the most important
risks in cryptocurrencies. Some mistakes and weaknesses have already been 
found, of which some have been fixed. In the original Bitcoin protocol,
uniqueness of transaction IDs was not guaranteed,
which would allow replay attacks.  This could be fixed
by including the block number in coinbase transaction by a soft protocol
change
(see the discussion in Subsect.~\ref{subsect:CorrectnessCorrectness}).
The size of the blocks of Bitcoin is no longer enough to cover
all transactions, a problem which is known, but which couldn't be
solved at time of writing this article: 
Bitcoin cash \cite{Wiki:2017:Bitcoincash} would have been a solution,
but became an alternative currency instead of being a corrected
form of Bitcoin; the SegWit soft fork \cite{Wiki:2017:SegWit} was successful, 
but not enough to fix the problem;
the second stage of SegWit, SegWit2x \cite{Wiki:2017:SegWit2x} 
was cancelled on 8 Nov 2017 \cite{belshe:CancellationOfseegwit2x}.

Therefore, it is important to fully verify cryptocurrency protocols.
A very high level of verification can be achieved by creating 
formal  models for cryptocurrency protocols,
and proving their correctness.
We haven't found a fully worked out complete formal model and 
correctness proof of cryptocurrency protocols, and the goal of this article 
is to take first steps in order to fill this gap.


The rise of smart contracts has led to new problems concerning the
security of financial investments. The up to now biggest
incident was the failure of the DAO \cite{Wiki:2017:DAO}, a form of
investor-directed venture capital fund based on smart contracts in the
cryptocurrency Ethereum. Malicious users exploited a vulnerability
in the DAO, at a time when the market value of the DAO had reached 
150 Million US-\$. The loss of the investor's money  was only avoided
 by making a hard 
fork, which deleted most transactions investing in the DAO. This hard
fork violated the principle that control over
the money owned by users in cryptocurrencies should only be 
governed by algorithms, without any human intervention. 
Therefore, there is need for
smart contracts, which are given together with a correctness argument, 
ideally a formal proof, which serves as a certificate. 
Such proofs need to rely on a model of the underlying 
cryptocurrency.

In this article we will develop, as a first step towards verifying the protocols of 
cryptocurrencies
and developing verified smart contracts, 
a model of Bitcoin in the 
interactive
theorem prover and dependently typed programming language Agda \cite{agdaWiki:home,agda:Documentation,stump2016verified}. We will formalise how notions of 
transaction trees, transactions, blocks, unspent transactions, and how
messages are signed. We will as well include aspects such as coinbase
transactions, giving the miner their rewards, and maturation time for
coinbase transactions. We will not include all aspects yet,
especially input and output scripts and the proof of work
will not be part of the model at this stage,
although we assume that including them would be relatively straightforward. 
We will indicate
how  correctness of Bitcoin can be specified, although
we will leave it to a future paper to actually carry out the precise
specification and proof of correctness. This seems at the moment more a matter
of using enough man power to carry out those proofs in an interactive
theorem prover, rather than solving fundamental problems.

What we will see is that the blockchain is from a theoretical point of view
a more complicated data structure than expected. Our definition will
simultaneously define the set of transaction trees and transactions
together with the unspent transactions, which will be an extended form
of an inductive-recursive definition 
\cite{dybjerjslinductionrecursion,dybjersetzer:2003:indrekjour,dybjerSetzer:indexedInductionRecursion:lncsversion}. This could be one reason
(apart from the fact that Bitcoin is rather new
-- the announcement and mining of the first block happened on 1 November 2008
\cite{nakamoto:AnnouncementBitcoins,nakamoto2008bitcoin}), why descriptions of Bitcoin are often very
difficult to understand. Many details are well described, but descriptions
of transactions which take into account that due to the reference
to unspent transaction outputs one obtains a tree like structure,
often lack clarity. 
We hope that the current paper will help to understand this data structure
better. We hope as well that it will help to popularise the concept
of induction-recursion, which we believe is a concept, which is not 
restricted to dependent type theory and Agda, but can be of great use in a 
more general 
mathematical and computer science setting.\par

\myparagraph{Content of this Article.} We will first give in 
the next Sect.~\ref{sectAgdaIntro}
a brief introduction into Agda. Then we show
in Sect.~\ref{sectWarnerLedger}, mainly following
\cite{warner:BitcoinTechnicalIntroduction:July2011}, how to develop,
starting from a simple model of a bank, a ledger based model of
Bitcoin.
A model in Agda based on this is presented in Sect.~\ref{sectLedgerAgda}.
This definition is much simpler than the final model, since we can define
ledgers and transaction separately, without the need of induction-recursion.
In Sect.~\ref{sectTransactionTreesTheoretical} we discuss the security problems
of this model, and introduce in theoretical form the notions of transaction
trees, transactions, and unspent transaction outputs. 
In Sect.~\ref{sectTreeLedger} we develop a model in Agda  which makes
use of transaction trees, trees, and unspent transaction outputs.
This model will be make use of inductive-recursion.
In Sect.~\ref{sectCorrectness} we discuss how one could specify and prove
the correctness of this model, and as well limitations of our approach
(not using input and output scripts, and the need to add the miners' proof
of work to the model). We finish by a short conclusion
including future work (Sect.~\ref{sectConclusion}).

\myparagraph{Related Work.} 
We haven't  found any research on developing the transaction tree as
a formal structure. A lot of research has been carried out
on modelling the Bitcoin and related protocols by modelling 
Bitcoin as interactions between agents.
Beukema \cite{beukema:FormalisingBitcoinProtocol} 
introduces a model of Bitcoin based on transactions referring to previous 
ones. It is formalised in the language mCRL2 as a transition
system between agents. An analysis is carried out using model checking
regarding the behaviour of the network, and how the consensus protocol
deals with corrupted messages and double spending attacks.
Andrychowicz et.~al.~\cite{Andrychowicz:ModellingBitcoinContracts}
model Bitcoin as timed automata and verify using UPPAAL correctness properties.
The automata formalise how agents interact regarding the original Bitcoin
protocol.
Chaudhary, et.~al.~
\cite{chaudhary:ModelingVerificationBitcoinProtocol} model the interactions
in Bitcoin as a state transition system between agents.
They use the model checker UPPAAL, and analyse the probability that a malicious
transaction is included in the longest chain, which
could give rise to double spending.
Bastiaan \cite{bastiaan:Preventing51percentattack} provides
a stochastic analysis of the Bitcoin mining protocol modelled
using continuous-time Markov chains. The focus of that paper is
on analysing mining and how to prevent the formation of 
large pools of miners.\par 
A lot of research has been carried out on verifying smart contracts.
Kosba et.~al.~\cite{kosbaEtal:HawkblockchainModelcrypography,kosbaEtal:HawkblockchainModelcrypography:ExtendedOnlineVersion} present a programming
language Hawk for writing smart contracts. They verify privacy properties using
formal methods. Their underlying model is a ledger model.
Luu et.~al.~\cite{Luu:2016:MSC:2976749.2978309} analyse vulnerabilities
in smart contracts. Their focus is directly on the language for smart
contracts. Bhargavan et.~al.~%
\cite{Bhargavan:VerificationSmartContractShort} 
translate  smart contracts in Ethereum into F$^\ast$ and analyse them.

\myparagraph{Acknowledgement.} We would like to thank Andrew Lawrence for 
inspiring us to start the project of formalising Bitcoin.
His idea led to a series of third year projects and MSc projects
under the
author's supervision.. 
The first one, in which the main basis 
was formed, was  carried out by 
Damon Jones \cite{damonjones:thirdyearproject:bitcoins},
which was followed\footnote{I hope this list is complete.}
by third year and MSc projects by
%
M. Zahid, 
Ifetayo Agunbiade,
Ehsan Alebrahim-Dehkordi,
Kieran Cullinan,
Dylan Maclennan,
Robert Locklan, and some projects,  which are in their initial phase.
%
%
%
These models were based on the brown bag talk by
Warner \cite{warner:BitcoinTechnicalIntroduction:July2011}. 
The first Agda model in this article (Sect.~\ref{sectLedgerAgda}) will be based on the models developed in those projects, with some enhancements such as 
adding the block reward, and the precise signatures for transactions. We want
to thank those students for invaluable
insights they added  to this project.

\myparagraph{Agda Code}.
Every line of Agda code in display format
provided in this paper has been type-checked by
Agda and rendered by the Agda \LaTeX-backend, and is
therefore type safe.  However, we mostly
omit administrative parts of the code such as modules and name space handling;
thus, the code as printed in this article will not be accepted by Agda
as-is.  The complete type checked code can be found in \cite{bitcoinAgda:Libraray}.  

\myparagraph{Disclaimer.}  This is a theoretical model, developed for scientific
purposes only, which models some aspects of blockchain technology and 
Bitcoin.
It is not intended as a tool to make business decisions. Business
decisions should be based on other sources. The legal disclaimer from 
bitcoin.org (\url{https://bitcoin.org/en/legal})
will apply to this article, with ``the website'' replaced by 
``the current article (Anton Setzer: Modelling Bitcoin in Agda)''.

\section{Introduction into Agda}
\label{sectAgdaIntro}

Agda~\cite{agdaWiki:home,agda:Documentation,stump2016verified} is a theorem prover based
on intensional Martin-L\"{o}f type theory~\cite{martin1984intuitionistic}.
It is closely related to the theorem prover Coq~\cite{coqWebpage}.
Furthermore, Agda is a total language, which is guaranteed by its
termination and coverage checker -- without it Agda would be
inconsistent.  The current version of Agda is
Agda 2, which was initially designed and implemented by Ulf Norell in
his PhD thesis~\cite{norell:PhD}, and has since been developed further
by the Agda community.

In Agda, there are infinitely many levels of types: the lowest one is
called for historic reasons $\AgdaPrimitiveType{Set}$
-- $\AgdaPrimitiveType{Set}$ stands for what is in other
languages called ``type''.
The main type constructions in Agda are dependent function types, inductive types, coinductive types (which don't occur in this article), and record types. 

Inductive data types are dependent versions of algebraic data types as
they occur in functional programming. They are given as sets
$\AgdaBound{A}$ together with constructors which are
strictly positive in $\AgdaBound{A}$. For instance, the even and odd
numbers are given by the simultaneous --- as denoted by the keyword
\AgdaKeyword{mutual} --- inductive data types:

\begin{minipage}{1.0\linewidth}
\lagdaFinnEvenOdd
\end{minipage} 

The expression $(n : \AgdaDatatype{ℕ}) \;\AgdaSymbol{→}\; A$ denotes a dependent
function type, which is similar to a function type, but  $A$ can
depend on the argument $n$.
The expression
$\AgdaSymbol{\{}\AgdaBound{n}\; \AgdaSymbol{:}
\;\AgdaDatatype{ℕ}\AgdaSymbol{\}}\;\AgdaSymbol{→}\; A$ is an implicit
version of the previous construct. Implicit
arguments can be omitted, provided they can be inferred by the type
checker. We can make an implicit argument explicit by writing, \eg,
$\AgdaInductiveConstructor{sucp}\; \{0\} \; \AgdaInductiveConstructor{0p}$.
We can as well write
$\AgdaInductiveConstructor{sucp}\; \{n=0\} \; \AgdaInductiveConstructor{0p}$,
which is useful when having several implicit arguments, of which
only some are given explicitly.
If there are several explicit or implicit dependent arguments in a type,
one can omit ``$\AgdaSymbol{→}$'', as illustrated in the following
example: $(a : A)(b: B) \;\AgdaSymbol{→}\; C$
instead of
$(a : A)\;\AgdaSymbol{→}\;(b: B) \;\AgdaSymbol{→}\; C$.
The elements of $(\AgdaDatatype{Even}\;n)$ and
$(\AgdaDatatype{Odd}\;n)$ are those that result from applying the
respective constructors. Therefore, we can define functions by case
distinction on these constructors using pattern matching, \eg

\begin{minipage}{1.0\linewidth}
\lagdaExampleFinFunPlus
\end{minipage} 

Here, $\AgdaSymbol{∀} a \;\AgdaSymbol{→}\; B$ is an abbreviation for
$(a : A) \;\AgdaSymbol{→}\;B$, where $A$ can be inferred by Agda.
$\AgdaSymbol{∀} \{a\} \;\AgdaSymbol{→}\; B$ is the same but for an
implicit argument, while $\AgdaSymbol{∀} \AgdaSymbol{\{}\AgdaBound{n}\
\AgdaBound{m}\AgdaSymbol{\}} \;\AgdaSymbol{→}\; B$ abbreviates
$\AgdaSymbol{∀} \AgdaSymbol{\{}\AgdaBound{n} \AgdaSymbol{\}} \;\AgdaSymbol{→}\;
\AgdaSymbol{∀} \AgdaSymbol{\{}\AgdaBound{m}\AgdaSymbol{\}} \;\AgdaSymbol{→}\; B$.
Agda supports mixfix operators, where ``$\_$'' denotes the position of the arguments.
For instance, $(\AgdaInductiveConstructor{0p} \;\AgdaFunction{+e}\; \AgdaBound{p})$
stands for
$(\AgdaFunction{\_+e\_}\;\AgdaInductiveConstructor{0p}\; \AgdaBound{p})$.
The combination of mixfix symbols together with
the availability of Unicode symbols
makes it possible to define Agda code which is very close to standard mathematical notation.

Nested patterns are allowed in pattern matching. The coverage checker verifies
completeness and the termination checker ensures that the recursive calls follow a schema
of extended primitive recursion.
In Agda, inductive types have been 
substantially extended. Agda supports a highly generalised
version of indexed inductive-recursive and inductive-inductive definitions.
This allows to define simultaneously several sets inductively while defining 
simultaneously  recursively functions. 

An important indexed data type is propositional equality $x\;\AgdaDatatype{≡}\;y$
(for $x,y : A$) which has as constructor a proof of reflexivity.
It expresses that propositional equality is the least
reflexive relation (modulo the built-in definitional equality of Agda):

\begin{minipage}{1.0\linewidth}
\EqualityModuleDefEquality
\end{minipage}

\noindent

An example of a record type is as follows:

\renewcommand{\agdaField}{\AgdaKeyword{field}} 
\begin{minipage}{1.0\linewidth}
\studentStudentRecord
\end{minipage} 
\renewcommand{\agdaField}{} 

We will in the following omit in this paper the keyword \AgdaKeyword{field},
and as well the part starting with \AgdaKeyword{open}
(which is needed in order to access the fields without prefixing them
with the name of the record type). 
The part starting with \AgdaKeyword{constructor} is optional, using it we can
introduce an element of Student by writing 
\AgdaFunction{exampleStudent}\;=\;\AgdaInductiveConstructor{student}\;
\AgdaString{"John"}\;\AgdaNumber{123456}.
We can access the fields of a record by writing for instance
\AgdaFunction{exampleStudent}\;\AgdaSymbol{.}\AgdaField{name}.
We can use as well the fields as functions, 
for instance \AgdaField{name}\;:\;\AgdaFunction{Student} \;$\rightarrow$\; \AgdaPostulate{String}.

Agda allows as well to assume the existence of an element of a type without
defining it, for instance  

\begin{minipage}{1.0\linewidth}
\studenthypotheticalStudent 
\end{minipage} 

This is consistent, as long as one could in  principle define an element of that
type. All occurrences of \AgdaKeyword{postulate} in this article are of this 
form. In general \AgdaKeyword{postulate} allows to make inconsistent 
definitions, and needs to be handled with care. An inconsistent example 
postulates the existence of an element of the empty set \AgdaDatatype{Fin}\;\AgdaNumber{0}:

\begin{minipage}{1.0\linewidth}
\lagdaFinnpostulateFalsity
\end{minipage}

\section{A Basic Ledger Based Model of Bitcoin}
\label{sectWarnerLedger}
\subsection{From a Model of a Bank to a Bitcoin Ledger} 
We will in the following repeat briefly the steps taking in the
brown bag talk by Warner 
\cite{warner:BitcoinTechnicalIntroduction:July2011} on how to 
derive a Bitcoin transaction tree starting from a simple model of a bank. 
From this model we will derive in Sect.~\ref{sectLedgerAgda}
a model of the ledger in Agda. In Sect.~\ref{sectTransactionTreesTheoretical} we will 
then adapt this model to take into account that inputs of transactions
are unspent transaction outputs rather than part of the amount contained
in the ledger. However, we will already here 
refine this model by adding coinbase transactions, a block reward function,
and defining the signatures for a transaction.


\begin{figure}[H]
\def\tabularxcolumn#1{m{#1}}
\begin{tabularx}{\linewidth}{@{}cXX@{}}
\begin{tabular}{ccc}
\includegraphics[height=2.5cm]{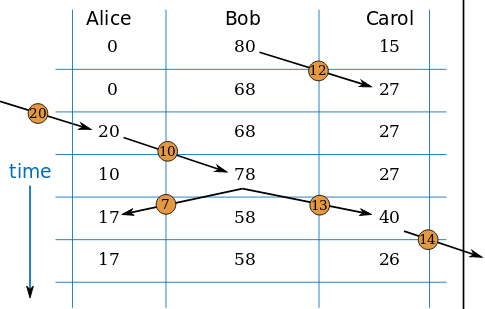}
&\includegraphics[height=2.5cm]{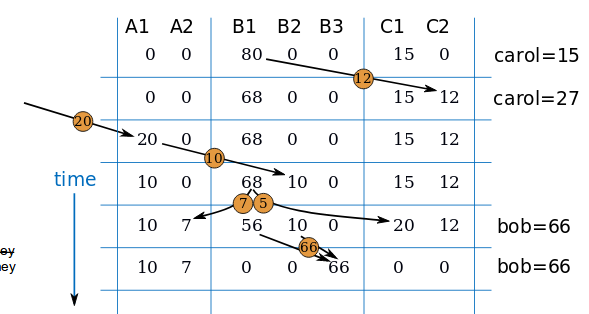}
&\includegraphics[height=2.5cm]{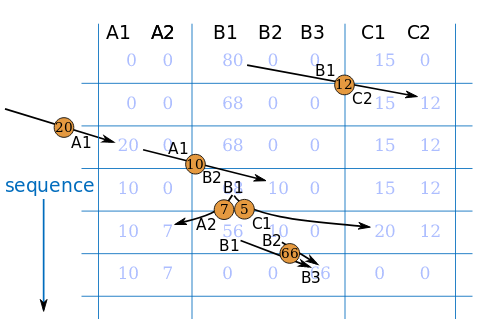}\\
1 (a) Model of a bank&%
1 (b) Replacing Users by&%
1 (c) Replacing Balance Log\\
&Public Keys&by Transaction Log
\end{tabular}
\end{tabularx}
\label{warnerpictures1}
\caption{Source:~\cite{warner:BitcoinTechnicalIntroduction:July2011} (Licensed under creative common license \url{https://creativecommons.org/licenses/by-sa/3.0/})}
\end{figure}

One starts (Fig.~1 (a))
with a simple model of a bank which consists of a
ledger and transaction.
The ledger is a table, which determines for each time (taken
as natural numbers) and account the amount in this account at that time.
Whereas in banks usually transactions are from one account to another account.
this is already at this time generalised to transactions having multiple
inputs and outputs.
There can as well be transactions without inputs,
which corresponds to money created by the central bank. 
In case of Bitcoin, it will
later be replaced by the block reward as given by a coinbase transaction.
Transaction can as well loose money, which means their sum of 
outputs is less than the sum of inputs, which will in the Bitcoin protocol
become transaction fees given to the next miner.\par 
The next step is to hide the account owners behind public/private key
pairs. The recipients of transactions are public keys, whereas the 
transaction needs to be signed  by the private keys of the input accounts. 
Therefore, transactions are now controlled by the 
clients, not by the bank, but bank is still needed as a central server
to control and synchronise the history, and is therefore a single point of
failure. One immediately replaces clients having one key by clients having
multiple keys. This makes it more secret since a big spender can hide
his/her identify behind many small accounts (Fig.~1 (b)). \par
The ledger can be derived from the transactions and the amount associated
with each key at time 0, and therefore only a transaction log is 
needed (Fig.~2 (c)).\par

\subsection{Refining the Model -- Block Rewards, Transaction Fees, Signatures}
\label{subsect:RefiningModel}
We will pause with the exposition by Warner.
However, we will already at this stage add as well the notions of a block,
and block rewards to the model, and discuss signatures and addresses.\par 
The proof of work (``miners' riddle'') will
be added to the model in a followup article.
We will here model the reward given to the miner.
There is a special transaction, called coinbase transaction,
which has no inputs, and where the miner decides about
the outputs. The sum of its outputs
needs to be equal to the sum of the block reward and the transaction fees. 
Transaction fees are the difference
between the amount in the inputs and the outputs of previous
transactions.\par 
At the time of writing, in order for
a transaction to be accepted in the next blockchain, a transaction fee
needs to be added since there is not enough space in a block to cover all
transactions.
The block reward is determined by a function
from time (which measures the number of blocks) to the amounts. 
This function is for Bitcoin the
function which starts at time zero with 50 bitcoins per block,
halves every 52,500 blocks,
and goes to zero from block 6,930,000 onwards.\par 

An address is the hash of a public key. In order to check
whether a message was signed, one needs therefore to provide the 
public key, which hashes to the address in question, and check
that the message was signed by the private key associated with the
public key  (see the discussion that hash actually
means double hashing in 
Subsect.~\ref{subsect:CryptographicAssumptions} below).

We finally discuss the signatures needed for transactions.
For each input, what needs to be signed is not only that 
the originator spends the amount in the given transaction. 
Otherwise, a hacker could easily make a replay attack:
if after a transaction there is still  enough money in the account,
one can create another transaction with the same input and signed by the same
signature,  which then goes to a different output. 
To avoid this, what needs to be signed is a message consisting
of both the input and and all outputs. 
In the Bitcoin protocol, the sender signs only his own input
together with the outputs, not the other inputs 
This way of signing doesn't prevent a replay attack,
as we will see in Sect.\ref{sectTransactionTreesTheoretical}.
There we develop the transaction based tree model which models more
accurately the Bitcoin model and prevents this form of replay attack.

\subsection{Cryptographic Assumptions}
\label{subsect:CryptographicAssumptions}

We will in this paper not implement cryptography, but just postulate
certain functions. We define messages, which can be formed from
natural numbers by using the sum of two messages and list of messages.
We assume a hash function which creates from a message, which 
can be encoded as a natural number, the hash.
(In the real Bitcoin protocol,
different hash functions are applied for different purposes, and usually
two hash functions are applied. By hashing we mean in the
following the application of one or two hash functions,
as it is used in the Bitcoin protocol for that instance). 
When proving properties
in later papers, we assume this function to be injective. This is not
really the case for the real hash functions,
but it is assumed that the probability of having
having a clash is very low. We assume as well a hash function
(again more precisely double-hash function)
from public  keys to addresses, and a predicate
which checks for a message, a public key, and a signature, whether
the message was signed by the private key corresponding to this public key.


\section{The Bitcoin Ledger in Agda}
\label{sectLedgerAgda}
We will in the following develop the model from Sect.~\ref{sectWarnerLedger}
in Agda. 
We start by introducing basic data types defining the \AgdaFunction{Time} (which is the number
of blocks taken up to now), \AgdaFunction{Amount} (amount of bitcoins being used), 
\AgdaFunction{Address}, transaction ids \AgdaFunction{TXID},
signatures \AgdaFunction{Signature}, public keys \AgdaFunction{PublicKey}.
As common in the literature, we use TX for transactions.
All of these types will be equated to the set of natural numbers (we define here
only the first one):\par 

\bitcoinVersFivebasicdef

We define a data type of messages, together with a hash function
which maps messages to natural numbers.
A message can be a natural number
(which might be a time, amount, address etc.), it may consist of two
messages, or it is a list of messages. 
Using this we can then form later a message encoding for instance a transaction
which consists of a list of inputs and outputs, each of which consists
of several components. The set of messages together with a postulated
hash function for messages is given as follows:

\bitcoinVersFiveMsg

We postulate a function which converts public keys into addresses 
(which corresponds in Bitcoin to another application of hash functions):

\bitcoinVersFivePublicKey

We assume a predicate \AgdaPostulate{Signed} which determines, whether a message was signed
by the private key corresponding to a given public key.
We introduce as well a set \AgdaRecord{SignedWithSigPbk}, 
consisting of a signature, a public key, the property that a 
message was signed by it, and that the address is the hashed 
version of the public key.
\AgdaRecord{SignedWithSigPbk} is an example of a record type which depends on 
parameters (here $msg$, $publicKey$ and $s$):

\bitcoinVersFiveSignature

An input or output of a transaction is given by the amount begin received
or sent, and the address of the sender or recipient: 

\bitcoinVersFiveTXField

We create a message for a field consisting of the amount and the address,
and extend it to a list of fields (\AgdaFunction{mapL} is the function
applying a function to each element in a list):

\bitcoinVersFiveTXFieldToMsg

The sum of inputs and outputs is computed using the following function

\bitcoinVersFiveTXFieldToTotalAmount

Here we use the function \AgdaFunction{sumListViaf} which sums up the
result of applying a function to each element of a given list:

\listLibsumListViaf

A transaction without signature consists of a list of inputs and outputs:

\bitcoinVersFiveTXUnsigned

We compute a message for it 

\bitcoinVersFiveTXToMsg

As discussed in Sect.~\ref{sectTransactionTreesTheoretical}, the 
message to be signed for an input consists of the input and the list of
outputs: 

\bitcoinVersFiveTXInputToMsg

The set of signatures for a transaction consists of a signature
for each input using the function \AgdaFunction{txInput2Msg}. 
Note that one needs to provide the public key, which hashes to the 
given address together with a signature for the message created:

\bitcoinVersFiveTXSign

A transaction is an unsigned transaction, for which the sum of
outputs is greater or equal than the sum of inputs, the
list of inputs and outputs is non-empty,  and which is signed
by the private key of the public keys, which in turn hashes to the input addresses:

\bitcoinVersFiveTX

A ledger gives for every time and address the amount associated with that 
address.
A ledger row (for brevity we use in Agda ledger instead of ledger row) is 
therefore a function from addresses to amount. We define as well
the empty ledger where the amount is \AgdaNumber{0}:

\bitcoinVersFiveLedger

We define the result  of updating a ledger by adding the values in a 
transaction, i.e.~the amount for the person given by the address in the field
is incremented by the amount in the field. This operation is then
extended to a list of transaction:

\bitcoinVersFiveaddTXFieldToLedger

The result of subtracting a ledger field is defined in the same way by replacing
\AgdaPrimitive{+} by \AgdaPrimitive{∸}.
Here \AgdaPrimitive{∸} is the cut off subtraction for natural numbers, replacing
negative results of subtraction by zero, e.g.~%
\AgdaNumber{3}\;\AgdaPrimitive{∸}\;\AgdaNumber{2}\; \AgdaSymbol{=} \;\AgdaNumber{1} and
\AgdaNumber{2}\;\AgdaPrimitive{∸}\;\AgdaNumber{3} \;\AgdaSymbol{=}\; \AgdaNumber{0}. We give here only the types of the functions:

\bitcoinVersFivesubtrTXFieldFromLedger

A ledger is updated by a transaction by first subtracting the inputs and then
adding the outputs:

\bitcoinVersFiveupdateLedgerByTX

An input is correct of a ledger is correct w.r.t.~a ledger row, if there is
enough money in the ledger to pay for it:

\bitcoinVersFivecorrectInput

A list of inputs of a ledger is correct if each input is correct, while
updating the ledger after each input, since there might be several inputs
for the same address. This function is then extended to a transaction:

\bitcoinVersFivecorrectInputs

An unmined block (which means there is no correctness check) is a list of transactions:

\bitcoinVersFiveUnminedBlock

The transaction fee of a transaction is the difference
between the sum of the inputs and sum of the outputs of the transaction.
The transaction fee of an unmined block is the sum of transaction fees of its
transactions.

\bitcoinVersFivetxtoTXfee

A block is correct if each transaction is correct, while updating the ledger
after each transaction. Note that Bitcoin allows to have
in one block transactions which make use of the output of another transaction
\cite{bitcoinStackexchange:outputSpentInSameBlock}. We determine as well
the result of updating the ledger after a block:

\bitcoinVersFivecorrectBlock

An unchecked block consists of an unmined block and the output given
to the miner. The miner adds one transaction to a block, 
called coinbase transaction, which
has no inputs, only outputs, such that the sum of outputs is
equal to the sum of the block reward and the transaction fees
of that block. We compute as well the transaction fee of a block:

\bitcoinVersFiveBlockUnchecked

A block is correct, if the unmined block is correct and the output
of the coinbase transaction is equal to the sum of
the block reward and the transaction fees:

\bitcoinVersFiveCorrectMinedBlock

After a block the ledger is updated by updating it using the unmined
block and adding the output of the coinbase transaction:

\bitcoinVersFiveupdateLedgerBlock

An unchecked blockchain is a list of unchecked blocks:

\bitcoinVersFiveBlockChainUnchecked

The correctness of a block depends on the block reward which
is determined by a block reward function, which determines the reward depending
on time (which is the number of blocks).
Assuming a start time and an
initial ledger (which for the final  blockchain will be 
set to \AgdaNumber{0} and \AgdaFunction{initialLedger}), the
correctness of a blockchain is defined as follows:

\bitcoinVersFiveCorrectBlockChain

The ledger at the end of the blockchain is defined as follows:

\bitcoinVersFiveFinalLedger

A blockchain is an unchecked blockchain which is correct :

\bitcoinVersFiveBlockchain

We determine the ledger at the end of a blockchain:

\bitcoinVersFiveblockChaintoFinalLedger

\section{From a Ledger to a Transaction Tree}
\label{sectTransactionTreesTheoretical}
The problem of the ledger is that transactions are not unique.
If for instance a user (as given by an address) has 10 bitcoins, and
transfers 4, somebody can make a replay attack
and repeat the same transaction, deducting 8 bitcoins from that account.
The data for both transactions 
is the same, so the message and signature will be the same. 

One might think that one could prevent this by adding for instance
the time to the message of a transaction. Then transactions in 
different blocks would be different, and therefore their messages
and signatures would be different, and a replay attack is not possible
in a different block. However, Bitcoin allows transactions which
make use of the output of a transaction in the same block
\cite{bitcoinStackexchange:outputSpentInSameBlock}, and therefore
transactions using the same input address should be allowed as well.
Transactions in the same block cannot be distinguished by their block
number; they could only distinguished by their physical time, which is
difficult to verify in a peer-to-peer network.\par 
Therefore in Bitcoin a different approach has been taken: a transaction
doesn't refer just to an address but it refers to the unspent output of a 
previous transaction. Should transactions be unique, this would
make a replay attack impossible. It turns out that this
was originally not the case in the Bitcoin protocol but has been fixed since, 
see the discussion in Sect.~\ref{sectCorrectness}.
Once transactions are unique, and one cannot infer
from a signature of one transaction the signature of another one,
replay attacks are no longer possible.\par 

An example is shown in Fig.~2 (a). In that diagram we have to the left a 
coinbase transaction with outputs 1, 6, 4.
Here the sum of outputs are equal to the input, which is the sum of 
previous transaction fees and block reward. All other transactions have a 
transaction
fee of 1. UTXO are the unspent transaction outputs, which can be used in 
future transactions.

\begin{figure}[H]
\def\tabularxcolumn#1{m{#1}}
\begin{tabularx}{\linewidth}{@{}cXX@{}}
\begin{tabular}{@{}c@{}c@{}c}
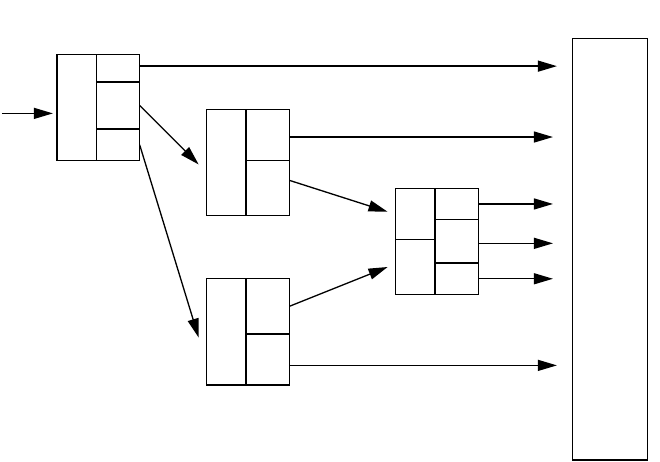
&\includegraphics[height=4.5cm]{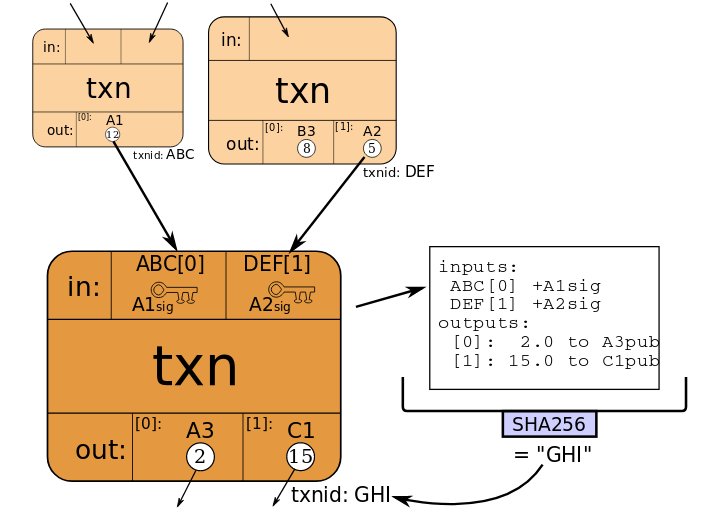}\\
Fig.~2 (a) Transaction Tree&%
Fig.~2 (b) Serialised Transaction\\
&Source:~\cite{warner:BitcoinTechnicalIntroduction:July2011} (License: as Fig.~1)
\end{tabular}
\end{tabularx}
\label{warnerpictures2}
\end{figure}

It follows that one needs to define the transactions, the transaction tree, 
and the
unspent transactions simultaneously. In order to determine the inputs for
coinbase transactions one needs as well to compute the transaction fees. 
This gives rise to an inductive-recursive 
\cite{dybjerjslinductionrecursion,dybjersetzer:2003:indrekjour,dybjerSetzer:indexedInductionRecursion:lncsversion}
definition (in an extended sense): one defines 
inductively the set of transactions and trees formed from
transactions, while recursively simultaneously defining
the list of unspent transaction outputs.\footnote{This is however
not a standard form of induction-recursion, since 
the unspent transactions are lists of transactions, which means
they refer to a set one is defining inductively simultaneously. 
It could be reduced to a simultaneous inductive definition
by defining first transaction trees, transactions, and transaction outputs
simultaneously inductively, without guaranteeing that only unspent 
transaction outputs are used as inputs. Then one could select those
transaction trees, in which only unspent transaction outputs are used. 
However, that would make the model much more difficult to understand and reason about.}
In order to  determine the inputs for coinbase transactions, one needs
to compute as well the set of transaction fees.\par 
When broadcasting transactions, one needs to reference previous transaction
outputs. This is done using Merkle trees \cite{Merkle:MerkleTreeArticle},
see Fig 2 (b): One computes, by recursion over the transaction tree, 
for each transaction an id. The id for a transaction is the
hashed message obtained as follows: for each input one takes
the transaction id of the transaction used together with the number of
the output used,
and the signature. The outputs consist of the amount
and the address of the recipient. All of this is then hashed, resulting
in the transaction id of the transaction, which can be used in future
transactions.

One last detail to be considered is the maturation time. Mining is used
in order to find a consensus which block chain is the correct one. We will 
not elaborate on this consensus protocol here, since we have still to include
mining into our model of Bitcoin. A malicious miner can create a false
blockchain, but that chain is expected to
die out fast, because of the consensus protocol. However, 
for a few blocks it might still be active. In order to avoid
that people spent money from malicious miners, coinbase transactions can only
be used after a certain number of blocks (currently 100) has been mined.
This number is called the maturation time, which we will take 
into account in our model.


\section{Transaction Trees in Agda}
\label{sectTreeLedger}
\subsection{Preliminaries}

The basic set up such as \AgdaFunction{Time}, \AgdaFunction{Amount}, etc, 
\AgdaDatatype{Msg}, \AgdaPostulate{hashMsg}, \AgdaPostulate{publicKey2Address},
\AgdaPostulate{Signed} are as before. We will in this section fix the
\AgdaPostulate{minerReward}. We define as well \AgdaPostulate{blockMaturationTime},
which determines the time after which the reward obtained by miners can
be used. In Bitcoin the value is currently 100. 


\bitcoinVersSixminerReward

The output field of a transaction is the same as the \AgdaRecord{TXField} in 
Sect.~\ref{sectTreeLedger}:

\bitcoinVersSixTXOutputfield

\subsection{Inductive-Recursive Definition of Transaction Tree, Transactions, and
Unspent Transaction Outputs}

We are going to give the inductive-recursive
definition (in an extended sense) of the transactions and unspent transaction outputs.
This is a relatively long simultaneous definition: We 
define inductively the set of transaction trees 
\AgdaDatatype{TXTree},
the transactions \AgdaDatatype{TX}, 
and the set of transaction outputs 
\AgdaRecord{TXOutput}, while recursively defining
the unspent transaction outputs \AgdaFunction{utxo}. In addition
we define simultaneously recursively
auxiliary functions, determining the type of possible transaction
inputs \AgdaFunction{TxInputs} for a transaction, 
the set of transaction outputs \AgdaFunction{tx2TXOutputs} 
of a transaction, and the number of outputs \AgdaFunction{nrOutputs}
of a transaction.\par 
The set \AgdaDatatype{TXTree} is essentially a sequence of transactions,
however the inputs of a transaction refer to previous transactions in 
that sequence. The tree \AgdaInductiveConstructor{genesisTree} is the
starting point of a blockchain, where there are no transactions yet:

\bitcoinVersSixTXTree

A transaction can be a normal transaction with inputs being previous 
transactions, and a coinbase transaction, where there are no inputs, but
we annotate the time (i.e.~the block number to which it belongs):

\bitcoinVersSixTX

We will record the set of unspent transaction outputs, where a
transaction output is given by a transaction belonging to a tree, and an output
number in that tree (since it occurs as part of a simultaneous inductive
definition, it is marked by the keyword \AgdaKeyword{inductive}):

\bitcoinVersSixTXOutput

Simultaneously we define recursively the set of unspent transactions
in a transaction tree, where we give the type now, but the definition later.
We define as well as an auxiliary definition 
\AgdaFunction{utxoMinusNewInputs} which determines the result of
deleting the inputs of a transaction from the list of unspent transaction.

\bitcoinVersSixutxo

The set of inputs of a transaction will be a sublist of the 
set of unspent transaction outputs.
A sublist of a list is given by a sequence  of indices from the original
list, but after each index deleting that element from that list:

\listLibSubList

We define result of deleting a sublist from a list and  a function converting
a sublist into the underlying list:

\listLiblistMinusSubList

The set of inputs will be a sublist of the list of unspent transaction, 
but we need to add as well the
public keys and signatures used. We don't add any checking
of the correctness of the signature -- this will be defined
later as an extra predicate. We first
introduce the notion
\AgdaDatatype{SubList+}, which is a sublist, with some extra information
(of type $Y$) added for each element:

\listLibSubListplus

The subtraction of \AgdaDatatype{SubList+} from a list is defined as before,
while ignoring the extra elements:

\listLiblistMinusSubListplus

The underlying list of a sublist contains as well the extra elements
of the sublist, defined similarly to before, having type:

\listLibsubListplustoList

We define as well two operations on sublists needed later. 
One is\\ \AgdaFunction{listMinusSubList+Index2OrgIndex} mapping indices from the list after subtracting
from it a sublist to indices of the original list:

\listLiblistMinusSubListplustoOrgIndex

The other function takes a sublist and gives a list of all the elements of the
sublist together with their indices in the original list:

\listLibsublistplustoIndicesOriginalList

With these definitions we can define the set of inputs as a sublist
of the set of unspent outputs, but adding a signature and a public key:

\bitcoinVersSixTxInputs

The function \AgdaFunction{utxoMinusNewInputs} deletes,
in case of a normal transaction, 
the inputs from the previous list of unspent transaction outputs,
using the function \AgdaFunction{listMinusSubList+}.
In case of a coinbase transaction, nothing needs
to be deleted:

\bitcoinVersSixutxoMinusNewInputs

The list of new outputs of a transaction to be added to \AgdaFunction{utxo}
is a list of pairs consisting of the new transaction and an index.
The indices are all elements of \AgdaDatatype{Fin}\;n where 
n is the number of outputs of the transaction. 
We first define the function \AgdaFunction{listOfElementsOfFin}\;n,
which returns a list containing all elements of \AgdaDatatype{Fin}\;n:

\listLiblistOfElementsOfFin

The list of outputs is obtained by mapping the elements of this list
to the corresponding output.
In the following
(\AgdaSymbol{λ}\;i \;\AgdaSymbol{→}\; \AgdaInductiveConstructor{txOutput}\;$tr$\;$tx$\;$i$)
denotes the definition of an anonymous function, mapping $i$
to \AgdaInductiveConstructor{txOutput}\;$tr$\;$tx$\;$i$:

\bitcoinVersSixTXOuptuts

The set of unspent transaction outputs is now defined as follows:
For \AgdaInductiveConstructor{genesisTree} it is empty. 
If we add a transaction to the transaction tree, we first
delete the  inputs spent in it and add 
its outputs:

\bitcoinVersSixutxodef

Finally, \AgdaFunction{nrOutputs} determines the number of outputs of a transaction, which is the length of the list of outputs:

\bitcoinVersSixnrOutputs 

This concludes the inductive-recursive definitions of  transaction trees,
transactions and unspent transactions. 

\subsection{Computing Sum of Inputs and Outputs of Transactions and Transaction Fees}

We need to define what it means that the sum of inputs of  transaction
is greater or equal than the sum of its outputs.
The sum of outputs is computed as follows:

\bitcoinVersSixtxtoSumOutputs

The inputs of a transaction can be mapped to a list of \AgdaRecord{TXOutput},
which consists of a transaction and an output number. This can be mapped to a 
\AgdaRecord{TXOutputfield}, which determines the amount being output and the
address of the recipient. The following functions extract
this information from an element of \AgdaRecord{TXOutput}:

\bitcoinVersSixtxOutputtoOutputfield

Now we lift the inputs of transaction to a list consisting of 
a \AgdaRecord{TXOutput}, a public key, and a signature.
We compute as well the sum of outputs of a transaction:

\bitcoinVersSixinputstoPrevOutputsSigPbk

The above will compute the sum of inputs for a normal transaction. For coinbase
transaction there is no explicit input, but the output needs to be equal to the 
transaction fees and the reward for the current block. So we need to compute the 
time of the current block, which determines the block reward.
The time for the current block is updated whenever we have a coinbase transaction.
(The correctness condition for a \AgdaRecord{TXtree} will guarantee that the
times in coinbase transactions are chosen correctly):

\bitcoinVersSixtxTreetoTimeNextTobeMinedBlock

Now we compute for simultaneously recursively the sum of inputs of a transaction
(which is in case of a coinbase transaction the sum of transaction fees and
the block reward) and the transaction fees in the next block to be mined (which is 
reset after a coinbase transaction to zero):

\bitcoinVersSixtxtoSumInputs

Coin transactions can only be used in a transaction after the maturation time
has passed. We therefore define the maturation time of an output of a transaction:

\bitcoinVersSixoutputToMaturationTime

\subsection{Computing Messages to be Signed and Transaction IDs}

We will compute the messages to be signed when an unspent transaction
output is used in a transaction, and the messages and corresponding transaction
ids for transactions. We first compute the massages for an output field and a list
of outputs of a transaction, which is a list of output fields:

\bitcoinVersSixtxOutputfieldtoMsg

We will now define simultaneously the messages and ids of transactions, and the
messages for unspent transactions. The latter is needed to construct
the message for a transaction, since the inputs are using unspent transactions.
One might think that one can compute the messages for unspent transaction outputs
directly from previous transaction ids. However, such a definition does not
termination check in Agda, since there is no structural reason that those
previous transaction ids were computed before the current transaction.
By defining simultaneously with the transaction messages and ids the messages
for unspent transaction, we overcome this problem. Later, one can show
that the messages of the unspent transactions are indeed equal to the combination
of the transaction id and output number.

The message for a normal transaction consists of the message for the outputs,
as defined before, and the message for the inputs. The message for the inputs
is computed by first computing indices in the list of unspent transactions
together with public keys and signatures, obtained from the inputs.
Then we construct from each input the message consisting of the message for
the corresponding unspent transaction, the public key and the signature.
Note that at this moment we haven't defined what it means for a transaction
to be correctly signed. 

For coinbase transactions the message consists
of the time and the message for the outputs. 
The fact that we include the time is a fix taken in Bitcoin
to guarantee that transactions are unique. This will be discussed in 
Subsect.~\ref{subsect:CorrectnessCorrectness}. 
The computation of the message for a transaction is as follows
(the part starting with \AgdaKeyword{where} starts auxiliary local definitions):\par 

\bitcoinVersSixtxtoMsg

We obtain the id of a transaction:

\bitcoinVersSixtxtoid

In order to compute the messages for the unspent transactions,
we compute first the messages for the unspent transactions
of the previous transaction tree  after deleting 
the inputs for the new transaction:

\bitcoinVersSixutxoMinusNewInputstoMsg

The new list of unspent transaction is the concatenation of two lists.
We define a function which, if one has two lists and functions mapping the 
indices of those
two lists to the elements in those lists, 
computes the indexing function which does the
same for the concatenation of both lists. It has type

\listLibconcatListIndextoOriginIndices

We use this function for calculating the messages for unspent transaction
by calculating these functions for the result of omitting the inputs from the
unspent transactions and for the new outputs of a transaction.
In the following definition there occur two more Agda features not discussed
before: The occurrence of () after \AgdaInductiveConstructor{genesisTree}
denotes the empty case distinction. 
\AgdaInductiveConstructor{genesisTree} has no unspent transactions,
therefore (\AgdaFunction{utxo}\;\AgdaInductiveConstructor{genesisTree})
is empty, and 
(\AgdaDatatype{Fin}\;(\AgdaFunction{length}\;(\AgdaFunction{utxo}\;\AgdaInductiveConstructor{genesisTree}))) is the empty set. So the case distinction
on this set is empty, and denoted by  (). 
Furthermore, we have an occurrence of 
\AgdaKeyword{module}\;\AgdaModule{utxo2Msgaux}\;\AgdaKeyword{where},
which is like a \AgdaKeyword{where} clause, but allows to reference
the local variables by using for instance
\AgdaFunction{utxo2Msgaux.f₁} for denoting \AgdaFunction{f₁}.
The definition of \AgdaFunction{utxo2Msg} is as follows:\par

\bitcoinVersSixutxotoMsg 

The message to be signed by the owner of the address in a transaction output
consists of the input  of the transaction and all the outputs of the transaction.
Interestingly, in Bitcoin only the current input is used, not the other ones.
Furthermore, for the input we cannot include the signature in the
message to be signed, because that signature is created from the message to be 
signed. So the input to be signed consists of 
the transaction id for that output, the output number, and the address
in that output:

\bitcoinVersSixmsgToBeSignedByInput

In order to define the correctness of a tree, we define first 
an operation expressing that for all elements of a list a given property P holds:

\listLibforallInList

Now we define the correctness of Transaction. 
A normal transaction is correct, if the
inputs and outputs are non-empty,
the sum of inputs is greater or equal to the sum of outputs,
for all inputs used the maturation time has passed, 
and for all inputs the public key given hashes to the address of the 
unspent transaction output  being used, and the signature given is 
actually a signature of the message to be signed for that input.
A  coinbase transaction is correct, if 
the outputs are non empty, the sum of outputs
is equal to the transaction fees obtained and the block reward, and
the time is the time of the currently to be mined block.
In the following definition 
\AgdaSymbol{λ}\AgdaSymbol{\{}($outp$\;\AgdaInductiveConstructor{,}\;
$pbk$\;\AgdaInductiveConstructor{,}\;
$sign$\AgdaSymbol{)}\AgdaSymbol{→} $\cdots $ \AgdaSymbol{\}}
denotes a local function definition, where we pattern match
on the argument, obtaining that it is of the form
\AgdaSymbol{(}$outp$\;\AgdaInductiveConstructor{,}\;
$pbk$\;\AgdaInductiveConstructor{,}\;
$sign$\AgdaSymbol{)}.

\bitcoinVersSixCorrectTX

A transaction tree is correct if all its transactions are correct,
and we define as well the notion of a correct transaction tree:

\bitcoinVersSixCorrectTxTree

We define as well the notion of a correct transaction

\bitcoinVersSixTXCorrect

We define operations for defining the genesis tree as a correct
transaction tree and adding a correct transaction to it. Alternatively,
we could as
well have defined \AgdaRecord{TXTreeCorrect} by having these operations
as constructors together with functions mapping it to the underlying 
transaction trees and correctness proofs:\par 

\bitcoinVersSixaddTxtreeCorrect

\subsection{Merkle Trees}
\label{subsect:MerkleTrees}
A Bitcoin transaction is not referring directly to previous transactions, but 
to the ids of those transactions. 
This form of forming ids
for the nodes of a tree, by 
determining for each node recursively first the hashes of the ids of its 
subtrees, 
and then  hashing the resulting data in order to obtain an
id for the node, is called a Merkle tree.
We call transactions which refer to transaction ids in the following
Merkle transactions. The outputs of a Merkle transaction are as before,
but the inputs are elements of \AgdaRecord{MerkleInputField}: 

\bitcoinVersSixMerkleInputField

Merkle transactions are coinbase transactions (which are the same as before)
or standard transactions, where the inputs are elements of
\AgdaRecord{MerkleInputField}: 

\bitcoinVersSixMerkleTX

We define a map from inputs of a transaction referring to a transaction
tree to Merkle inputs:

\bitcoinVersSixtxOutputsSignedtoMerkleInput

A Merkle transaction corresponds to a normal transaction, if
the inputs and outputs are the same, modulo the maps defined before:

\bitcoinVersSixMerkleTXCorrespondstoTX

A correct Merkle transaction is a Merkle transaction which corresponds to a 
correct transaction:

\bitcoinVersSixMerkleTXCor

Finally we show that by adding correct Merkle transaction to a correct
transaction tree we obtain a correct transaction tree:

\bitcoinVersSixupdateTXTree

What needs to be shown in the correctness proof, which will be presented
in a follow up paper, is that for a correct Merkle tree, the message
obtained from the Merkle tree is the same as the message of the 
corresponding \AgdaRecord{TX}, that this message is unique, and that we can 
actually
decide whether a Merkle transaction is correct.\par

\section{Correctness and Limitations of the Model}
\label{sectCorrectness}
\subsection{Limitations}

While we have dealt with coinbase transaction, including transaction fees,
block rewards, and maturation times, we haven't formalised mining yet. 
It should be relatively easy to define the message of a block
and define abstractly the proof of work (``the miner's riddle''), 
and even the adjustment
of difficulty should be relatively easy. However, formalising the consensus
protocol, which guarantees that the chain with the main computational power 
wins, would be more challenging.

At the moment outputs are outputs to one single recipient. Bitcoin
allows the outputs to be proper scripts. When using an output
one needs to provide an input script which if concatenated with the output
script results in the truth value true. Formalising the script language
and adding input and output scripts to our model 
is probably not very difficult, since the script language of Bitcoin is very simple.

\subsection{Correctness and Decidability}
\label{subsect:CorrectnessCorrectness}
In this article we have provided a basic model of Bitcoin, but we haven't
proved any correctness or decidability properties.\par 
If one starts with decidable cryptographic functions, all properties
used in this article should be decidable. Proving it in Agda might be
more challenging, especially in the Subsect.~\ref{subsect:MerkleTrees},
where one needs to determine for a Merkle transaction a corresponding
normal transaction it belongs to -- a pen and paper proof is relatively
easy (check whether we have in the list of unspent
transaction outputs those having the transaction id and output number needed).

Regarding correctness, it should be relatively easy to prove a simple
property, namely that the sum of unspent transaction outputs is equal to the
sum of the bitcoins mined. What is more challenging are the following properties:
\begin{itemize} 
\item The signatures for transactions are unique. This requires
the assumption that the hash functions are 
injective. Although this doesn't hold,  the 
probability of a clash is very low.
\item If a transaction had an output, and this output is not
used in a different transaction, it is in the unspent transaction list.
\item If a transaction output has been spent, it cannot be used  again.
\item Any normal transaction input was the output of a transaction.
\end{itemize} 
All these properties rely on the most important property:
\begin{itemize} 
\item Prove that transactions messages and ids are unique.
\end{itemize} 
All the above properties are provable on pen and paper. Proving them
in Agda will take some time.

We want to note that originally in the Bitcoin protocol
time (which corresponds to the block number) was not included in
the message for coinbase transactions, and therefore
identical coinbase transactions could be created, namely by having the same
outputs. It actually occurred
\cite{reddit:WhatIsATXID,github:HandleHistoricTransactionsWithDuplicateIDs}.
The duplicate messages were the coinbase transaction in blocks 91842 and 91812
\footnote{The first (coinbase) transactions in the following blocks coincide:\\
\url{https://blockexplorer.com/block/00000000000743f190a18c5577a3c2d2a1f610ae9601ac046a38084ccb7cd721}
\url{https://blockexplorer.com/block/00000000000271a2dc26e7667f8419f2e15416dc6955e5a6c6cdf3f2574dd08e}}
and  as well in blocks 91880 and 91722\footnote{The coinbase transactions in the following blocks coincide:\\
\url{https://blockchain.info/block-index/106662}\\
\url{https://blockchain.info/block-index/106692}}
As a fix the height of the transaction 
was required to be added in the optional part of a coinbase transaction
\cite{github:bip-0034.mediawiki}.\par

p
\section{Conclusion}
\label{sectConclusion}
We have reviewed how to obtain from a model of a bank first a model of
the blockchain based on a simple ledger, and then in a second step
a model of the blockchain based on transaction trees. Both models
have been modelled in Agda, where we included transaction fees, block rewards,
signatures, and transactions based on Merkle trees.
We saw that transactions, transaction trees and unspent transaction outputs
form an extended form of an inductive-recursive definition. \par 
When studying descriptions of blockchain technology, 
we found that when it comes to transactions referring to previous unspent
transactions, often the material is not presented in a clear way.
It seems that the underlying reason is the fact that one needs
to deal in some form with the underlying transaction tree.
This definition is inductive-recursive  in nature, and therefore
difficult to understand.  We hope that this article will help to 
obtain a better understanding of the underlying transaction trees.

\myparagraph{What is a bitcoin?} This is a frequently asked question. 
The correct question is what bitcoins are. If we look at our model we see
that the money which can be used are the unspent transaction outputs
which can be computed from the current blockchain. So the bitcoins
are the unspent transaction outputs of the blockchain, 
which can be computed from it, where the blockchain is stored in a 
peer-to-peer network with consensus obtained by the mining protocol.

\myparagraph{Future work.} We have already indicated
some aspects still be to be added (mining, input and output scripts in
transactions), and indicated what needs to be shown in order to guarantee
that the protocol is correct and to prove decidability.
Future work would be to use our work with Abel and Adelsberger
on writing user interfaces in Agda \cite{ooAgda:Library,JFP:ooAgda}
to write a simulator for Bitcoin in Agda. What would be 
very challenging is to write a cryptocurrency in Agda, where 
cryptographic functions make use of the foreign language interface to
Haskell available in Agda. One could think of having a language
for smart contracts as well, which are verified in Agda.
This would allow to have a cryptocurrency
in which correctness is proved in the same language as the implementation,
and which allows to define verified smart contracts.



\bibliography{refstypes2017main}

\end{document}